%% file: WJHJ18_TWC_practical_aspects.tex
\PassOptionsToPackage{}{graphicx}
\PassOptionsToPackage{usenames,dvipsnames,table}{xcolor}
\documentclass[onecolumn,12pt,draftcls]{IEEEtran}  % IEEE arxiv Style: Transaction of Signal Processing 

\usepackage{amsmath}
\usepackage{etex} 

\usepackage{amssymb}
\usepackage{amsxtra,amsthm}
\usepackage{wasysym}% lightning                                          
\usepackage{bbold}
\usepackage{amsfonts} 
\usepackage{mathdots}
\usepackage{mathtools}
\usepackage{thmtools}
\usepackage{pdfcolmk}
\usepackage[pdftex]{graphicx}                                   % Eps Grafiken für PostScript Drucker
\usepackage{xcolor}
\usepackage{xifthen}
\usepackage{algorithm}
\usepackage{algpseudocode}
\usepackage[framed,numbered,autolinebreaks]{mcode} % needs to be manual copied in root folder

\usepackage{mathrsfs}
\usepackage[utf8]{inputenc}                                         % -> damit man sie auch gleich im Code eingeben kann
\usepackage[T1]{fontenc}
\usepackage[pdfborder={0 0 0},colorlinks,citecolor=blue,hyperindex=true,hypertexnames=false,extension=pdf]{hyperref} 
\usepackage{subcaption} 
\usepackage{verbatim}
\usepackage{float}
\usepackage{srcltx}
\usepackage[english]{babel}

\usepackage[shortlabels]{enumitem}
\usepackage[all]{xy}

\usepackage{epstopdf}   % convert eps to pdf files automatically. Eps file will remain.
\usepackage{rotating}
\usepackage{tocloft}
\usepackage{import}

\usepackage[style=ieee,    % https://www.ctan.org/tex-archive/macros/latex/contrib/biblatex-contrib/biblatex-ieee
            firstinits=true, % vornamen = erste buchstabe
            url=false,
            backend=bibtex,
            doi=false,
            maxbibnames=99,
            isbn=false,
            natbib=true,
            hyperref=true,
            bibencoding=utf8]{biblatex}
\AtEveryBibitem{\clearfield{note}}   % damit mein komentare nicht geprintet werden.
\DefineBibliographyStrings{english}{%
  references = {},% replace "references" with "bibliography"  for `book`/`report`
}

\def\pdfdir{} 
\ifthenelse{\isundefined{\iftikz}}{
\newif\iftikz\tikzfalse
}{
\tikzfalse
}

\include{Header_Philipp}

\newcommand{\seefor}[1]{\!}
\newcommand{\seeintern}[1]{\!}
\newcommand{\subfigref}[1]{(\subref{#1})}

\newcommand{\pilot}{\ensuremath{\text{Pilot}}}
\newcommand{\guard}{\ensuremath{\text{Guard}}}

\newsavebox\cpsave

\renewcommand*{\arraystretch}{0.7}

\DeclareUnicodeCharacter{FB01}{fi}

\ifthenelse{\isundefined{\ifdetails}}{
\newif\ifdetails\detailsfalse
}{\detailsfalse}

\ifthenelse{\isundefined{\ifold}}{
\newif\ifold\oldfalse
}{\oldfalse}

\ifthenelse{\isundefined{\ifalt}}{
\newif\ifalt\altfalse
}{\altfalse}

\ifx\unsure\undefined
 \newcommand{\unsure}[1]{}
\else
 \renewcommand{\unsure}[1]{} 
\fi
\ifx\wrong\undefined
  \newcommand{\wrong}[1]{}
\else
  \renewcommand{\wrong}[1]{}
\fi
  
\ifx\unsurenolist\undefined
  \newcommand{\unsurenolist}[1]{}
\else
  \renewcommand{\unsurenolist}[1]{}
\fi

\ifx\unsurenote\undefined
  \newcommand{\unsurenote}[2]{}
\else
  \renewcommand{\unsurenote}[2]{}
\fi

\ifdetails
  \newcommand{\detail}[1]{\color{brown}{#1}\color{black}}
\else
\newcommand{\detail}[1]{}
\fi
\newtheorem{thm}{Theorem}

\ifx\definition\undefined
\newtheorem{definition}{Definition}         %%%%%%%%%%%%%%%%%%%%%%%%%%%%%%
\fi
\ifx\corrolary\undefined
\newtheorem{corrolary}{Corrolary} %%%%%%%%%%%%%%%%%%%%%%%%%%%%%% 
\fi
\ifx\conjecture\undefined
\newtheorem{conjecture}{Conjecture}         %%%%%%%%%%%%%%%%%%%%%%%%%%%%%%
\fi
\ifx\thm\undefined
\newtheorem{thm}{Theorem}         %%%%%%%%%%%%%%%%%%%%%%%%%%%%%%
\fi
\ifx\lemma\undefined
\newtheorem{lemma}{Lemma}
\fi
\ifx\question\undefined
\newtheorem{question}{\color{red}{Question}}         %%%%%%%%%%%%%%%%%%%%%%%%%%%%%%
\fi
\ifx\hypothesis\undefined
\newtheorem{hypothesis}{Hypothesis}
\fi

\ifx\remark\undefined
\newenvironment{remark}{\par\vspace{1.5ex}\noindent{\em Remark\/}.}{\par\vspace{1.5ex}}
\fi

\ifx\example\undefined
\newcounter{theexample}
\newenvironment{example}{\par\vspace{1.5ex}\noindent{\em Example\/ }\arabic{theexample}.
\stepcounter{theexample}}{\par\vspace{1.5ex}}
\fi
\ifx\approach\undefined
\newcounter{theapproach}
\newenvironment{approach}[1][]{\par\vspace{1.5ex}\noindent{\em Approach\/ }\arabic{theapproach}. #1
\stepcounter{theapproach}}{\par\vspace{1.5ex}}
\fi

\altfalse
\oldtrue

\usepackage{scalerel} %https://tex.stackexchange.com/questions/109947/how-do-you-get-a-scriptscriptstyle-sized-prime 
\usepackage{multirow}
\def\minus{%
  \setbox0=\hbox{-}%
  \vcenter{%
    \hrule width\wd0 height \the\fontdimen8\textfont3%
  }%
}
\newif\iflong\longfalse
\newif\ifextras\extrasfalse % extras, not on arxiv version and not on Wcom journal

\newif\ifwrong\wrongfalse
\newif\ifall\allfalse
\newif\ifarxiv\arxivfalse
\newif\ifphilipp\philippfalse
\newif\ifcode\codefalse % if matlab code for ACPC implementation

\begin{document}
    %
    % paper title:can use linebreaks \\ within to get better formatting as desired
    %

  \title{MOCZ for Blind Short-Packet Communication: Some Practical Aspects}

    % author names and affiliations
    % use a multiple column layout for up to three different
    % affiliations
  \author{
    \IEEEauthorblockN{Philipp Walk\IEEEauthorrefmark{1}, Peter Jung\IEEEauthorrefmark{2}, 
    Babak Hassibi\IEEEauthorrefmark{3}, and Hamid Jafarkhani\IEEEauthorrefmark{1}\\}
    \IEEEauthorblockA{\IEEEauthorrefmark{1}Dept. of Electrical Engineering \& Computer Science, UCI, Irvine, CA 92697\\}
    \IEEEauthorblockA{\IEEEauthorrefmark{2}Communications \& Information Theory, TU Berlin, 10587 Berlin\\}
    \IEEEauthorblockA{\IEEEauthorrefmark{3}Dept. of Electrical Engineering, Caltech, Pasadena, CA 91125\\}
    {\small Emails: \{pwalk,hamidj\}@uci.edu, peter.jung@tu-berlin.de,  hassibi@caltech.edu}} 

\maketitle   % make the title area

\begin{abstract}
    We will investigate practical aspects for a recently introduced blind (noncoherent) communication scheme, called modulation on
    conjugate-reciprocal zeros (MOCZ), which enables reliable transmission of sporadic and short-packets at ultra-low
    latency in unknown wireless multipath channels, which are static over the receive duration of one packet.  Here
    the information is modulated on the zeros of the transmitted discrete-time baseband signal's $z-$transform.  Due to
    ubiquitous impairments between transmitter and receiver clocks a carrier frequency offset (CFO) will occur
    after a down-conversion to the baseband, which results in a common rotation of the zeros. To identify fractional rotations of the
    base angle in the zero-pattern, we propose an oversampled  direct zero testing decoder to identify the most likely
    one.  Integer rotations correspond to cyclic shifts of the binary message, which we compensate by a cyclically
    permutable code (CPC).  Additionally, the embedding of CPCs into cyclic codes, allows to exploit additive error
    correction which reduces the bit-error-rate tremendously. Furthermore, we use the trident structure in the
    signals autocorrelation to estimate timing offsets and the channels effective delay spread. We
    finally demonstrate how these impairment estimations can be largely improved by receive antenna diversity, which
    enables extreme bursty reliable communication at low latency and SNR.
\end{abstract}

\section{Introduction}

We introduced in \cite{WJH18a,WJH18b} a novel blind (noncoherent) communication scheme for the physical layer, called
modulation on conjugate-reciprocal zeros (MOCZ), to reliably transmit sporadic short-packets of fixed size over unknown
wireless multipath channels with bandwidth $W$ at an incredible low-latency.  Here the information of the packet is
modulated on the zeros of the transmitted discrete-time baseband signal's $z-$transform. We will call the discrete-time
baseband signal a \emph{MOCZ symbol}, similar to an orthogonal frequency division multiplexing (OFDM) symbol, which is a
finite length sequence of complex-valued coefficients. These coefficients will then modulate a continuous-time pulse
shape at a sample period of $T=1/W$ to generate the continuous-time baseband waveform. Since the
MOCZ symbols (sequences) are neither orthogonal in time nor frequency domain, the MOCZ design can be seen as a non-orthogonal
multiplexing scheme.  
After up-converting to the desired carrier frequency, the transmitted passband signal will propagate in space such that,
due to reflections, diffractions, and scattering, different delays of the attenuated signal will interfere at the
receiver. Hence, multipath propagation causes a time-dispersion which results in a frequency-selective fading channel
\cite{TV05}. 
Due to ubiquitous impairments between transmitter and receiver clocks a \emph{carrier frequency offset}
(CFO) will be present after a down conversion to the baseband. Doppler shifts due to relative velocity causes additional
frequency dispersion which can be also approximated in first order by a CFO.  This is a known weakness in many
multi-carrier modulation schemes, such as OFDM \cite{TV05,Moo94,ZGX10,LLTC04}, and various approaches have been
developed to estimate or eliminate the CFO effect.  A common approach for OFDM systems is to learn the CFO in a training
phase or from blind estimation algorithms, such as MUSIC \cite{LT98} or ESPRIT \cite{TLZ00}.  Furthermore, due to the
unknown distance and asynchronous transmission, a \emph{timing offset} (TO) of the received symbol has to be determined
as well, which will otherwise destroy the orthogonality of the OFDM symbols \cite[5.1]{CKYK10},\cite{PKPKKH06}. By
``sandwiching'' the data symbol between two training symbols a timing and frequency offset can be estimated
\cite{SC97},\cite{SC96}.  By using antenna arrays at the receiver, antenna diversity of a single-input-multiple output
(SIMO) system can be exploited to improve the performance \cite{ZGX10}.

Whereas OFDM is typically used in long frames, consisting of many successive OFDM symbols and hence much longer signal
lengths, we consider here only one single symbol transmission with a very short signal length. This places high demands
on such a bursty signaling scheme, since timing and carrier frequency offsets have to be addressed from only one
received symbol. Here our MOCZ scheme will be a promising solution.   Since any communication will be scheduled and
timed on the MAC layer by a certain bus, running with a known bus clock-rate, timing-offsets of the symbols can be
assumed as fractions of the bus clock-rate.  We will introduce here an improved receiver design for a coded binary MOCZ
(BMOCZ) scheme
and demonstrate by bit-error-rate (BER) simulations the robustness against these impairments.  

In the MOCZ design, a CFO will result in an unknown common rotation of all received zeros. Since the angular zero
spacing in a BMOCZ symbol of length $K+1$ is given by a base angle of $2\pi/K$, a fractional rotation can  be
easily obtained at the receiver by an oversampling during the post-processing to identify the most likely transmitted
zeros (zero-pattern).  
Rotations, which are integer multiples of the base angle, correspond to cyclic shifts of the binary message word. By
using a \emph{cyclically permutable code} (CPC) for the binary message, the BMOCZ symbol becomes invariant against any
cyclic shift and hence against any CFO.  This prevents any further symbol transmissions for estimating the CFO, which
will reduce overhead, latency, and complexity. As a byproduct, this has the appealing feature of providing a CFO
estimation from the decoding process of a single BMOCZ symbol.  Furthermore, due to the embedding into
a cyclic code, such as BCH codes, we can use their error correction capabilities to improve the BER and moreover the \emph{block
error-rate} (BLER) performance
tremendously. 
By measuring the energy  of the expected symbol length with a sliding window in the received signal, we can identify
arbitrary TOs at the receiver. We will show the robustness of the TO estimation analytically, which reveals another
strong property of the MOCZ design. 

At last, we will combine CFO and TO with error correction over multiple receive antennas and demonstrate antenna
diversity of the SIMO system. By simulating BER over the received SNR for various average
power delay profiles, with constant and exponential decay as well as random sparsity constraints, we will demonstrate the
performance in various indoor and outdoor scenarios by using the simulation framework Quadriga \cite{JRBT14}.  

\subsection{Notation} We will use small letters for complex numbers in $\C$.  Capital Latin letters denote natural
numbers $\N$ and refer to fixed dimensions, where small letters are used as indices.  Boldface small letters denote row
vectors and capitalized letters refer to matrices. Upright capital letters denote complex-valued polynomials in $\C[z]$.
We will denote the first $N$ natural numbers in $\N$ as $[N]:=\{0,1,\dots,N-1\}$. For $K\in\N$ we denote by
$K+[N]=\{K,K+1,\dots K+N-1\}$ the $K-$shift of the set $[N]$. The Kronecker-delta symbol is given by $\del_{nm}$ and is
$1$ if $n=m$ and $0$ else. For a complex number $x=a+\im b$, given by its real part $\Re(x)=a\in\R$ and imaginary part
$\Im(x)=b\in\R$ with imaginary unit $\im=\sqrt{-1}$, its complex-conjugation is given by $\cc{x}=a-\im b$ and its
absolute value by $|x|=\sqrt{x\cc{x}}$.  For a vector $\vx\in\C^N$ we denote by $\cc{\vx^-}$ its complex-conjugated
time-reversal or \emph{conjugate-reciprocal}, given as $\cc{x_k^-} = \cc{x_{N-k}}$ for $k\in[N]$. We use
$\vA^*=\cc{\vA}^T$ for the complex-conjugated transpose of the matrix $\vA$. For the $N\times N$ identity  matrix we
write $\id_N$ and for a $N\times M$ matrix with all elements zero we write $\vzero_{N,M}$.   By $\vD_{\vx}$ we refer to
the diagonal matrix generated by $\vx\in\C^N$.  The $N\times N$ unitary Fourier matrix $\Fmatrix=\Fmatrix_N$
is given entry-wise by $f_{l,k}=e^{-\im 2\pi lk/N}/\sqrt{N}$ for $l,k\in[N]$. By $\vT\in\R^{N\times M}$  denote the
elementary Toeplitz matrix given element-wise as $\del_{n-1 m}$.  The all one and all zero vectors in dimension $N$ will
be denoted by $\eins_N$ and $\zero_N$, respectively.  The $\ell_p$-norm of a vector $\vx=(x_0,\dots,x_{N-1})\in\C^N$ is
given by $\Norm{\vx}_p=(\sum_{k=0}^{N-1}|x_k|^p)^{1/p}$ for $p\geq 1$.  If $p=\infty$ we write
$\Norm{\vx}_{\infty}=\max_k |x_k|$ and for $p=2$ we set $\Norm{\vx}_2=\Norm{\vx}$.  The expectation of a random variable
$x$ is denoted by $\Expect{x}$.  

\section{System Model and Requirements}

We are interested in a blind and asynchronous transmission of a short {\bfseries single MOCZ symbol} at a designated
bandwidth $W$.  In this ``one-shot'' communication we assume no synchronization and no packet scheduling between
transmitter and receiver. Such extreme sporadic, asynchronous, and ultra short-packet transmissions are required, for
example, in critical control applications, exchange of channel state information (CSI), signaling protocols, secret
keys, authentication, commands in wireless industry applications, or initiation, synchronization and channel probing
packets to prepare for longer or future transmission phases.
By choosing the carrier frequency, transmit sequence length, and bandwidth accordingly, a receive duration in the order
of the channel delay spread can be obtained, which pushes the latency at the receiver to the lowest possible.  Since the
next generation of mobile wireless networks aims for large bandwidths with carrier frequencies beyond $10$Ghz, in the so
called \emph{mmWave} band,  the transmitted signal duration will be in the order of nano seconds. Hence, even at
moderate mobility, the wireless channel in an indoor or outdoor scenario can be considered as approximately
time-invariant over such a short time duration.  On the other hand, wideband channels are highly frequency selective,
which is due to the superposition of different delayed versions (echos) of the transmitted signal at the receiver. This
makes equalizing in time-domain very challenging and is commonly simplified by using OFDM instead. But conventional OFDM
requires an additional cyclic prefix to convert the frequency-selective channel to parallel scalar channels and in
coherent mode it requires additional pilots (training) to learn the channel coefficients. This will increase the latency
for short messages dramatically. 

For a communication in mmWave band massive antenna arrays are exploited to overcome the large attenuation, which
increases the complexity and energy consumption in estimating the huge amount of channel parameters and
becomes the bottleneck in mmWave MIMO systems, especially for mobile scenarios. 
However, in a sporadic communication only one symbol will be transmitted and a next symbol may follow at an unknown time
later. In a random access channel (RACH), a different user may transmit the next symbol from a different location, which
will therefore experience an independent channel realization. Hence, the receiver can barely use any channel information
learned from past communications. OFDM systems approach this by transmitting many successive OFDM symbols as a long
frame, to estimate the channel impairments, which will cause a considerable overhead and latency if only a few data-bits
need to be communicated.  Furthermore, to achieve orthogonal subcarriers in OFDM, the cyclic prefix has to be at least
as long as the channel impulse response (CIR) length, resulting in signal lengths at least twice as the CIR length
during which the channel also needs to be static. Using OFDM signal lengths much longer than the coherence time might be
not feasible for fast time-varying block-fading channels. Furthermore, the maximal CIR length needs to be known at the
transmitter and if underestimated will lead to a serious performance loss. This is in high contrast to our MOCZ design,
where the signal length can be chosen for a single MOCZ symbol independently from the CIR length. 
%
%Let us stress at this point, that the linear-time invariance assumption of the channel is only for the reception time
%of one MOCZ symbol, which is defined by the maximal channel delay $LT$ and symbol duration $(K+1)T$. Since the sampling
%rate $T=1/W$ is defined by the bandwidth, wideband channels will have extreme short symbol duration,  and 
%
The goal in this work is to address the ubiquitous impairments of the MOCZ design under such ad-hoc communication
assumptions and signal lengths in the order of the CIR length.% Let us first formulate the channel model.

After up-converting the MOCZ symbol, which is a discrete-time complex-valued baseband signal
$\vx=(x_0,x_1,\dots,x_K)\in\C^{K+1}$ of two-sided bandwidth $W$, to the desired carrier frequency $f_c$, the transmitted
passband signal will propagate in space. Regardless of directional or omnidirectional antennas, the signal will be
reflected and diffracted at point-scatters, resulting in different delays of the attenuated signal which interfere at
the receiver if the maximal delay spread $T_d$ of the channel is larger than the sample period $T=1/W$. Hence,
the multipath propagation causes  time dispersion resulting in a frequency-selective fading channel. 
Due to ubiquitous impairments between transmitter and receiver clocks an unknown \emph{frequency offset}
$\Deltaf$ will be present after the down-conversion to the received continuous-time baseband signal
\begin{align}
  \tilr(t)=r(t)e^{\im 2\pi t\Deltaf}.
\end{align}
By sampling $\tilr_n=\tilr(nT)$ at the sample period $T$, the received discrete-time baseband signal can be represented
by a \emph{tapped delay line} (TDL) model. Here the channel action is given as the convolution of the MOCZ symbol $\vx$
with a finite impulse response $\{h_\ell\}$, where the $\ell$th complex-valued channel tap $h_\ell$  describes the
$\ell$th averaged path over the bin $[\ell T,(\ell+1)T)$, which we model by a circularly symmetric Gaussian random
  variable in $\CN(0, s_{\ell}p^{\ell})$ for $l\in[L]$ and zero elsewhere.  The average \emph{power delay profile} (PDP)
  of the channel can be sparse and exponentially decaying, where $s_{\ell}\in\{0,1\}$ defines the sparsity pattern of
  $S=|\supp{\vh}|=\sum_{\ell=0}^{L-1} s_{\ell}=\Norm{\vs}_1$ non-zero coefficients and $\pdp\leq 1$ the exponential
  decay rate.
%such that the average power of the $\ell$th channel tap is $\Expect{|h_{\ell}|^2}=s_{\ell}  p^{\ell}$.
% 
To obtain equal average transmit and average receive power we will eliminate in our analysis the overall channel gain by
normalizing the CIR realization $\vh=(h_0,\dots,h_{L-1})$ by its average energy $\sum_{l=0}^{L-1} s_lp^l$ (for a given
sparsity pattern), such that $\Expect{\Norm{\vh}^2}=1$.  The convolution output is then additively distorted by Gaussian
noise $w_n\in\CN(0,N_0)$ of zero mean and variance (average power density) $N_0$ for $n\in\N$ as
\begin{align}
  \begin{split}
  \tilr_{n} %= e^{\im 2\pi n\frac{\Deltaf}{W}} r_n
  &= e^{\im n\phi} \sum_{k=0}^{K} x_{k} h_{n-\toff-k} \!+\! e^{\im n\phi}
  w_n %= e^{jn\phi}(\vx*\vh)_{n-\tau_0} + \tw_n\\
  =\sum_{k=0}^K e^{\im k\phi}x_k e^{\im (n-k)\phi}h_{n-\toff-k} \!+\! \tw_n %=(\vtx*\vth)_{n-\toff}+\tw_n. 
  =\sum_{k=0}^K \tx_k\tilh_{n-\toff-k} \!+\! \tw_n.
\end{split}\label{eq:receivedsamples}
\end{align}
Here $\phi=2\pi\Deltaf/W \mod 2\pi\in[0,2\pi)$ denotes the \emph{carrier frequency offset} (CFO) and $\tau_0\in\N$ the
  \emph{timing offset} (TO), which marks the delay of the first symbol coefficient $x_0$ via the first channel path
  $h_0$, measured in integer multiples of the sample time $T$.  The modulated MOCZ symbol $\vtx\in\C^K$ will have
  rotated coefficients $\tx_k=e^{jk\phi}x_k$ as well as the channel $\tilh_{\ell}=e^{\im (\ell+\tau_0)\phi}h_{\ell}$,
  which will be also effected by a \emph{global phase} $\tau_0\phi$.  Since the channel taps have a uniform independent
  phase the distribution does not change.  By the same argument, the Gaussian noise distribution is not alternated by
  the phase, hence we have $\tilde{w}_n\in\CN(0,N_0)$ for any $n$ and $\phi$.

In \cite{WJH18b,WJH18a} a good signal-codebook is given for Binary MOCZ (BMOCZ) for the set of normalized \emph{Huffman
sequences} $\Code(R,K)=\set{\vx\in\C^{K+1}}{\va(R,K)=\vx*\cc{\vx^-},x_0>0}$, i.e., by all $\vx\in\C^{K+1}$ with positive
first coefficient%
%
%\footnote{Note, the trivial ambiguity of the autocorrelation is a global phase $e^{\im \phi}$ which we set to
%$\phi=0$.}
%
and ``impulsive-like`` autocorrelation \cite{HUf62}, given by
\begin{align}
  \va=\va(R,K)=\vx*\cc{\vx^-} =(-\eta,\zero_{K},1,\zero_K,-\eta) 
  \quad \text{with}\quad\eta=1/(R^K+R^{-K})
  \label{eq:huffmanauto}
\end{align}
for some $R>1$. The absolute value of \eqref{eq:huffmanauto} forms a \emph{trident} with one main peak at the center,
given by the energy $\Norm{\vx}^2=1$, and two equal side-peaks of $\eta\in[0,1/2)$, see
  \figref{fig:HuffmanTridentFrame}.  From an analytical and empirical investigation \cite{WJH18a}, the BMOCZ symbols are
  most robust against noise if
\begin{align}
  R=R(K)=\sqrt{1+\sin(\pi/K)}>1  \quad,\quad K\geq 2.
  \label{eq:optimalR}
\end{align}
Hence, the BMOCZ codebook (constellation set) $\Code=\Code(K)$ is only determined by the number $K$.  Each BMOCZ symbol
(constellation, Huffman sequence) defines the coefficients of a polynomial of degree $K$, where the $K$ zeros are
uniformly placed on a circle of radius $R$ or $R^{-1}$, selected by the message bits $\vm=(m_1,\dots,m_K)\in\{0,1\}^K$
as
\begin{align}
  \uX(z)=\sum_{k=0}^{K} x_k z^k=x_K \Pro_{k=1}^K (z -R^{2m_k-1} e^{\im 2\pi (k-1)/K})=x_K \Pro_{k=1}^K
  (z-\alp_k^{(m_k)})\label{eq:Xzeros}
\end{align}
see also \figref{fig:cfo_fractional}.  Hence, the BMOCZ encoder  is defined iteratively for $q=2,\dots,K$ by its
\emph{zero codeword} $\valp(\vm)=(\alp_1^{(m_1)},\dots,\alp_K^{(m_K)})\in\Zero\subset\C^K$ as
\begin{align}
  \vx_q = (0, \vx_{q-1}) - (R^{2m_q-1} e^{\im \frac{2 \pi q}{K} } \vx_{q-1})
  \quad\text{with}\quad \vx_1=(-R^{2m_1-1}e^{\im \frac{2\pi}{K}},1),
\end{align}
where we normalize after the last iteration step $\vx=\vx_K/\Norm{\vx_K}$.
From the received $N=L+K$ noisy signal samples (no CFO and TO) 
\begin{align}
  y_n=\sum_{k=0}^K x_k h_{n-k} + w_n=(\vx*\vh)_n+w_n,\label{eq:convolutionoutput}
\end{align}
the decoder is given as a \emph{Direct Zero Testing} (DiZeT) of the received polynomial $\uY(z)=\sum_{n=0}^{N-1}y_n z^n$
at the $2K$ possible zero positions as
\begin{align}
  \hm_k = \begin{cases} 
    1,& |\uY(R e^{\im 2\pi \frac{k-1}{K}})| <R^{N-1}|\uY(R^{-1} e^{\im 2\pi \frac{k-1}{K}})|\\
    0,& \text{else}
  \end{cases}\label{eq:dizetpoly} \quad,\quad k=1,\dots,K,
\end{align}
see \cite{WJH18a,WJH18b}.
A global phase in $\uY(z)$ will have no affect to the DiZeT decoder and to the received zeros. But the CFO $\phi$
modulates the BMOCZ symbol in \eqref{eq:receivedsamples} and causes a rotation\footnote{The CFO would
rotate the zeros in any scheme of modulation on zeros, but we will consider here for simplicity only the BMOCZ scheme.}
of its zeros by $-\phi$ in \eqref{eq:Xzeros}, which will destroy the hypothesis test of the DiZeT decoder. Hence, one
needs to either estimate $\phi$ or use an outer code for BMOCZ to be invariant against an arbitrary rotation of the
entire zero codebook $\Zero$, which we will introduce in \secref{sec:frequencyoffset}. 
However, before we can apply the DiZeT decoder, we have to identify the timing offset of the symbol which yields to the
convolution output in \eqref{eq:convolutionoutput}. 
%
%We will first address the estimation of the TO from the received baseband samples \eqref{eq:receivedsamples} in
%\secref{sec:timingoffset} and then introduce in \secref{sec:frequencyoffset} a cyclic outer code for BMOCZ together
%with an oversampled DiZeT decoder to obtain robustness against CFO.  

\input{timingoffset.tex}

\input{frequencyoffset}

% still need here the simulations and check that this makes sense
%\input{CFOforPMOCZ}

\input{simulations}

\section{Conclusion}

We proposed a timing-offset and carrier frequency offset estimation for the novel BMOCZ modulation scheme in wideband
frequency-selective fading channels.   The CFO robustness is realized by a cyclically permutable code, which allows to
identify the integer CFO. An oversampled DiZeT decoder allows to estimate the fractional CFO.  The CPC code construction
with cyclic BCH codes allow to correct additional bit errors which enhances the performance of the BMOCZ design for
moderate SNRs.  Furthermore, we used a novel simulation software
Quadriga version 2.0, to generate random CIR
%used in a mmWave band
at a bandwidth of $150$Mhz.  
Due to the low-latency of BMOCZ the CFO and TO estimation from one single BMOCZ symbol, this blind scheme is ideal for
control-channel applications, where few critical and control data need to be exchanged while at the same time, channel
and impairments information need to be communicated and estimated. Coded BMOCZ with ACPC is therefore a promising scheme to enable
low-latency and ultra-reliable short-packet communications over unknown wideband channels.

\section{Acknowledgements}

The authors would like to thank Richard Kueng, Anatoly Khina, and Urbashi Mitra for many helpful discussions. 
%We like to thank the SURF student Mattia Carrera for his support during a summer program at Caltech. 
%Most of the work by
%Philipp Walk was done during a two year postdoc fellowship at Caltech, which was sponsored by the DFG WA 3390/1.
Peter Jung is supported by DFG grant JU 2795/3.

\section*{References}
\vspace{-3ex}
%\printbibliography[heading=bibintoc]
\printbibliography
% Need maybe later: 
%  \appendices
%  \input{app_chernoffbounds}
%  \newpage
\end{document}

%% file: Header_Philipp.tex
\usepackage{tensor} % noetig für einige indizes. Sieht sonst scheisse aus.

\usepackage{cleveref} % referenzing mulitple equations package,http://www.ctan.org/tex-archive/macros/latex/contrib/cleveref/ 

\newcommand{\diag}{\operatorname{diag}}

\let\forallalt\forall
\renewcommand{\forall}{\;\forallalt\;}

\let\refalt\ref
\renewcommand{\ref}[1]{(\refalt{#1})}

\newcommand{\alp}{\ensuremath{\alpha}}
\newcommand{\bet}{\ensuremath{\beta}}
\newcommand{\Del}{\ensuremath{\Delta}}
\newcommand{\del}{\ensuremath{\delta}}
\newcommand{\eps}{\ensuremath{\epsilon}}
\newcommand{\Gam}{\ensuremath{\Gamma}}
\newcommand{\gam}{\ensuremath{\gamma}}

\newcommand{\lam}{\ensuremath{\lambda}}

\newcommand{\tht}{\ensuremath{\theta}}

\newcommand{\vGam}{\ensuremath{\mathbf{\Gam}}}
\newcommand{\vm}{\ensuremath{\mathbf m}}
\newcommand{\vdel}{\ensuremath{\boldsymbol \delta}}

\newcommand{\cp}{{\ensuremath{\text{CP}}}}
\newcommand{\cpone}{{\ensuremath{\text{CP}^{(1)}}}}
\newcommand{\cptwo}{{\ensuremath{\text{CP}^{(2)}}}}
\newcommand{\vxcpone}{\ensuremath{\vx_{\cp}^{(1)}}}
\newcommand{\vxcptwo}{\ensuremath{\vx_{\cp}^{(2)}}}

\newcommand{\va}{{\ensuremath{\mathbf a}}}

\newcommand{\vc}{{\ensuremath{\mathbf c}}}
\newcommand{\vd}{{\ensuremath{\mathbf d}}}

\newcommand{\ve}{{\ensuremath{\boldsymbol{\del}}}} % better and new version
\newcommand{\vg}{{\ensuremath{\mathbf g}}}
\newcommand{\vh}{{\ensuremath{\mathbf h}}}                         % Vektorwertige Funktion
\newcommand{\vi}{{\ensuremath{\mathbf i}}}                         % Vektorwertige Funktion
\newcommand{\vr}{{\ensuremath{\mathbf r}}}                         % Vektorwertige Funktion
\newcommand{\vs}{{\ensuremath{\mathbf s}}}

\newcommand{\vv}{{\ensuremath{\mathbf v}}}
\newcommand{\vw}{{\ensuremath{\mathbf w}}}
\newcommand{\vx}{{\ensuremath{\mathbf x}}}
\newcommand{\vy}{{\ensuremath{\mathbf y}}}

\newcommand{\vtc}{\ensuremath{\tilde{\mathbf c}}}

\newcommand{\vth}{\ensuremath{\tilde{\mathbf h}}}

\newcommand{\vtr}{\ensuremath{\tilde{\mathbf r}}}

\newcommand{\vtv}{\ensuremath{\tilde{\mathbf v}}}
\newcommand{\vtw}{\ensuremath{\tilde{\mathbf w}}}
\newcommand{\vtx}{\ensuremath{\tilde{\mathbf x}}}

\newcommand{\vty}{\ensuremath{\tilde{\mathbf y}}}

\newcommand{\hvd}{{\ensuremath{\hat{\mathbf d}}}}

\newcommand{\hvy}{{\ensuremath{\hat{\mathbf y}}}}

\newcommand{\uC}{{\ensuremath{\mathrm C}}}

\newcommand{\uG}{{\ensuremath{\mathrm G}}}

\newcommand{\uI}{{\ensuremath{\mathrm I}}}
\newcommand{\uR}{{\ensuremath{\mathrm R}}}
\newcommand{\uM}{{\ensuremath{\mathrm M}}}
\newcommand{\uS}{{\ensuremath{\mathrm S}}}
\newcommand{\uX}{{\ensuremath{\mathrm X}}}
\newcommand{\uY}{{\ensuremath{\mathrm Y}}}

\newcommand{\vA}{{\ensuremath{\mathbf A}}}

\newcommand{\vD}{{\ensuremath{\mathbf D}}}

\newcommand{\vG}{{\ensuremath{\mathbf G}}}
\newcommand{\vH}{\ensuremath{\mathbf H }}                         % Vektorwertige Funktion
\newcommand{\vM}{\ensuremath{\boldsymbol \Omega}}
\newcommand{\vP}{\ensuremath{\mathbf P }}                         % Vektorwertiges Ma{\ss}

\newcommand{\vR}{\ensuremath{\mathbf R}}
\newcommand{\vS}{\ensuremath{\mathbf S }}                         % Vektorwertiges Ma{\ss}
\newcommand{\vT}{\ensuremath{\mathbf T}}

\newcommand{\vtG}{{\ensuremath{\tilde{\mathbf G}}}}
\newcommand{\vtH}{{\ensuremath{\tilde{\mathbf H}}}}

\newcommand{\valp}{\ensuremath{\boldsymbol{ \alp}}}

\newcommand{\zero}{{\ensuremath{\mathbf 0}}}

\newcommand{\vzero}{{\ensuremath{\mathbf O}}}
\newcommand{\tilh}{\ensuremath{\tilde{h}}} %\tg ist schon definiert

\newcommand{\tilr}{\ensuremath{\tilde{r}}}

\newcommand{\tw}{\ensuremath{\tilde{w}}}
\newcommand{\tx}{\ensuremath{\tilde{x}}}
\newcommand{\ty}{\ensuremath{\tilde{y}}}

\newcommand{\tH}{\ensuremath{\tilde{H}}}

\newcommand{\tN}{\ensuremath{\tilde{N}}}

\newcommand{\talp}{\ensuremath{\tilde{\alp}}}

\newcommand{\hm}{\ensuremath{\hat{m}}}

\newcommand{\hs}{\ensuremath{\hat{s}}}

\newcommand{\htht}{\ensuremath{\hat{\tht}}}

\newcommand{\htau}{\ensuremath{\hat{\tau}}}

\newcommand{\Calg}{\ensuremath{\mathscr C}}
\newcommand{\Code}{\Calg}

\newcommand{\Galg}{\ensuremath{\mathcal G}}

\newcommand{\Sset}{\ensuremath{\mathscr S}}

\newcommand{\tvm}{\ensuremath{\tilde{\vm}}}

\newcommand{\G}{\ensuremath{\mathcal{G} }}						% Grundzust{\"a}nde im therm. Limes
\newcommand{\Galois}{\ensuremath{\mathbb{F}} }
\newcommand{\Ftwo}{\ensuremath{\mathbb{F}_2} }

\newcommand{\C}{{\ensuremath{\mathbb C}}}

\newcommand{\R}{{\ensuremath{\mathbb R}}}
\newcommand{\N}{{\ensuremath{\mathbb N}}}

\newcommand{\CN}{\ensuremath{\C^N}}

\newcommand{\Fmatrix}{{\ensuremath{\mathbf F}}}

\newcommand{\Fmatrixa}{{\ensuremath{\mathbf F}^*}}

\newcommand{\id}{{\ensuremath{\mathbf I}}}

\newcommand{\eins}{{\ensuremath{\mathbf 1}}}

\newcommand{\LRA}{\ensuremath{\Leftrightarrow} }

\newcommand{\dmin}{{\ensuremath{d_{\text{\rm min}} } }}

\newcommand{\Pro}{\prod}

\definecolor{gray}{rgb}{0.3,0.3,0.3}
{\color{black}}                      % change to gray if you want see any changes
{\color{black}}

{\color{black}}                      % change to gray if you want see any changes
{\color{black}}
\ifx\algref\undefined                         % Algorithm undefined
\newcommand{\algref}[1]{Algorithm~\ref{#1}}             % Enumerations (i) etc..
\else
\renewcommand{\algref}[1]{Algorithm~\ref{#1}}             % Enumerations (i) etc..
\fi
\newcommand{\codref}[1]{Code~\ref{#1}}             % Enumerations (i) etc..

\newcommand{\secref}[1]{Section~\ref{#1}}

\newcommand{\figref}[1]{Figure~\ref{#1}}

\newcommand{\noi}{\noindent}

\DeclareMathOperator{\support}{supp}

\DeclareMathOperator*{\argmin}{argmin}
\DeclareMathOperator*{\argmax}{argmax}

\newcommand{\supp}[1]{{\ensuremath{\support(#1)}}} % Circulant Matrix generated by #1
\newcommand\tr{\operatorname{tr}} 
\newcommand{\Norm}[1]{\ensuremath{ \left\|#1\right\| }}

\newcommand{\Expect}[1]{{\ensuremath{\mathbb E}[#1]}}

\newcommand{\cc}[1]{{\ensuremath{\overline{#1}}}} % complex conjugation

\makeatletter
  \newcommand{\set}[2]{\ensuremath{%
  \setbox0=\hbox{\ensuremath{#2}}
  \dimen@\ht0
  \advance\dimen@ by \dp0
  \left\{\left.#1\rule[-\dp0]{0pt}{\dimen@}\;\right|\;#2\right\} }}
\makeatother

\renewcommand{\argmin}{\operatornamewithlimits{argmin}}

\ifx \@paragraph \@empty
\makeatletter
\renewcommand\paragraph{\@startsection
{paragraph}{4}{\z@}{-3.5ex plus-1ex minus-.2ex}%
{1.3ex plus.2ex}{\normalfont\itshape}}
\makeatother
\fi

\newcommand{\tG}{\ensuremath{\widetilde G }}                         % universal ambient group
\renewcommand{\Re}{\ensuremath{\operatorname{Re}}}
\renewcommand{\Im}{\ensuremath{\operatorname{Im}}}
\DeclareFontFamily{U}{matha}{\hyphenchar\font45}
\DeclareFontShape{U}{matha}{m}{n}{
      <5> <6> <7> <8> <9> <10> gen * matha
      <10.95> matha10 <12> <14.4> <17.28> <20.74> <24.88> matha12
      }{}
\DeclareSymbolFont{matha}{U}{matha}{m}{n}
\DeclareFontSubstitution{U}{matha}{m}{n}

\DeclareFontFamily{U}{mathx}{\hyphenchar\font45}
\DeclareFontShape{U}{mathx}{m}{n}{
      <5> <6> <7> <8> <9> <10>
      <10.95> <12> <14.4> <17.28> <20.74> <24.88>
      mathx10
      }{}
\DeclareSymbolFont{mathx}{U}{mathx}{m}{n}
\DeclareFontSubstitution{U}{mathx}{m}{n}

\DeclareMathDelimiter{\vvvert}{0}{matha}{"7E}{mathx}{"17}

\ifx\oast\undefined % MdSymbol defines this. IF not loaded, set it manually.
\newcommand{\oast}{\ensuremath{\circledast}}
\fi

\newcommand{\im}{\ensuremath{j}} % imaginary number

\ifdefined\Zero
\renewcommand{\Zero}{\ensuremath{\mathscr Z}}
\else
\newcommand{\Zero}{\ensuremath{\mathscr Z}}
\fi
\newcommand{\vtD}{\ensuremath{\tilde{\mathbf D}_{\vp}}} % symbol time 
\ifdefined\Dset
\renewcommand{\Dset}{\ensuremath{\mathfrak D}} % decoding sets 
\else
\newcommand{\Dset}{\ensuremath{\mathfrak D}} % decoding sets 
\fi
\ifdefined\Sset
\renewcommand{\Sset}{\ensuremath{\mathfrak S}} % sector sets 
\else
\newcommand{\Sset}{\ensuremath{\mathfrak S}} % sector sets 
\fi
\ifdefined\hvd
\renewcommand{\hvd}{\ensuremath{\hat{\vd}}}
\else
\newcommand{\hvd}{\ensuremath{\hat{\vd}}}
\fi

\newcommand{\vGin}{\ensuremath{{\mathbf G}_{\text{in}}}}
\newcommand{\vGout}{\ensuremath{{\mathbf G}_{\text{out}}}}
\newcommand{\vtGout}{\ensuremath{{\mathbf \tG}_{\text{out}}}}

\newcommand{\vtHout}{\ensuremath{{\mathbf \tH}_{\text{out}}}}
\newcommand{\uGin}{\ensuremath{{\uG}_{\text{in}}}}
\newcommand{\uGout}{\ensuremath{{\uG}_{\text{out}}}}

\renewcommand{\alp}{\ensuremath{\alpha}}

\renewcommand{\CN}{\ensuremath{\mathcal{CN}}}

\newcommand{\Epeaks}{\ensuremath{E_{\text{sides}}}}
\newcommand{\Ecenter}{\ensuremath{E_{\text{center}}}}

\newcommand*\xbar[1]{%
   \hbox{%
     \vbox{%
       \hrule height 0.5pt % The actual bar
       \kern-0.1ex%         % Distance between bar and symbol
       \hbox{%
         \kern-0.0em%      % Shortening on the left side
         \ensuremath{ #1}%
         \kern-0.1em%      % Shortening on the right side
       }%
     }%
   }%
} 
\newcommand*\sxbar[1]{%
   \hbox{%
     \vbox{%
       \hrule height 0.5pt % The actual bar
       \kern-0.3ex%         % Distance between bar and symbol
       \hbox{%
         \kern-0.0em%      % Shortening on the left side
         \ensuremath{\scriptstyle #1}%
         \kern-0.1em%      % Shortening on the right side
       }%
     }%
   }%
}

\newcommand{\Nx}{{\ensuremath{K+1}}}
\newcommand{\Nalp}{{\ensuremath{K}}}
\newcommand{\Nxdpsk}{{\ensuremath{K_{\text{dif}}}}}
\newcommand{\Ndpsk}{{\ensuremath{N_{\text{dif}}}}}
\newcommand{\nx}{{\ensuremath{k}}}

\newcommand{\pdp}{\ensuremath{p}} % power delay profile

\renewcommand{\ve}{\ensuremath{{\mathbf e}}}

\newcommand{\tuY}{{\ensuremath{\tilde{\uY}}}}
\newcommand{\rSNR}{{\ensuremath{\text{rSNR}}}}
\newcommand{\Ebno}{{\ensuremath{E_{\text{b}}/N_0}}}

\newcommand{\tvalp}{{\ensuremath{\tilde{\valp}}}}
\newcommand{\uvR}{{\ensuremath{\underline{\vR}}}}

\newcommand{\CodeCPC}{\ensuremath{\Code_{\text{CPC}}}}
\newcommand{\CodeACPC}{\ensuremath{\Code_{\text{ACPC}}}}

\newcommand{\Codeout}{\ensuremath{\Code_{\text{out}}}}
\newcommand{\Codein}{\ensuremath{\Code_{\text{in}}}}
\newcommand{\CodeCRC}{\ensuremath{\Code_{\text{CR}}}}

\newcommand{\Peter}[1]{\textcolor{red}{#1}}
\renewcommand{\Peter}[1]{{}}

\renewcommand{\vM}{\ensuremath{\mathbf M}}
\newcommand{\pvy}{\ensuremath{\tilde{\mathbf y}}}
\newcommand{\puY}{\ensuremath{\tilde{\uY}}}
\newcommand{\Deltaf}{\ensuremath{\Delta f}} % frequency offset
\newcommand{\toff}{\ensuremath{\tau_0}} % timing offset
\newcommand{\Tend}{\ensuremath{N_{\text{obs}}}}
\newcommand{\Nobs}{\ensuremath{N_{\text{obs}}}}
\newcommand{\Huframe}{\ensuremath{\boldsymbol{\varkappa}}}
\newcommand{\Hufbracket}{\ensuremath{\boldsymbol{\varkappa}}}
\newcommand{\aHufbracket}{\ensuremath{\breve{\va}}}

\newcommand{\vxint}{\ensuremath{\dot{\vx}}}
\newcommand{\vxbraint}{\ensuremath{\dot{\breve{\vx}}}}
\newcommand{\vtxint}{\ensuremath{\dot{\vtx}}}
\newcommand{\vxext}{\ensuremath{\mathring{\vx}}}
\newcommand{\vtxext}{\ensuremath{\mathring{\vtx}}}
\newcommand{\aHuframe}{\ensuremath{\breve{a}}}
\newcommand{\vaHuframe}{\ensuremath{\breve{\va}}}
\newcommand{\Ncut}{\ensuremath{N_{\text{eff}}}}

\newcommand{\Nnoise}{\ensuremath{N_{\text{noise}}}}
\newcommand{\Lcut}{\ensuremath{L_{\text{eff}}}}
\newcommand{\Leff}{\ensuremath{L_{\text{eff}}}}
\newcommand{\vhcut}{\ensuremath{\vh_{\text{eff}}}}
\newcommand{\vttw}{\ensuremath{\tilde{\vtw}}}
\newcommand{\hatt}{\ensuremath{\hat{t}}}

\newcommand{\Ecut}{\ensuremath{E_{\text{eff}}}}
\newcommand{\NIM}{\ensuremath{K_{\text{IM}}}}

\iftikz
\usepackage{filecontents}
\makeatletter
\@ifpackageloaded{tikz}{
\usetikzlibrary{positioning,decorations.pathreplacing}

  \setcounter{MaxMatrixCols}{18}

}{}
\makeatother
\fi % end tikz

%% file: timingoffset.tex
% Timing Offset Synchronization
\section{Timing Offset and Effective Delay Spread for BMOCZ}\label{sec:timingoffset}

%Since the transmitter broadcasts the short-packets in a sporadic and asynchronous manner, such as in a random access
%channel (RACH), the receiver needs to estimate blindly the timing offset, or more precisely, identify the first
%symbol coefficient which arrives via the line-of-sight (LOS) path or the first significant path.

In an asynchronous communication, the receiver does not know when a packet from a transmitter (user) will arrive.
Hence, at first the receiver has to detect a transmitted packet, which is already one bit of information.  We will
assume that the receiver decide correctly, that in an observation window of $\Tend=\Nnoise+K+L$ received samples, one
single MOCZ packet of length $K$ with  maximal channel length of $L$ is captured. By assuming a maximal length $L$ and a
known or a maximal $K$ at the receiver, the observation window can be chosen, for example, as $\Nobs=2N$. From the noise
floor knowledge at the receiver, a simple energy detector with a hard threshold over the observation window can be used
for a packet detection.  Then, an unknown TO $\tau_0\in[\Nnoise]$ and CFO $\phi\in[0,2\pi)$ will be present in the
  observation window 
\begin{align}
  \tilr_n =e^{\im n\phi}(\vx*\vh)_{n-\tau_0} + \tilde{w}_n = (\vtx* \vth\vT^{\toff})_n + \tilde{w}_n\quad,\quad
  n\in[\Nobs].\label{eq:receivedsamplesNobs}
\end{align}
%
%, where we introduced the \emph{modulation} $\vM_K^\phi$ and \emph{translation} matrix
%$\vT_{L,\Nobs}^{\tau_0}$ given by
%%
%\begin{align}
%  \vM^{\phi}_K= \begin{pmatrix} 
%    e^{\im\phi\cdot 0} & 0 &\dots & 0\\
%    0 &e^{\im\phi\cdot 1} & \dots & 0\\
%    \vdots  & & \ddots & \vdots\\
%    0 & 0 &\dots  &e^{\im\phi\cdot K}\\
%  \end{pmatrix}
%\end{align}
%
The challenge here is to identify $\tau_0$ and the efficient channel length which contains most of the energy of the
instantaneous CIR realization $\vh$.  The estimation of these \emph{Timing-of-Arrival} (TOA) parameters are usually done by
observing the same channel under many symbol transmission, to obtain a sufficient statistic of the channels PDP
\cite{GGKST03}, \cite{CWM02}.
Since we only have one observation available, a good estimation is very challenging.  

\begin{figure}[t]
  \begin{center}
  \begin{subfigure}[b]{0.49\textwidth}
    \hspace{-3ex}    \includegraphics[width=1.1\textwidth]{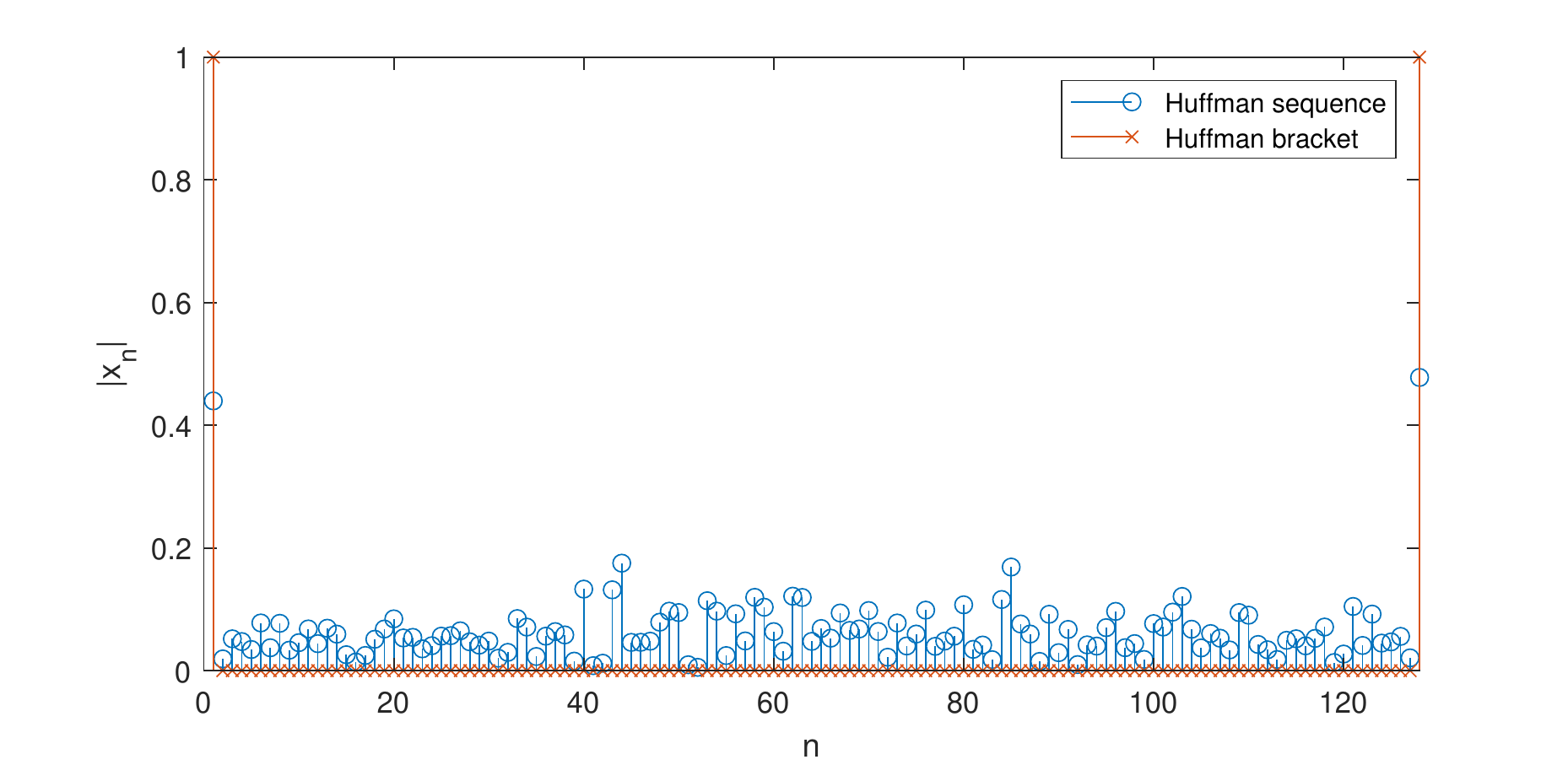}
    %\vspace{-2ex}
    %\caption{} \label{fig:HuffmanTridentFrameA}
  \end{subfigure}
  \begin{subfigure}[b]{0.49\textwidth}
    \includegraphics[width=1.1\textwidth]{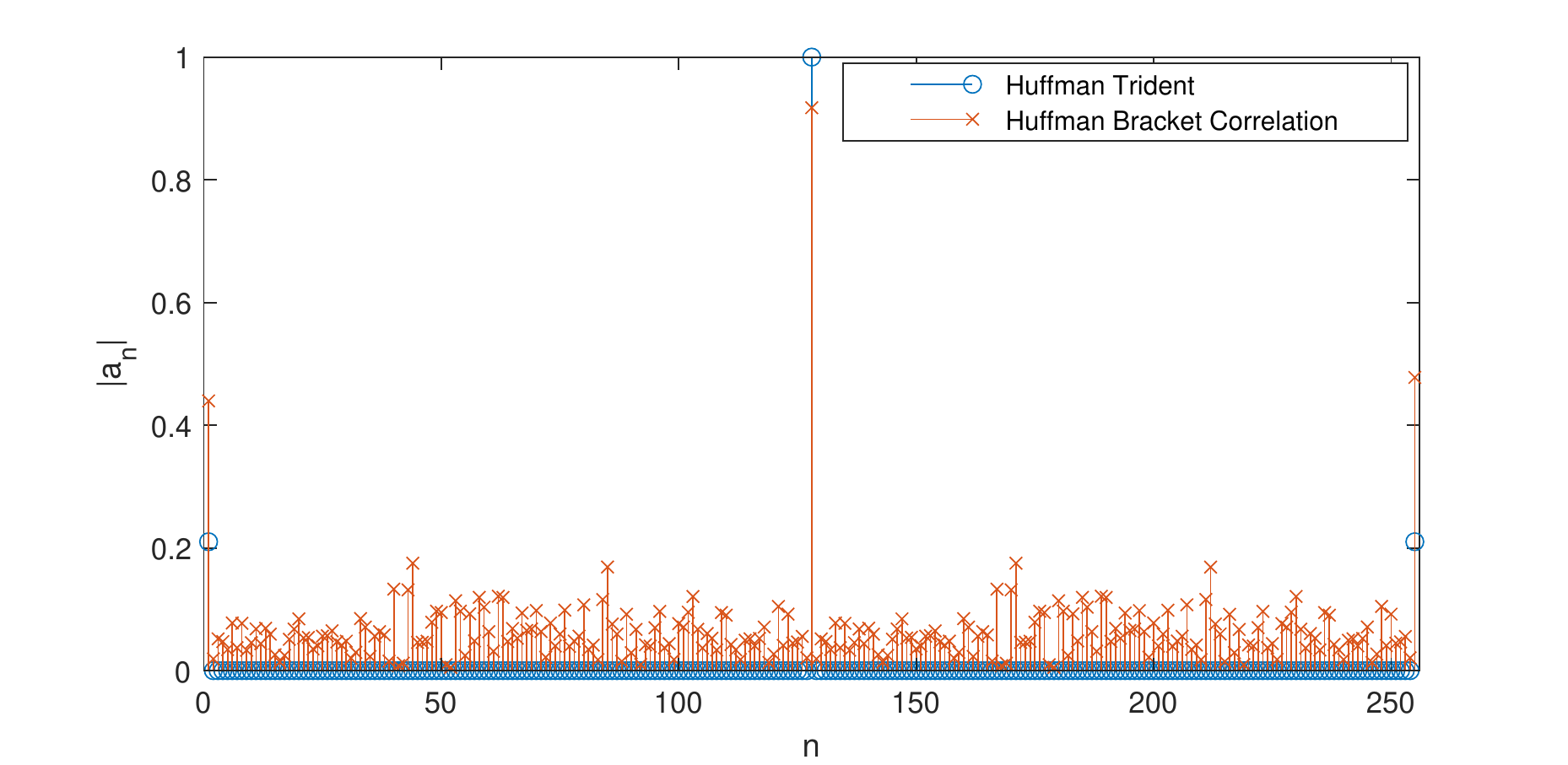}
    %\vspace{-2ex}
    %\caption{} \label{fig:HuffmanTridentFrameB}
  \end{subfigure}
\end{center}
\vspace{-3ex} \caption{Left figure shows the magnitudes of a Huffman sequence for $K=127$ in blue and the absolute
  Huffman bracket in red. The
right figure shows the autocorrelation magnitudes as  well as the correlation between bracket and Huffman sequence,
which both reveal a trident.
}\label{fig:HuffmanTridentFrame} 
\end{figure}

The efficient (instantaneous) channel length $\Leff$, defined by an energy concentration window, will be much less than
the maximal channel length $L$, due to blockage and attenuation by the environment, which might also cause a sparse,
clustered, and exponential decaying power delay profile.
For the MOCZ scheme, it is essential to correctly identify in the window \eqref{eq:receivedsamplesNobs} the first
received sample $h_0x_0$ from the transmitted symbol $\vx$, or at least do not miss it, since it will carry most of the
energy if $h_0$ is the \emph{line of sight} (LOS) path. It was shown in \cite{WJH18b} that for the optimal radius in
BMOCZ, $x_0$ carries in average $1/4$ to $1/5$ of the BMOCZ symbol energy, see also \figref{fig:HuffmanTridentFrame}. On
the other hand, an overestimated channel length $\Leff$ will reduce the overall bit-error performance because the
receiver collects unnecessary noise samples.

Since we assume no CSI at the receiver, the channel characteristic, i.e., the instantaneous power delay profile, has to
be determined entirely from the received MOCZ symbol. We will introduce here an efficient approach for the BMOCZ design,
by exploiting the radar properties of the Huffman sequences, to obtain excellent estimation of the timing offset and
the effective channel delay in moderate and high SNR.  

Huffman sequences have an impulsive autocorrelation \eqref{eq:huffmanauto}, originally designed for radar applications,
and are therefore very suitable to measure the channel impulse response \cite{GG05}. Since the transmitted Huffman
sequence is still unknown at the receiver, we can not correlate the received signal with the correct Huffman sequence to
retrieve the CIR. Instead, we will use an approximative universal Huffman sequence, which is just the first and last
peak of a typical Huffman sequence, expressed by the impulses $(\vdel_0)_k=\del_{0k}$ and $(\vdel_K)_n=\del_{Kk}$ for $k\in[K+1]$ as
\begin{align}
  \Hufbracket_{\psi}=\vdel_0 + e^{\im\psi}\vdel_{K}\in\C^{K+1}
\end{align}
which we call the \emph{$K-$Huffman bracket of phase $\psi$}. %
%\footnote{Let us note here, that
%  $\pm\frac{1}{\sqrt{2}}\Hufbracket_{\psi}$ are actually the only normalized Huffman sequences for $R=1$ with
%  real-valued first coefficient, where all zeros are on the unit circle rotated by $\psi$. If the peaks are different,
%  then the interior is vanishing if and only if all zeros are outside or inside the unit circle, .}.
%
Since the first and last coefficients are
\begin{align}
  x_0= \sqrt{R^{2\Norm{\vm}_1-K}\eta}\quad,\quad x_K=-\sqrt{R^{-2\Norm{\vm}_1+K}\eta}
  \label{eq:x0xK}
\end{align}
see \cite{WJH18b}, typical Huffman sequences, i.e., having same amount of ones and zeros, will have 
\begin{align}
  |x_0|^2=|x_K|^2=\eta.
\end{align}
By correlating the modulated Huffman sequence $\vtx$ with the \emph{Huffman bracket} $\Huframe_{\psi}$ we keep the
locational properties of the Huffman autocorrelation (trident)
\begin{align}
  \vaHuframe=\cc{\Hufbracket_{\psi}^-}*\vtx=
  %\begin{pmatrix} e^{-\im \psi}x_0\\ e^{-\im\psi}\vtxint \\ x_0 + e^{\im(K\phi-\psi)}x_K  \\ \vtxint \\ e^{\im K\phi}x_K\end{pmatrix}
   ( e^{-\im \psi}x_0, e^{-\im\psi}\vtxint, x_0 + e^{\im(K\phi-\psi)}x_K, \vtxint, e^{\im K\phi}x_K)
  \quad,\quad \vtx=\vtxext+\vtxint\label{eq:ahufbracket},
\end{align}
where $\vxext=x_0 \vdel_0 + x_K\vdel_K$ denotes the \emph{exterior signature} and $\vxint=(0,x_1,\dots,x_{K-1},0)$ the
\emph{interior signature} of the Huffman sequence $\vx$, see \figref{fig:HuffmanTridentFrame}. Here, the interior
signature can be seen as the data noise floor distorting the trident $\va$ in \eqref{eq:huffmanauto}. Taking the
absolute-squares in \eqref{eq:ahufbracket}, we get for the three peaks of the approximated trident  
\begin{align}
  |\aHuframe_0|^2= |x_0|^2, |\aHuframe_K|^2=|x_0|^2+ |x_K|^2 - 2|x_0 x_K|\cos(K\phi-\psi),
  |\aHuframe_{2K}|^2=|x_K|^2,\label{eq:aHuffpeaks}
\end{align}
where the side-peaks have energy
\begin{align}
  \Epeaks= |x_0|^2+|x_K|^2= (R^{2\Norm{\vm}_1-K} + R^{-2\Norm{\vm}_1+K})\eta \leq \Norm{\vx}^2=1.
\end{align}
Since $\eps=2\Norm{\vm}_1-K\in[-K,K]$ we get by \eqref{eq:huffmanauto} and $2\leq R^\eps+R^{-\eps}\leq R^K+R^{-K}$ that
$2\eta\leq \Epeaks \leq 1$, where the lower bound is achieved for typical sequences with $\eps=0$ (having the same
amount of ones and zeros) and the upper bound for $\eps=\pm K$ (all ones or all zeros). If $\eta=1/2$ then the two
coefficients (the exterior signature $\vxext$) will carry all the energy of the Huffman sequence. But then also $R=1$
and the only Huffman sequences (real valued first and last coefficient) are given by $x_0=\pm1/\sqrt{2}, x_K=\pm
1/\sqrt{2}$ and $x_k=0$ for $k$ else, which are the coefficients of polynomials with $K$ uniform zeros on the unit
circle, see \cite{WJH17a}. For $R$ given by \eqref{eq:optimalR} the autocorrelation side-lobe $\eta$ is exponentially
decaying in $K$ but is bounded to $\eta\geq1/5$ for $K=1,2,\dots,512$. Hence, $\Epeaks \geq 0.4$, such that almost half
of the Huffman sequence energy is always carried in the two peaks. \ifextras In average, we can calculate with the binomial
theorem
\begin{align}
  \Expect{\Epeaks}= \frac{1}{2^K(1+R^{2K})} \sum_{b=0}^{K} \binom{K}{b} R^{2b} + \binom{K}{K-b} R^{2b} =
  \frac{(1+R^2)^K}{2^{K-1}(1+R^{2K})}
\end{align}\fi
If the CFO would be known, we can set $\psi=K\phi-\pi$ and
get for the center peak in \eqref{eq:aHuffpeaks}
\begin{align}
  \Ecenter=|x_0|^2+|x_K|^2+2|x_0||x_K|\geq4\eta\geq0.8,
\end{align}
i.e., the energy of the center peak is roughly twice as large as the energy of the side-peaks, and reveals the trident in the
approximated Huffman autocorrelation $\vaHuframe$.   But , since we do not know the CFO and $K\phi-\psi\approx 2n\pi$ for some $n\in\N$
then we get $x_0+x_K\approx 0$ for typical Huffman sequences, such that the power of the center peak will vanish.
Hence, in the presence of an unknown CFO the center peak does not always identify the trident.  We will therefore
correlate the \emph{positive Huffman bracket} $\Hufbracket_0$ with the absolute-square value of $\vx$ or in presence of
noise and channel with the absolute-square of the received signal $\vtr$, which will result approximately in  
\begin{align}
  \vd=\cc{\Huframe_0^-}*|\vtr|^2= \cc{\Hufbracket_0^-}*|\vtx*\vth\vT^{\toff}|^2 + \cc{\Hufbracket^-_0}*|\vtw|^2 
  \approx  |\aHufbracket|^2*|\vth\vT^{\toff}|^2 + |\vttw|^2\in\R_+^{\Nobs+K}
  \label{eq:d}
\end{align}
where $\vtw$ and $\vttw$ are colored noise and 
\begin{align}
  |\aHufbracket|^2=(|x_0|^2, |\vxbraint|^2,\Epeaks,|\vxbraint^-|^2,|x_K|^2)\label{eq:absquarea}
\end{align}
denotes the noisy trident which collects three times the \emph{instant power delay profile} $|\vh|^2$ of the shifted
CIR. These three echos of the CIR will be separated if we have $K\geq L$.
%, which we will assume in our simulations.  Here, the strongest path gain $|h_{s+\tau_0}|^2$ will multiply with the
%signal \eqref{eq:absquarea} and produce the strongest trident structure in $\vd$ which we will use to identify the
%timing-offset and the RMS delay spread of the channel, see \figref{fig:timingoffset}. 
%%
%
The approximation in \eqref{eq:d} can be justified by the isometry property of the Huffman convolution.  Briefly, $L<K$, the
generated (banded) $L\times N$ Toeplitz matrix $\vT_{\vx}=\sum_{k=0}^K x_k\vT^k$, for any Huffman sequence
$\vx\in\Code(K)$, is a stable \emph{linear time-invariant} (LTI) system, since the energy of the output satisfy for any
CIR realization $\vh\in\C^L$
\begin{align}
  \Norm{\vx*\vh}^2= \Norm{\vh\vT_{\vx}}^2=\tr(\vh\vT_{\vx}\vT_{\vx}^*\vh^*)=\tr(\vh^*\vA_{\vx}\vh)=
  \Norm{\vx}^2\tr(\vh^*\vh)=\Norm{\vx}^2\Norm{\vh}^2.
\end{align}
Here, $\vA_{\vx}=\vT_{\vx}\vT_{\vx}^*$ is the $L\times L$ autocorrelation matrix of $\vx$, which is the identity scaled
by $\Norm{\vx}^2$ if $L<K$.  Hence, each normalized Huffman sequence, generates an isometric operator $\vT_{\vx}$ having
the best stability among all discrete-time LTI systems, as studied in \cite{WJP15}.

%The energy-window detects the maximal instant
%SNR, given as the sum of the signal $P$ and channel $H$ power
%%
%\begin{align}
%  \iSNR_{\text{dB}}=10\log_{10} \frac{\Norm{\vx*\vh}_2^2}{\Norm{\vw}_2^2} = \frac{P + H}{\tilde{N}_0}.
%\end{align}
%%
%If the CIR energy is concentrated in the first $K-1$ taps, then the timing-offset detection works very well. 
\subsection{Timing Offset Estimation}

The delay of the strongest path $|h_s|=\Norm{\vh}_{\infty}$ can be identified from the maximum in \eqref{eq:d} 
\begin{align}
  \hatt=\left(\argmax_{t\in[\Nobs+K]} d_t\right) -K=\left(\argmax_{t\in[\Nobs-K]+K}d_t\right)
  -K,\label{eq:centerpeak_estim} 
\end{align}
where the last equality follows from the fact that both peaks in $\Hufbracket_{0}$ are contributing  between $K$ and
$\Nobs$.  If the CIR has a LOS path, then $s=0$ and we immediately have found an estimate for the timing-offset by
${\htau}_0=\hatt$.  In case of NLOS or if the first paths are equally strong, we have to go further back and identify
the first significant peak above the noise floor, since the convolution sum of the CIR with the interior signature might
produce a significant peak. Let us note here, that this might result in a misidentification of the tridents center peak
by \eqref{eq:centerpeak_estim}, for example if $|h_s|/|h_0|\gg1$. Therefore we will use as a peak threshold
\begin{align}
  \rho=\rho(K,\vd)=\frac{1}{K+1}\sum_{n=\hs+\htau_0}^{\hs+\htau_0+K} d_{n},
\end{align}
which is the average power of the Huffman sequence distorted by the channel and noise. By comparing to the noise power
$N_0$ we found empirically to set the noise-dependent threshold to
\begin{align}
  \rho_0=\max\{\rho(K,\vd)/10,\dots, \rho(K,\vd)/100,10\cdot N_0\} 
\end{align}
to ensure with high probability to be above the instantaneous noise energy. By using an iterative back stepping in
\algref{alg:cpbs}, we will stop if the sample power falls below the threshold $\rho_0$, which finally yields an estimate
$\htau_0$ of the timing-offset.
%  Then we will set 
%%
%\begin{align} \hat{\tau}_0=\min\set{\tau\in\hatt-[K/2]}{|r_{t}|^2>\rho_{0} \text{ for } t\in\{\tau,\dots,\hatt\}}
%\end{align}
%%
%
In line $3$ of \algref{alg:cpbs} we update the timing-offset estimate, if the sample power is larger than the
threshold and the average power of the preceding samples is larger than the threshold divided by $1+(\log b)/3$, which
will be weighted by the amount $b$ of back-steps.

\subsection{Efficient Channel Length Estimation}

Since the BMOCZ design does not need any channel knowledge at the receiver, it is also well suited for estimating the
channel itself at the receiver. Here, a good channel length estimation is essential for the performance of the decoder,
if the power delay profile (PDP) is decaying. At some extent, the channel delays will fade out exponentially and the
receiver can cut-off the received signal by using a certain energy ratio threshold. %
Let us recall the average \emph{received SNR} 
\begin{align}
  \rSNR= \frac{\Expect{\Norm{\vx*\vh}^2}}{\Expect{\Norm{\vw}^2}} = E\frac{\Expect{\Norm{\vh}^2}}{N\cdot N_0}=
  \frac{1}{N_0}\label{eq:rsnr}
\end{align}
where $E=\Norm{\vx}^2=N$ is the energy of the BMOCZ symbols, which is constant for the codebook.  If the power delay
profile is \emph{flat}, then the collected energy will be uniform and the SNR will not change if we cut the channel
length at the receiver. However, the additional channel zeros will increase the confusion for the DiZeT decoder and 
reduce the BER performance. Therefore, the performance will decrease for increasing $L$ at a fixed symbol length $K$, see
simulations in \cite{WJH18b}.  For the most interesting scenario of $K\approx L$ the BER performance loss is only $3$dB
over $E_b/N_0$, but will
increase dramatically if $L\gg K$. The reason for this behaviour is the collection of many noise taps, which will lead
to more distortion of the transmitted zeros. Since in most realistic scenarios the PDP will be decaying, most of the
channel energy will be concentrated in the first channel taps. 
Hence, if we cut the received signal length to $\Ncut=K+\Lcut$, we will reduce the channel length to $\Lcut<L$ and improve the \rSNR\
for \emph{non-flat} PDPs with $p<1$, since it holds
\begin{align}
  \frac{1}{N_0}=\frac{N\Big(\sum_{l=0}^{\Lcut-1} p^l + p^{\Lcut}\sum_{l=0}^{L-\Lcut} p^l\Big)}{N\cdot N_0}
  \approx \frac{N\sum_{l}^{\Lcut-1} p^l}{(K+L)N_0} < \frac{N\sum_{l}^{\Lcut-1} p^l}{(K+\Lcut)N_0}
  =\frac{N\Expect{\Norm{\vhcut}^2}}{\Ncut N_0}\label{eq:averageCIRpower}.
\end{align}
Since $1/\Expect{\Norm{\vhcut}^2}\ll N/\Ncut$ we obtain a significant gain in SNR if $L\gg K$ and $p<1$. 
Hence, by cutting the received signal to the effective channel length, given by a certain energy concentration, we can
improve the SNR and reduce at the same time the amount of channel zeros, which we will
demonstrate by simulations in \secref{sec:TOundCutsim}.
 
\begin{figure}[ht]
  \begin{center}
  \begin{subfigure}[b]{0.39\textwidth}
    \hspace{-2ex}
    \vspace{-1.5ex}
    \includegraphics[width=1.14\textwidth]{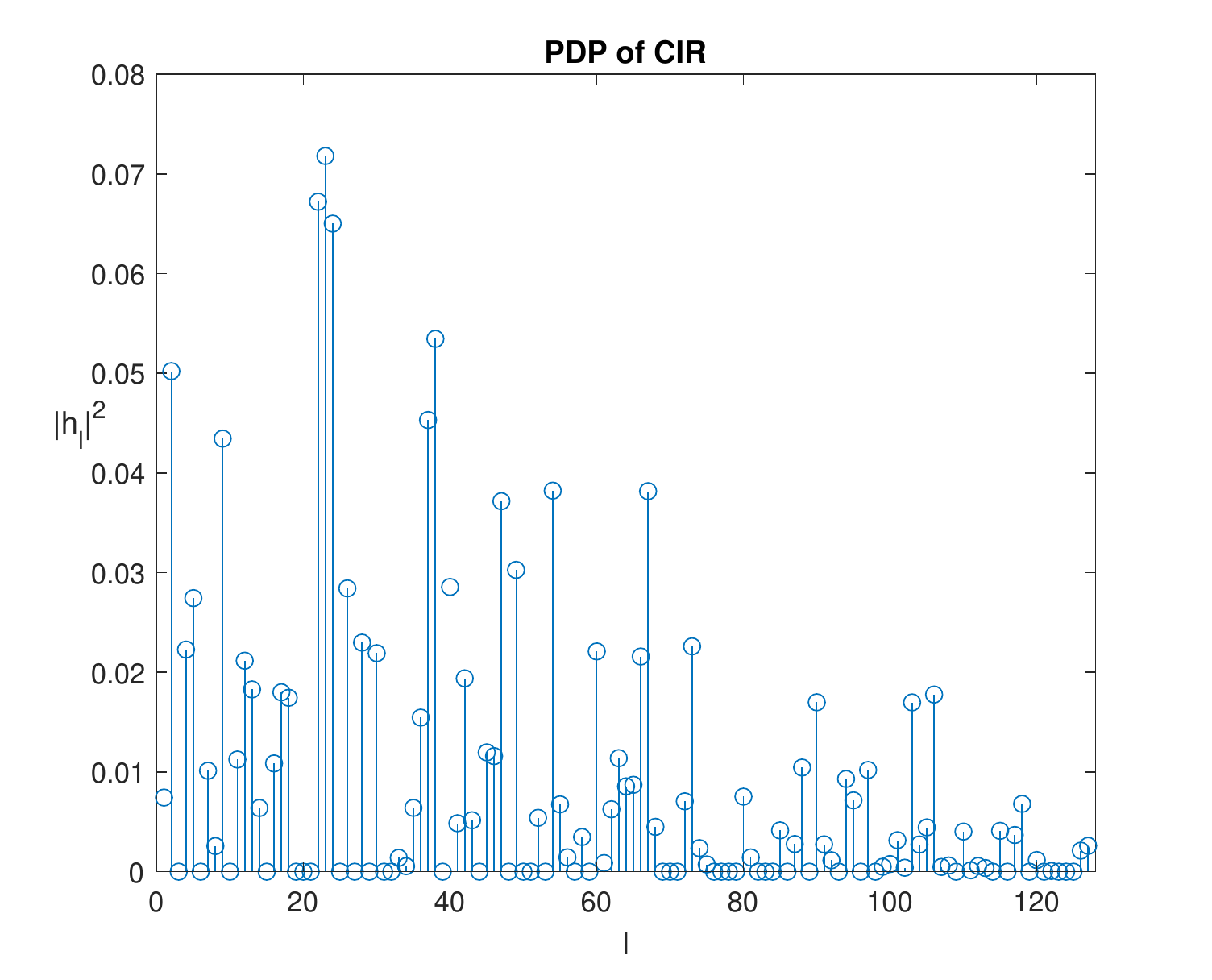}
    \caption{} \label{fig:ChannelPDPL128}
  \end{subfigure}
  \begin{subfigure}[b]{0.59\textwidth}
    \hspace{2ex}
    \vspace{-1.5ex}
    \includegraphics[width=1.09\textwidth]{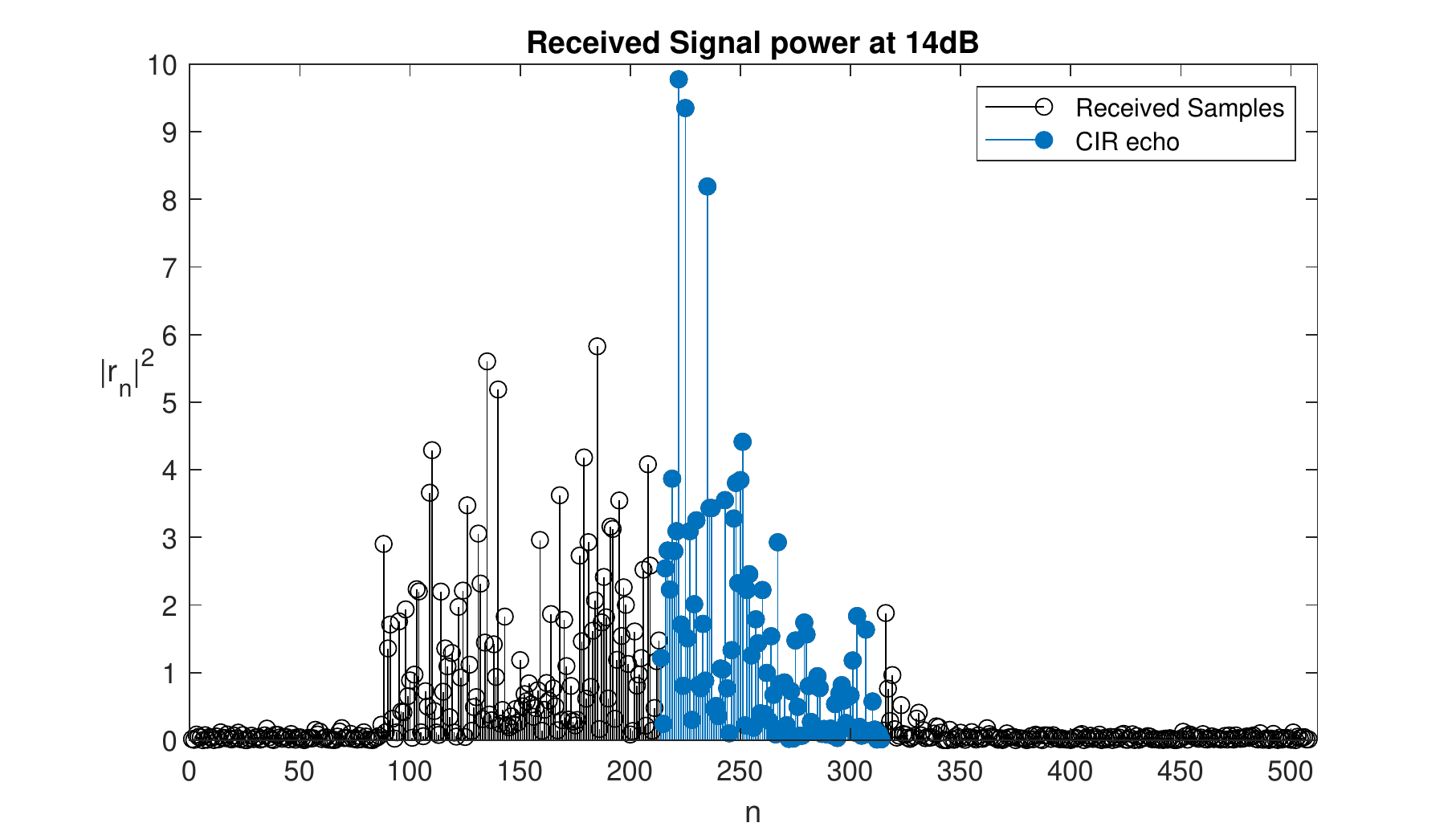}
    \caption{} \label{fig:receivedpower}
  \end{subfigure}
\end{center}
\begin{center}
  \vspace{-1.5ex}
\begin{subfigure}[b]{0.39\textwidth}
  \hspace{-2ex}
  \vspace{-1.5ex}
  \includegraphics[width=1.14\textwidth]{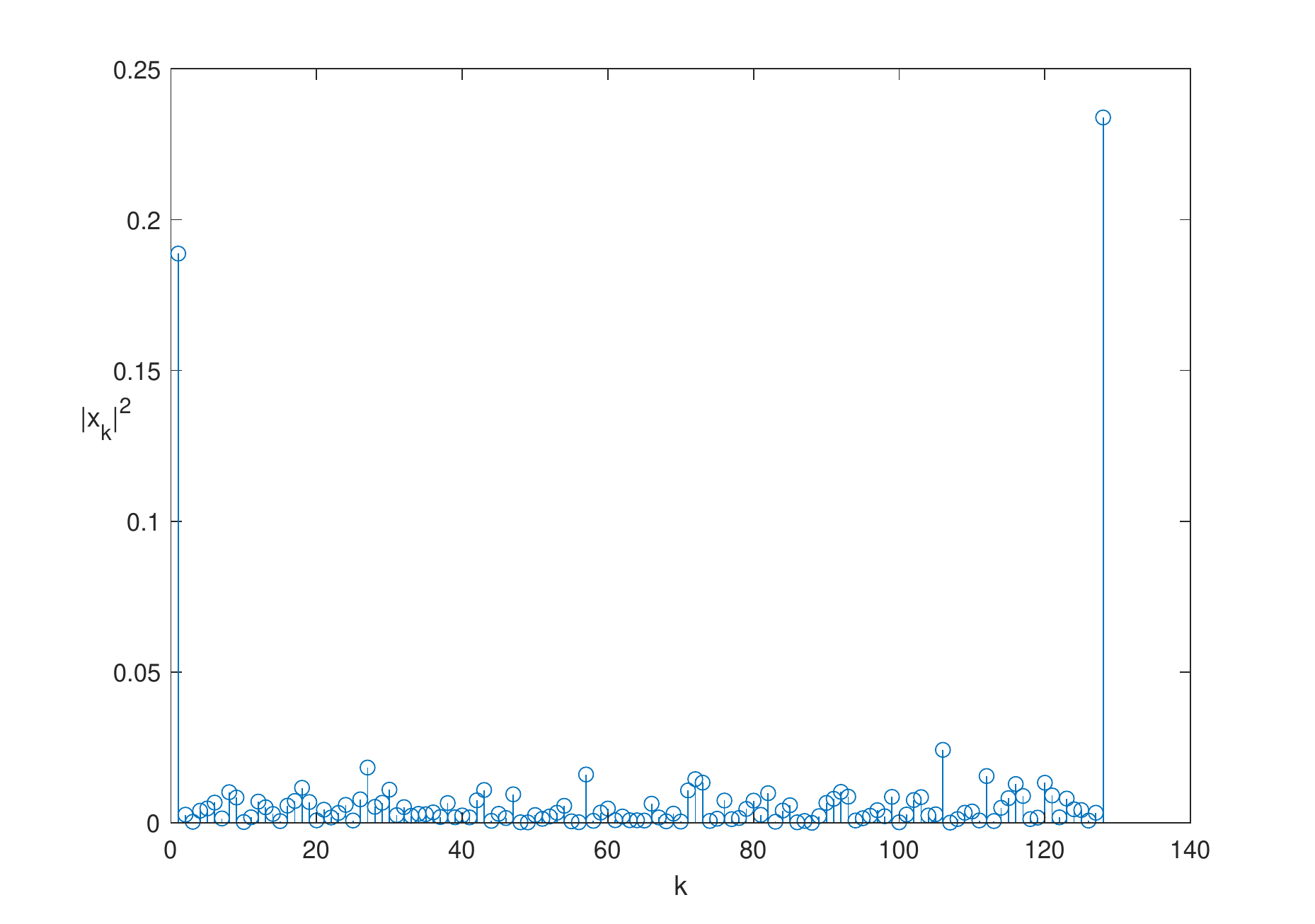}
  \caption{} \label{fig:huffmanK127}
\end{subfigure}
\begin{subfigure}[b]{0.59\textwidth}
  \hspace{2ex}
  \vspace{-1.5ex}
  \includegraphics[width=1.09\textwidth]{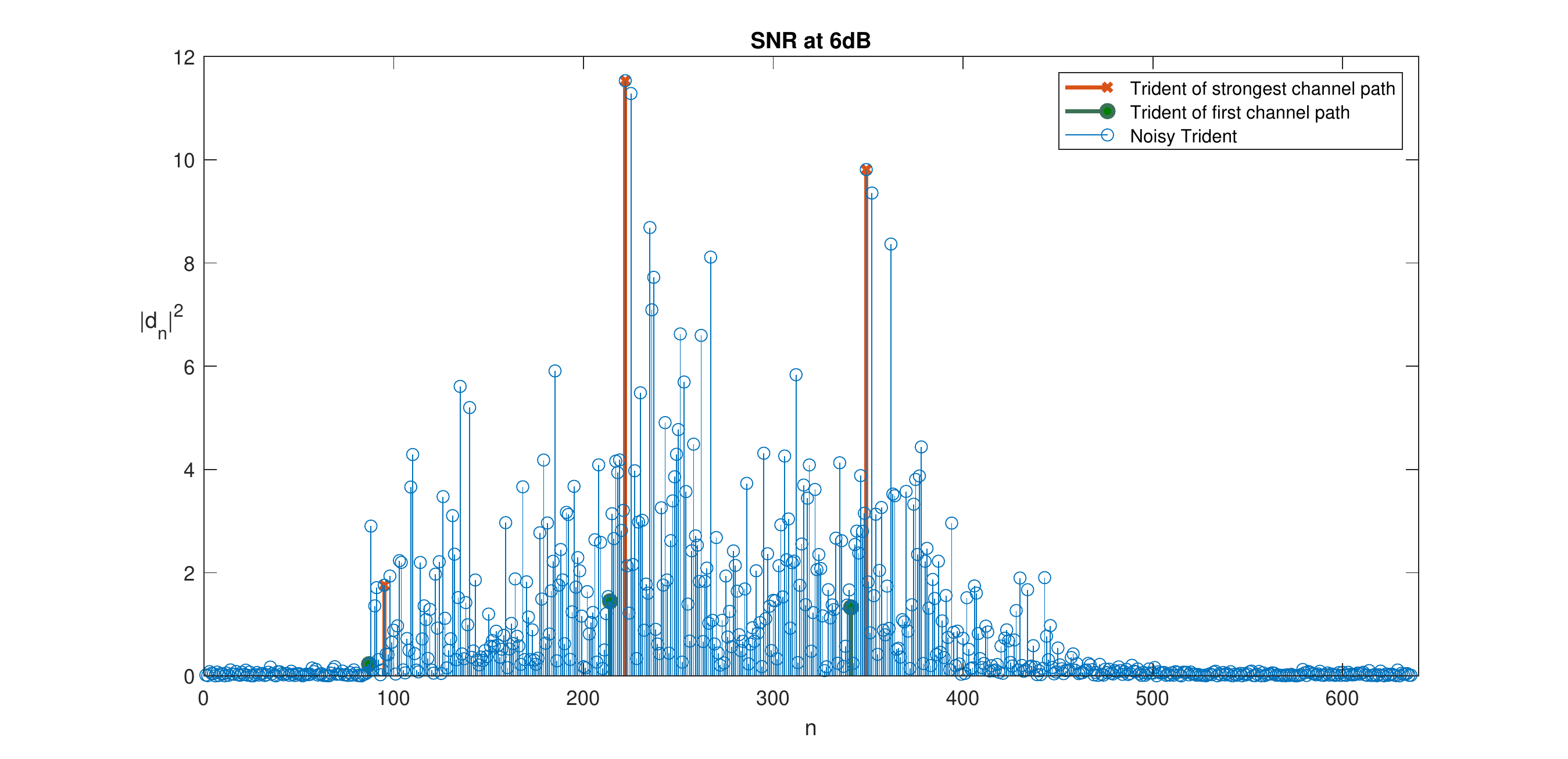}
  \caption{} \label{fig:NoisyTrident}
\end{subfigure}
\end{center}
\vspace{-3ex}
\caption{Absolute-squares of \subfigref{fig:ChannelPDPL128} a CIR realization with NLOS for $L=127,S=83,$ and $p=.98$ 
\subfigref{fig:huffmanK127} normalized BMOCZ symbol with $K=127$ and $68$ outer zeros  \subfigref{fig:receivedpower}
received samples which echoes the distorted CIR with efficient length $\Lcut=100$ and \subfigref{fig:NoisyTrident} the correlation of
the received samples with the Huffman bracket at  $\rSNR=14dB$ with the echo of the trident multiplied by the
strongest path in red and the first path in green at a timing-offset $\tau_0=86$.}\label{fig:timingoffset}
\end{figure}

Assuming the knowledge of the noise
floor $N_0$ at the receiver, a cut-off time can be defined as the window time which, for example, contains $95\%$ of the
received energy.
The estimation of the efficient channel length $\Lcut$ can be done after the detection of the timing-offset $\tau_0$
with \algref{alg:cpbs}. We
assume here that the maximal channel delay is $L$. Since the BMOCZ symbol length is $K+1$, we know that the samples
$\vr_h=(r_{\tau_0+K+1},\dots,r_{\tau_0+K+L})$ of the received time-discrete signal in \eqref{eq:receivedsamples}, which
is the CIR correlated by shifts of $\vx$ and distorted by additive noise (we ignore here the CFO distortion since it
will be not relevant for the PDP estimation), see \figref{fig:ChannelPDPL128}-\subref{fig:receivedpower}.  We therefore
need to determine $\Lcut$ by an energy concentration threshold, which depends on the \emph{instantaneous SNR} of $\vr_h$.
%% I dont know what that should be?
%\begin{align}
%  \iSNR=\frac{\Norm{\vr_h}^2}{K+1}-1.
%\end{align}
%%
%If $\Norm{\vx*\vh}^2/N\approx 1$, the $\iSNR$ is equal to the noise power density $N_0$, which we used in our
% simulations.  
We know, that the last channel tap $h_{\Lcut}$ will be multiplied by $|x_K|^2$, which is as strong as $|x_0|^2$ in
average.
%From the received signal
%we can determine the maximal sample $\rSNR$, by the noise floor $N_0$ at the receiver and the maximum peak
%power $\Norm{\vr}_{\infty}^2=|r_n|^2$. 
% 
There are many signal processing methods to detect the  efficient energy window
$\Ncut$ in the received samples, like total variation  smoothing \cite{BV04}, or regularized least-square methods
\cite{BV04,FR13} which promotes short window sizes (sparsity).  We propose in \algref{alg:CIRenergy} an iterative increasing
of $\Lcut$ starting at $1$ and increase until enough channel energy is collected.
Here we set the estimate channel/signal energy to 
\begin{align}
  E_r=\Norm{\vr_h}^2-LN_0
\end{align}
where we start with the maximal CIR length $L$.
By assuming a path exponent of $p$ we can calculate a threshold for the effective energy $\Ecut\simeq \mu E_r$ with
$\mu=p^{LN_0/2E_r}$. The algorithm then collects as many samples $\Lcut$ of $\vr_h$ until the energy $\Ecut$ is achieved
and sets $\Ncut=\Nalp+\Lcut$. The  extracted modulated signal is
then given by
\begin{align}
  \vty=(\tilr_{\htau_0},\tilr_{\htau_0+1}\dots,\tilr_{\tau_0+\Ncut}),
\end{align}
which will processed further for a CFO detection and final decoding. 
%
%We used for the simulations an
%iterative hard-thresholding algorithm by reducing the sample powers $|r_n|^2$ with multiples of the noise power $N_0$
%%
%\begin{align}
%  z_n=\max\{0,|r_n|^2-c N_0\} \quad,\quad n\in[\Tend]
%\end{align}
%%
%and find the local maxima of the noise reduced signal $\vz=(z_0,\dots,z_{\Tend-1})$.
%By increasing the threshold $cN_0$ we will reduce the amount of local maxima until only two positions remain, which
%will define the start $\htau_0$ and end point $\htau_0+\Ncut$. 
%\begin{figure}
\begin{figure*}[ttt!]
%  \centering
  \begin{minipage}[t]{0.5\textwidth}
  %\centering
    \begin{algorithm}[H]
  \caption{center-peak-back-step (CPBS)}
    \label{alg:cpbs}
    \begin{algorithmic}[1]
      %\Require Received magnitude-square $|\vr|^2$, threshold $\rho_0$, number of zeros $K$, initial estimation $\hatt$
      \Require $|\vr|^2$, $\rho_0$,  $K$, and $\hatt$
  \For{$b=1$ to $\lfloor K/2 \rfloor$}
  \If{$|r_{\hatt-b}|^2>\rho_0$ $\&$  $\sum_{n=1}^{b}|r_{\hatt-n}|^2 >\frac{b\rho_0}{1+\frac{\log b}{3}}$}
        \State $\htau_0=\hatt-b$
     \EndIf
  \EndFor\\
  \Return{$\htau_0$}
\end{algorithmic}
\end{algorithm}
\end{minipage}
\hfill
\begin{minipage}[t]{0.49\textwidth}
  \begin{algorithm}[H]
  \caption{CIR energy-detection (CIRED)}
    \label{alg:CIRenergy}
    \begin{algorithmic}[1]
      %\Require Receive signal $|\vr|^2$, TO $\htau_0$, noise $N_0$, number of zeros $K$, max CIR length $L$, PDP $\pdp$
      \Require $|\vr|^2$,  $\htau_0$,  $N_0$,  $K$,  $L$, and $\pdp$
      \State $E_r=\sum_{\ell=1}^{L} |r_{\htau_0+K+1+\ell}|^2-L N_0$;
      \State $\Ecut=|r_{\htau_0+K+1}|^2$; $\Lcut=1$; $\mu=\pdp^{L N_0/2E_r}$;
      \While{$\Ecut<\mu\cdot E_r$ }
       \State $\Lcut=\Lcut+1$;
       \State $\Ecut=\Ecut+|r_{\htau_0+K+\Lcut}|^2$;
     \EndWhile\\
   \Return{$\Lcut$}
\end{algorithmic}
\end{algorithm}
\end{minipage}
\end{figure*}

%% file: frequencyoffset.tex
% Frequency offset problem
\section{Carrier Frequency Offset}\label{sec:frequencyoffset}

\begin{figure}[t]
  \centering \def\svgwidth{0.5\textwidth} \footnotesize{ 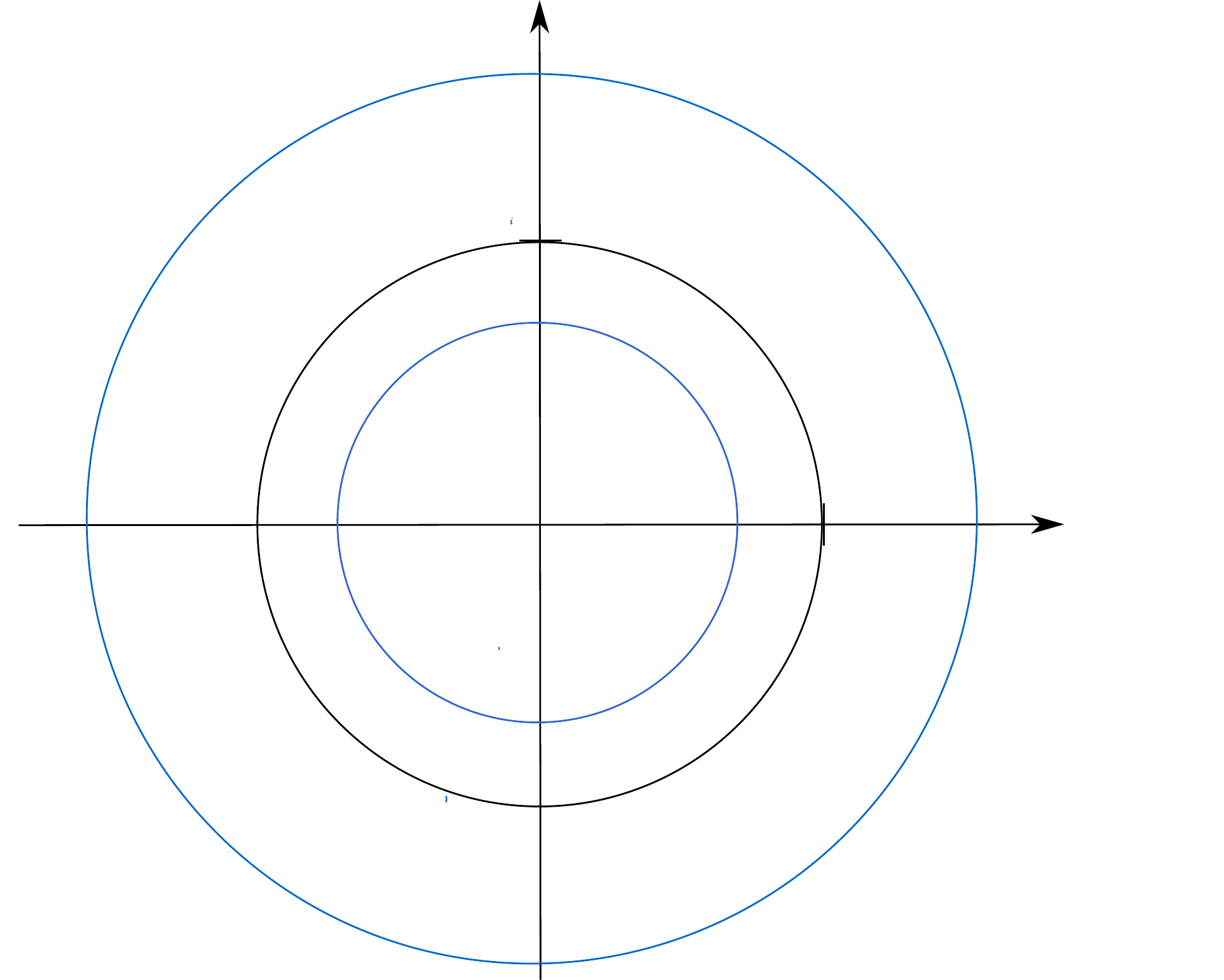 }
  \caption{Frequency-offset estimation via oversampling the DiZeT decoder for $K=6$ data zeros and without channel zeros
  by a factor of $Q=K=6$.  Blue circles denote the codebook $\Zero(6)$, solid blue circles the transmitted zeros
  $\valp_k$ and red the received rotated zeros $\talp_k$.}
  \label{fig:cfo_fractional}
\end{figure}
We assume now, that the down-converted baseband signal in \eqref{eq:receivedsamplesNobs} has no further timing-offset and captured all
path delays up to $N=K+L$. The signal will experience an unknown CFO of $\phi\in[0,2\pi)$
\begin{align}
  \ty_n=e^{\im n\phi}(\vx*\vh)_n+w_n \quad,\quad n\in[N]. \label{eq:CFOy}
\end{align}
This is a common problem in many multi-carrier systems, such as OFDM, which therefore require CFO estimation algorithms
\cite{Moo94,LLTC04,ZGX10}.
For a bandwidth of $W=1/T$, the relative frequency offset is 
\begin{align}
    \eps=\Del f \cdot T =\Del f/W.
\end{align}
Let us consider, for example, a carrier frequency of $f_c=1$GHz with a drastic frequency offset of $\Del f \in [-1,1]$MHz and
bandwidth $W=1$Mhz.
This would result in a relative frequency offset  $\eps\in[-1,1]$,
which is able to rotate all zeros by any $\phi=2\pi \eps\in [0,2\pi)$ in the $z$-plane.
Hence, the received polynomial (noiseless) will experience a rotation of all  its $N-1$ zeros by the angle $\phi$  
\begin{align}
  \puY(z)&=\sum_{n=0}^{N-1} \ty_n z^n = \sum_{n=0}^{N-1} y_n e^{\im n \phi}z^n = \uY(e^{\im \phi} z) = y_{N-1}\Pro_{n=1}^{N-1} (e^{\im \phi} z - \gam_n)\\
  &= e^{\im \phi(N-1)}
  y_{N-1} \Pro_{n=1}^{N-1} (z-\gam_n e^{-\im\phi})= e^{\im \phi(N-1)} x_K h_{L-1} \Pro_{k=1}^K (z-\alp_k e^{-\im \phi})
  \Pro_{l=1}^{L-1} (z-\bet_l
  e^{-\im \phi}).\notag
\end{align}
%
\if0 {\bfseries Problem:} The DiZeT decoder will need to know the rotation $\phi$ to sample $\tuY(z)$ at the right zero
positions of $\uX(z)$, otherwise we will obtain a large demodulation error. The rotation  does not need to include the
cyclic shift of the zeros, which is a integer multiple of $2\pi/K$. This indeed can be compensated by using rotational
invariant codes, developed for $K$ dimensional $K-$PSK signal constellations. Here, we can see the BMOCZ block $\vx$ in
the $z-$domain as $K$ successive transmitted $K-$PSK symbols $\alp_k\in S:=\set{R^m e^{\im 2\pi
  \frac{n}{K}}}{k=0,1,2,\dots,K-1, m=-1,1}$ . Hence $\valp$ is a $K$-dimensional (block) $K-$PSK codeword. RI codes on
  bases of Trellis-code-modulation (TCM) exist and where developed in the early 90ties, see for example \cite{NA03}.
  \fi
%
\noi As illustrated in \figref{fig:cfo_fractional}, we have to ensure that each rotated zero (red) $\talp_k=e^{-\im
\phi} \alp_k$ does not leave the zero-codebook set (blue)
$\Zero=\Zero(K):=\set{R^{\pm 1} e^{\im 2\pi\frac{k}{K}}}{k\in[K]}$.
To apply the DiZeT decoder, we have to find $\tht$ such that $e^{\im\tht}\tvalp\in \Zero$, i.e., we need to ensure that all
the $K$ data zeros will lie on the uniform grid. Hence, for $\tht_K=2\pi/K$ the CFO can be split in 
\begin{align}
  \phi=l\tht_K + \tht 
\end{align}
for some $l \in [K]$ and $\tht\in[0,\tht_K)$, where $l$ is called the \emph{integer} and $\tht$ the \emph{fractional
CFO}, which are also present in OFDM systems \cite[Cha.5.2]{CKYK10}.  Only if $\tht=0$ (or correctly compensated), the
DiZeT decoder, will sample at correct zero positions and decode, due to the unknown integer shift $l$, a cyclic permuted
bit sequence $\tvm=\vm\vS^l$, which we will correct in \secref{sec:cpc} by an cyclically permutable code.

\renewcommand{\vtD}{\ensuremath{\tilde{\vD}}}
\subsection{Decoding BMOCZ via FFT}
The DiZeT decoder for BMOCZ allows also a simple hardware implementation at the receiver.
Let us scale the received samples $y_n$ with the radius powers $R^n$ respectively $R^{-n}$
\begin{align}
  \vy\vD_{\uvR}:=\vy\begin{pmatrix}
    1 & 0  & \dots &0\\
    0 & R & \dots &0\\[-0.5em]
    \vdots & & \ddots& \vdots\\
    0 & 0&\dots & R^{N-1}\end{pmatrix} \label{eq:scalingDR}.
\end{align}
By applying the $\tN-$point unitary IDFT matrix $\Fmatrixa$ on the $N_0$ zero-padded scaled signal, where $\tN=QK$ with
$Q:=\lceil N/K\rceil$, we get the samples of the $z-$transform\footnote{An even more efficient FFT
  calculation with $N\log(N)$ could be achieved if $\tN=2^{n'}$ for some $n'\in\N$.}  by 
\begin{align*}
  \sqrt{N}\vy \begin{pmatrix}\vD_{\uvR} & \zero_{N,QK-N}\end{pmatrix}  \Fmatrixa 
 % = \begin{pmatrix}\sum_{n=0}^{N-1} y_n  R^n e^{\im 2\pi \frac{0\cdot n}{\tN}}\\ \vdots\\
 %   \sum_{n=0}^{N-1} y_n R^n  e^{\im 2\pi \frac{(\tN-1)\cdot n}{\tN}}\end{pmatrix}^T
    = \left(\sum_{n=0}^{N-1} y_n  R^n e^{\im 2\pi \frac{0\cdot n}{\tN}},\dots,\sum_{n=0}^{N-1} y_n R^n 
    e^{\im 2\pi \frac{(\tN-1)\cdot n}{\tN}}\right)
  =\uY(\valp_Q^{(1)})
%  ,\quad \sqrt{N} \vy\Fmatrixa  \begin{pmatrix}\vD_{\uvR^{-1}} \\ \zero_{N_0,N}\end{pmatrix}  = \uY(\valp_Q^{(0)})
\end{align*}
where $\alp^{(m)}_{Q,k}=\alp_k^{(m)} (e^{0},\dots,e^{\im 2\pi \frac{Q-1}{QK}})\in\C^{Q}$. Hence, the \emph{DiZeT
decoder} simplifies to %
\begin{align}
  \hm_{\nx} = \begin{cases} 1&, |\vy\vtD_{\uvR}\Fmatrixa|_{Q(k-1)}< R^{K-1}|\vy\vtD_{\uvR^{-1}} \Fmatrixa|_{Q(k-1)}\\
    0 &, \text{ else}
  \end{cases} \quad,\quad k=1,\dots,K.\label{eq:dizetdecoder}
\end{align}
Here, $Q\geq 2$ can be seen as an oversampling factor of the IDFT, where we pick each $Q$th sample point to obtain the zero
sample values.
Hence, the decoder can be fully implemented by a simple IDFT from the delayed amplified received signal, by
using for example FPGA or even analog front-ends.
We can also rewrite the diagonal scaling matrix \eqref{eq:scalingDR} in the symmetric form 
\begin{align}
  \vD_R:=\diag(R^{(N-1)/2},\dots,R^{-(N-1)/2})=R^{-(N-1)/2}\vD_{\uvR},
\end{align}
such that $\vD_R^{-1}=\vD_{R^{-1}}$ corresponds to a time-reversal of the diagonal, which brings us to 
\begin{align}
  |\vy \vtD_R |_{Qk} \Fmatrix^*_{QK}\leq |\cc{\vy^{-}} \vtD_R \Fmatrix^*_{QK}|_{Qk}\quad,\quad k\in[K],
\end{align}
since the absolute values cancel the phases from a circular shift $\vS$ and the conjugate-time-reversal
$\cc{\vy^-}=\cc{\vy}\vS\vGam$, where $\vGam=\Fmatrix^2$ is the circular time-reversal, can be rewritten by using
$\cc{\Fmatrixa\vGam}=\cc{\Fmatrix}=\Fmatrixa$.  

\subsection{Fractional CFO estimation via Oversampled FFTs}\label{sec:fractionalCFO}

\begin{figure}[t]
  \centering
\includegraphics[width=0.8\textwidth]{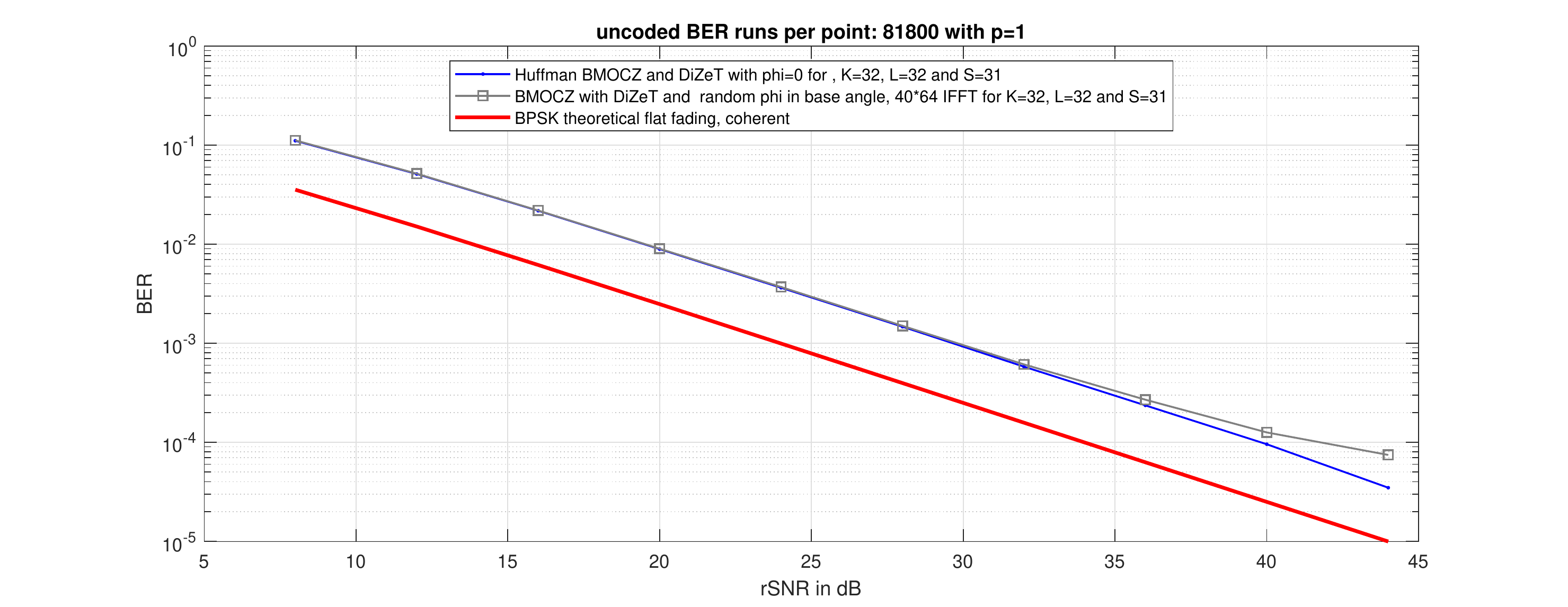}
\vspace{-2ex}
\caption{BER over rSNR for BMOCZ with DiZeT for random $\phi\in[0,\tht_K/2)$ for $L=K=32$.}
    \label{fig:fo_BER_IFFT}
\end{figure}

To estimate the factional frequency offset, we will oversample by choosing $Q>\lceil N/K\rceil$ to add $Q$ further $K$ zero
blocks to $\vD_R$. This leads to an oversampling factor of $Q$ and allows to quantize $[0,\tht_K)$ in $Q$ uniform bins
  with separation $\phi_Q=\tht_K/Q$  for a \emph{base angle} $\tht_K=2\pi/K$. Hence, the  absolute values of the sampled
  $z-$transform in \eqref{eq:CFOy} of the rotated codebook-zeros are given by
\begin{align}
  |\puY(e^{\im  q \phi_Q}\alp_k^{(1)})|= |\pvy\vtD_R \Fmatrix^*_{QK} |_{Qk\oplus_{\tN} q}\quad,\quad 
  |\puY(e^{\im  q \phi_Q}\alp_k^{(0)})|= |\cc{\pvy^-}\vtD_R \Fmatrix^*_{QK} |_{Qk\oplus_{\tN} q} 
\end{align}
for each $q\in [Q]$ and $k\in[K]$, where $\oplus_N$ is addition modulo $N$.
To estimate the fractional frequency offset of the base angle, we will sum the $K$ smaller sample values and
select the fraction corresponding to the smallest sum
\begin{align}
  \hat{q}&= \arg\min_{q\in[Q]} \sum_{k=1}^K 
  \min\{|\vty\vtD_R \Fmatrix_{QK} |_{Qk\oplus_{\tN} q}, |\cc{\vty^-}\vtD_R \Fmatrix_{QK} |_{Qk\oplus_{\tN} q}\}
  \quad\text{ and }\quad  \hat{\tht} = \frac{\hat{q}}{Q}\tht_K.
  \label{eq:fractionalCFOestimater}
%  \label{eq:hatfractionalCFO}
\end{align}
Then the recovered signal $\hvy=\pvy\vM^{-\htht}$ will have the data zeros on the constellation grid $\Zero_K$.
See \figref{fig:fo_BER_IFFT} for a random fractional CFO $\tht$ and \figref{fig:cfo_fractional} for
a schematic picture.\\

\subsection{Using Cyclically Permutable Codes}\label{sec:cpc}
To be robust against rotations which are integer multiple of the base angle, we will need an outer block code in
$\Ftwo^K$ for the binary message $\vm\in\Ftwo^B$, which is invariant against cyclic shifts, i.e., a bijective
mapping on the Galois Field $\Ftwo=\{0,1\}$ 
\begin{align}
  \Galg \colon \Ftwo^B\to \Ftwo^K,   \vm\mapsto \vc=\G(\vm)
\end{align}
%\footnote{We will from now on regard words of length $n$ as
%  row vectors and write them in vector notation $\va=(a_0,\dots,a_{n-1})\in\Galois_q^n$}.%
%
such that $\Galg^{-1}(\vc\vS^l)=\vm$ for any $l\in[K]$. We will use the common notation  for the code length $n=K$.
%Here $\Ftwo=\{0,1\}$ is the \emph{Galois Field} with characteristic $2$.
Such a block code is called a \emph{cycling
register code} (CRC) $\CodeCRC$ \cite{Gol67}, which can be constructed from the linear block code $\Ftwo^n$, by
separating it in all its \emph{cyclic equivalence classes} 
\begin{align}
  [\vc]_{\text{CRC}}=\set{\vc\vS^{l}}{l\in[n]}\quad,\quad \vc\in\Ftwo^n
\end{align}
where $\vc$ has \emph{cyclic order} $\nu$ if $\vc\vS^\nu =\vc$
for the smallest possible $\nu\in\{1,2,\dots,n\}$. To make coding one-to-one,  each equivalence class can be
represented by the codeword $\vtc$ with smallest decimal value \cite{RW75}\footnote{The authors call the cycling register codes
as cyclically permutable codes, which are nowadays defined differently. Furthermore, they claim that CRCs are also
comma-free codes, which is not true by definition.}. Then $\Ftwo^n$ is given by the union of all its equivalence class
representatives and its cyclic shifts, i.e.
\begin{align}
  \Ftwo^n=\bigcup_{i=1}^{M_{\text{CRC}}} \set{\vtc_i\vS^{l}}{l=1,\dots,\nu(i)}
\end{align}
This will generate in a systematically way a look-up table for the cycling register code.
Unfortunately, the construction is non-linear and combinatorial difficult.  However, the cardinality of such a code is
proven explicitly for any positive integer $n$ in \cite[Thm.VI.3]{Gol67} to be (number of cycles in a cycling
register)
\begin{align}
  |\CodeCRC(n)| =  \frac{1}{n}\sum_{d|n}\Phi(d) 2^{n/d}\label{eq:CReuler}
\end{align}
where $\Phi(d)$ is the \emph{Euler function}, which counts the number of elements $t\in[d]$ coprime to $d$.
For $n$ prime, we obtain 
\begin{align}
  |\Code_{\text{CR}}(n)|=\frac{1}{n} (2^{n} + (n-1)2) \geq \frac{2^n}{n}= 2^{n-\log_2 n}\label{eq:primitivelinear} 
\end{align}
which would allow to encode at least $B= n-\lceil\log n\rceil$bits. For $n=K=31$ this would result in a loss of only $5$bits and is similar to
the loss in a BCH-$(31,26)$ code, which can correct $1$bit error.  Note, the cardinality of a cycling register code
$\Code_{\text{CR}}$ is minimal if $n$ is prime. This can be seen by acknowledging the fact that the cyclic order of a
codeword is always a divisor of $n$, see for example \cite[Lem.1]{RL-NA18}. Therefore, if $n$ is prime we only have the
trivial orders $1$ and $n$, where the only codewords with order $\nu=1$ are the all one $\eins$ and all zero $\zero$
codeword and all other codewords have the \emph{maximal or full cyclic order} $\nu=n$.  Hence, extracting the CRC from
$\Ftwo^n$ obtains the same cardinality \eqref{eq:primitivelinear}.  Taking from a block code of length $n$ only the
codewords of maximal cyclic order and selecting only one representative of them defines a \emph{cyclically permutable
code} (CPC) and was first introduced\footnote{Let us emphasize at this point, that comma-free codes, introduced by
  Golomb et al. \cite{GGW58} are a class of codes with smaller size than CPCs or CRCs. Altough, the combinatoric is very
  similar and was proven to be \eqref{eq:CReuler} if the Euler function is substituted by the Möbius function, see for
  example \cite{Lev04}.} by Gilbert in \cite{Gil63}. If $n$ is prime, only two equivalence classes in $\CodeCRC$ are not
  of maximal cyclic order, hence the cardinality for a CPC if $n$ is prime is at most $(2^n-2)/n$.

However, the construction from $\Ftwo^n$, even for $n$ prime, is a combinatorial problem, especially the decoding.
Hence, to reduce the combinatorial complexity, many approaches starting from cyclic codes and extract all codewords with
maximal cyclic order \cite{KT06,SP08,L-NR14}.  Since a cyclic $(n,k,\dmin)$ code corrects up to $(\dmin-1)/2$ bit
errors, any CPC code extraction will inherit the error correction capability.  

We will follow an approach from \cite{KT06} to construct a CPC from a binary cyclic $(n,k,\dmin)$ code by still obtaining
the best possible cardinality $(2^{k}-2)/n$ if $n$ is a Mersenne prime. By reasons which will be clear later, we can
extract from this CPC an affine subcode whith maximal dimension, which we will call an  \emph{affine cyclically
permutable code} (ACPC). This allows a linear encoding of $B=n-\lfloor \log n \rfloor$ bits by a generator matrix and an
additive non-zero row vector, which defines the affine translation.

\longtrue % for ME!
\subsubsection{CPC construction from cyclic codes}
Cyclic codes exploit efficiently the algebraic structure of  Galois fields $\Galois_q$, given by a finite set having a
prime power cardinality $q=p^{m'}$.  A linear block code over $\Galois_q$ is a \emph{cyclic code} if each cyclic shift
of a codeword is a codeword. It is a \emph{simple-root cyclic code} if the characteristic $p$ of the field $\Galois_q$
is not a divisor of the block length $n$. %\cite{Cha98}. 
If the block length is of the form $n=q^m-1$, we call it a
primitive block length and if the code is cyclic we call it a \emph{primitive cyclic code}, \cite[Def.5.3.1]{Bla03}. We
will investigate here binary cyclic codes of length $n=2^m-1$ with $q=p=2$, which are simple-root and primitive cyclic
codes.  Due to its linearity, cyclic codes can be encoded and decoded by a generator $\vG$ and check matrix $\vH$ in a
systematic way.  The cardinality of a binary cyclic $(n,k)$ code is always $M_c=2^k$. Hence, by partitioning the cyclic
code in equivalence classes of maximal cyclic order and selecting one codeword as their representative leaves us with a
maximal cardinality of
\begin{align}
  |\CodeCPC|\leq \frac{2^k-1}{n}.\label{eq:optimalCC}
\end{align}
for any extracted CPC.  Note, the zero codeword is always a codeword in a linear code but has cyclic order one and hence
is not an element of a CPC.
%
%However, for our purpose, we always can add the zero-codeword such that we will have a cycling register of cardinality
%up to $2^{k-\log n}$ which allows to encode $k-\log n$ bits.
%
To exploit the cardinality most efficiently, the goal is to find cyclic codes such that each non-zero codeword has
maximal cyclic order.  We will follow a construction of Kuribayashi and Tanaka in \cite{KT06} for prime code lengths of
the form $n=2^m-1$, also known as \emph{Mersenne primes}. For $m=2,3,5,7$ this applies to $K=n=3,7,31,127$, which are relevant
signal lengths for binary short-messages. Furthermore, we will only consider cyclic codes which have $\eins$ as a
codeword. We will show later that this is indeed the case. Since $n$ is prime, each codeword, except $\zero$ and $\eins$
has maximal cyclic order. Hence, each codeword of a\footnote{There are multiple cyclic $(n,k)$ codes, which differ by
the choice of the generator polynomial.} cyclic $(n,k)$ code has maximal cyclic order and since it is a cyclic code all
its cyclic shifts must be also codewords. Hence the cardinality of codewords having maximal order is exactly $M =2^k-2$.
Therefore, we only need to partition the cyclic code in its cyclic equivalence classes, which will leave us with
$|\CodeCPC|=M/n =(2^k-2)/n$. Note, this number must be indeed an integer, by the previous mentioned properties, see also
\cite[Lem.2]{KT06}. 
The main advantages of the  Kuribayashi-Tanaka (KT) construction is the systematic code construction and the inherit
error-correcting capability of the underlying cyclic code, from which the CPC is constructed. Furthermore, combining
error-correction and cyclic-shift corrections will be ideal for BMOCZ. 

The code construction is two-folded. We have an inner $\Codein-(k,k-m)$ and outer $\Codeout-(n,k)$ cyclic code, where
the inner cyclic codewords will be affine translated by the Euclidean vector $\ve_1=(1,0,\dots,0)\in\Ftwo^k$.  In this
sense, the CPC construction is \emph{affine}.  This can be realized by an affine mapping from $\Ftwo^{k-m}$ to
$\Ftwo^n$, which can be represented by the BCH generator matrices $\vGin$ and $\vGout$ together with the affine
translation $\ve_1$ as
\begin{align}
  \begin{split}
  \Galg\colon \Ftwo^{k-m}&\to \Codein+\ve_1\to\Codeout\subset \Ftwo^n \\
  \vm &\mapsto  \vi=\vm\vGin+\ve_1 \mapsto \vc=\vi\vGout=\Galg(\vm).
\end{split}\label{eq:affineGalg}
\end{align}
To derive the generator matrix we can exploit the algebraic structure of the cyclic codes, given by its Galois fields.
By definition of cyclic codes, we have to factorize the polynomial 
\begin{align}
  x^n-1=(x-1)\Pro_{s=1}^S \uG_s(x)
\end{align}
in irreducible polynomials $\uG_s(x)$ of degree $m_s$, which must be divisors of $m$. If $n$ is prime, $m$ must be also
prime\footnote{The Mersenne prime number $n=2^m-1$ can be seen as a definition for prime numbers $m$. Assume, $m$ is not prime,
  than it is a composite number $m=ab$ for some integers $a,b>1$. But since the geometric series
  $\sum_{n=0}^{b-1}2^{an}= (2^{ab}-1)/(2^a-1)=n/(2^a-1)$ is an
integer, $n$ can not be prime.} 
and hence all the irreducible polynomials  are primitive and of degree $m_s=m$, except one of them, $\uG_0(x)=x-1$, has degree
one.
%Note, their ordering is arbitrary.
%
Hence, it must hold $S=(n-1)/m$.
As outer generator polynomial $\uGout(x)=\Pro_{s=S-J+1}^{S} \uG_s(x)=\sum_{i=0}^{Jm} g_{\text{out},i} x^k$ we choose the
product of the last $1\leq J\leq S$ primitive polynomials $\uG_s(x)$ yielding to a degree $Jm=n-k$, see
\cite[(23)]{KT06}. Each \emph{codeword polynomial} of degree less than $n$ is then given by
\begin{align}
  \uC(x)=\uI(x)\uGout(x)
\end{align}
where $\uI(x)=\sum_{a=0}^{k-1} i_a x^a$ is the \emph{informational polynomial} of degree less than $k$ and represented by the
binary information word $\vi=(i_0,i_1,\dots,i_{k-1})\in\Ftwo^k$, which we call the inner codeword. Similar, the codeword
polynomial $\uC(x)$ is represented by the CPC codeword $\vc\in\Ftwo^n$. 
By \cite[Thm.2]{KT06} a \emph{message polynomial} $\uM(x)=\sum_{b=0}^{k-m-1}m_b x^b$ of degree less than $k-m$ and $\uR(x)=1$
will be mapped to the information polynomial by the inner generator polynomial
\begin{align}
  \uI(x)&=\uM(x)\uGin(x) +1.
\end{align}
In \cite{KT06} the authors map all possible cyclic equivalence classes to $\uM(x)$, by separating the inner code in
$S-1$ separated inner codes. However, the remaining $S-2$ codes will only map $<2^{k-m}$ more message polynomials to
codeword polynomials and therefore be not enough to encode an additional bit. Hence, we will just omit these other inner
codewords.  This has the advantage, that we can write with \eqref{eq:affineGalg} the subset of the CPC as an
\emph{affine cyclically permutable code} (ACPC),  which is given by the polynomial multiplication over $\Ftwo$ 
\begin{align}
  \uC(x)&=\uI(x)\uGout(x)=(\uM(x)\uGin(x) +1 )\uGout(x)=\uM(x)\uG(x)+\uGout(x). \label{eq:polynomialenc} 
\end{align}
Here, we introduced a third generator polynomial $\uG(x)=\uGin(x)\uGout(x)$ which will be affine translated by the
polynomial $\uGout(x)$.  This generator polynomial  $\uG(x)$ will map surjective $\Ftwo^{k-m}$ to a cyclic code in
$\Ftwo^n$ and can therefore be expressed in matrix form as 
\newcommand{\vgout}{\ensuremath{\vg_{\text{out}}}} \newcommand{\goutzero}{\ensuremath{g_{\text{out},0}}}
\newcommand{\goutnk}{\ensuremath{g_{\text{out},n-k}}}
\begin{align}
   \vm \vG + \vgout\in\CodeACPC\subset\Codeout \label{eq:CPCenc} % = \vm \vGin\vGout + \ve_1\vGout=\vm\vGin\vGout + \vg_L
\end{align}
where $\vgout=(\goutzero,\dots,\goutnk,0,\dots,0)\in\Ftwo^n$ % and $\vg_L=(g_{L,0},\dots,g_{L,n-k}, 0 ,\dots,0)\in\Ftwo^n$ 
and the generator matrix\footnote{The notation is flipped compared to \cite{Bla03} to match the ordering of words
and polynomials. Note also, $\vc=\va\vG \LRA\vc^T=\vG^T\va^T$.} (Toeplitz) is given by
\begin{align}
  \vG&=\begin{pmatrix}
    g_{0} &  g_{1} &\dots  &g_{k-(k-m)-1} & g_{k-(k-m)} & 0 &\dots & 0\\
    0       &  g_{1} & \dots & g_{k-(k-m)-2} & g_{k-(k-m)-1} & g_{k-(k-m)} &\dots  & 0\\
    \vert & \vdots  &&& && \diagdown & \vdots\\
    0 &0 &\dots&&&  & \dots & g_{m}\end{pmatrix} 
  \in\Ftwo^{k-m\times k}.
\end{align}
For linear codes, we can use the Euclidean algorithm, to compute $x^{n-i}= Q_i(x)\uG(x) + \uS_i(x)$ for $i=1,\dots,k-m$,
%
% \label{eq:euclideanalg}
%
where the remainder polynomials $\uS_i(x)=\sum_j s_{i,j}x^j$ will have degree less than $n-k=n-(n-m)=m$, which allows
one to
rewrite the Toeplitz matrices as systematic matrices, see \cite[pp.112]{Bla03}.
Here, the check symbols $s_{i,j}$ define  the $k-m\times n$ \emph{systematic generator} and  $n\!-\!k\!+\!m\times n$
\emph{check matrix}  
\begin{align}
\renewcommand*{\arraystretch}{0.7}
\setlength\arraycolsep{1.5pt}
\vtG\!=\! [-\vP \ \id_{k-m} ],\ \vtH \!=\![ \id_{n-k+m}\ \vP^T ] \text{ with }\vP\!=\!\begin{pmatrix} 
    s_{k-m,0} & \dots & s_{k-m,n-k-1} & \zero_m \\
    s_{k-m-1,0} & \dots & s_{k-m-1,n-k-1} &\zero_m\\
    \vdots & & \vdots &\vdots\\
    s_{1,0} & \dots & s_{1,n-k-1} &\zero_m
  \end{pmatrix}%\in\Ftwo^{k-m\times n-k}
  \label{eq:systG}
\end{align}
such that $\vtG\vtH^T=\vzero_{k-m,n-k+m}$. Here we needed zero padding to embed the code in $\Ftwo^n$.
%
%Note that we rewrote  $-s_{ij}$ to $s_{ij}=-s_{ij} \mod 2$.
%
\ifcode See Matlab-\codref{cod:systematicGH}.\fi  This allows one to write the cyclic outer code as
$\Ftwo^{k-m}\vG=\Ftwo^{k-m}\vtG$. Of course, each $\vm\vG$ and $\vm\vtG$ will be mapped to different codewords, but this
is just a relabeling.  Since $\Ftwo^{k-m}\vtG$ defines a cyclic $(n,k-m)$ code and $n$ is prime, each codeword must have
$n$ distinct cyclic shifts.  The affine translation $\vgout$ separates than each of these $n$ distinct cyclic shifts by
mapping them to representatives of distinct cyclically equivalence classes of maximal order.  This gives us a very
simple {\bfseries encoding rule} for each $\vm\in\Ftwo^B$:
\begin{align}
  \vc=\vm\vtG+\vgout=\Galg(\vm)\in\CodeACPC\label{eq:sysCPCenc},
\end{align}
and {\bfseries decoding rule}. Here an ACPC codeword $\vc$ can be decoded by just subtracting the affine
translation $\vgout$ and cut-off the last $B=k-m$ binary letters to obtain the message word $\vm\in\Ftwo^B$, see
\eqref{eq:systG}. However, we will observe a cyclic shifted codeword $\vv=\vc\vS^l$ and by construction we know that
only one cyclic shift will be an element of \CodeACPC\ and consequently it holds 
\begin{align}
  \forall j\not=l \mod (n-1) \colon \vtc_j=\vv\vS^{-j} -\vgout\not \in\Ftwo^{k-m}\vtG\quad \LRA\quad
  \vtc_j\vtH^T\not=\zero\label{eq:decshift}.
\end{align}
Hence, we only need to check all $n$ cyclic shifts of the sense-word $\vv$ to identify the correct cyclic shift $l$,
which is given if $\vtc_l\vtH^T=\zero$.  If there is an additive error $\ve$ we can use the {\bfseries error correcting} property of
the \emph{outer cyclic code} \Codeout in \eqref{eq:polynomialenc} to repair the codeword.
%
%However, if the cyclic shift is not known, it will introduce a much larger
%error vector, which will be in most cases not correctable. Hence, we will need the \emph{outer cyclic code \Codeout}
%embedding to correct the additive error.  
%
Here, we can use again the systematic matrices
$\vtGout$ and $\vtHout$ to represent the outer cyclic code in a systematic way. Note, that each information message
$\vi$ which corresponds to a CPC codeword $\vc$ will be an element of $\Ftwo^{k}\vtGout$. Furthermore, all its cyclic
shifts $\vc\vS^l$ will be outer cyclic codewords. Hence, by observing the sense-word 
\begin{align}
  \vtv=\vc\vS^l+\ve
\end{align}
and determining its syndrome $\vs=\vv\vtHout^T$, which we look-up int the syndrome table $\vT_{\text{synd}}$  of
$\vtHout$  to identify the corresponding coset leader (error-word)
$\ve$, which recovers the shifted codeword (we assume maximal $t$ errors, bounded-distance decoder) as
\begin{align}
  \vv=\vtv-\ve,
\end{align}
see for example \cite[p.59]{Bla03}.  This allows to correct up to $t=(n-k-1)/2$ errors of the cyclic codeword. From the
additive error-free sense-word $\vc$ we can, as described previously, identify the correct shift $\hat{l}$ from
\eqref{eq:decshift} and by taking the last $B=k$ letters the original message $\vm$.  If the error vector $\ve$
introduces more than $t$ bit flips, the error correction will fail and the chance is high that we will mix up the ACPC codewords
and experience a block (word) error. Therefore, we will need low coding rates for the outer cyclic code $\vGout$ to
prevent such a catastrophic error.
The identified cyclic shift and fractional CFO estimation \eqref{eq:fractionalCFOestimater} yields then the estimated CFO 
\begin{align}
  \hat{\phi}=\hat{\tht} + \hat{l}\tht_K.
\end{align}

% end ifcode

\if0 % redundant
\noi {\bfseries Decoding:}
%
First we perform an additive error-correction in the outer code $\Codeout$, which works as for cyclic codes via the
syndrome-table $\vT_{\text{synd}}$ and the check matrix $\vtHout$, see \cite[p.59]{Bla03}.  First, the syndrome
$\vs=\vv\vtHout^T$ is computed and looked-up in the syndrome-table to identify the corresponding coset leader (error-word)
$\ve$, which yields the recovered shifted codeword (we assume maximal $t$ errors, bounded-distance decoder) as
%
\begin{align}
  \vtc=\vv-\ve.
\end{align}
%
From the additive error-free but circular shifted codeword $\vtc$ we construct all $n$ shift, subtract the constant
shift vector $\va=\vgout$, and apply the check matrix $\vtHout$ to find the one shift which is most likely the ACPC
codeword\ifcode, see Matlab-\codref{cod:ACPCdecoder}\fi.
\fi
\ifcode
\begin{lstlisting}[caption={ACPC Encoder},captionpos=b,label=cod:ACPCencoder]
function cw=ACPCencoder(mw,tG,a)
    cw=rem(mw*tG,2);     % cyclic codeword (inner codeword)
    cw=mod(cw+a,2);      % affine shift, ACPC codeword
end
\end{lstlisting}

\begin{lstlisting}[caption={ACPC Decoder},captionpos=b,label=cod:ACPCdecoder]
function mw=ACPCdecoder(vw,tHout,Tsynd,tG,tH,a,n,B)
s=rem(vw*tHout',2);                 % syndrome 
ew=Tsynd(1+bi2de(s,'left-msb'),:);  % error vector
tcw=mod(vw+ew,2);                   % additive error correction
serror=zeros(1,n);
for l=0:1:n-1
    serror(l+1)=sum(rem(mod(circshift(tcw,l)+a,2)*tH',2)); % shift error value
end
[~,sopt]=min(serror);               % optimal shift error position
mw=mod(circshift(tcw,sopt-1)+a,2);  % shift error correction
mw=mw(n-B+1:n);                     % message word, last B letters  
end
\end{lstlisting}
\fi  % end ifcode
% EXAMPLE

\begin{example}
  The most non-trivial\footnote{For $m=2$ only $\eins$ and $\zero$ are the BCH codewords, which
  would be removed for the CPC. Note, $m$ has to be prime.} example of ACPC is for $m=3$ and $l=1$, which gives 
  \begin{align}
    n=2^3-1=7,\quad k=7-3=4, \quad B=4-1=1
  \end{align}
  This is also the Hamming $(7,4)$ code with minimal distance $\dmin=3$ and hence can correct $1$ bit error. The code is
  also perfect.
\end{example}
\begin{example}
For $K=n=31$ we get for $t=2$ error
corrections a message length of $k=21$ in a cyclic BCH-$(31,21)$ code from which we can construct a CPC of cardinality
\cite[(25),(30)]{KT06}
\begin{align}
  |\CodeCPC|=\sum_{i=1}^4 2^{5(i-1)+1}=2^1+2^6+2^{11}+2^{16}=6750\geq 2^{16}.
\end{align}
This allows to encode $B=16=k-m$bits. The cardinality is optimal for any cyclic $(31,26)$ code \eqref{eq:optimalCC}
\begin{align}
  |\CodeCPC|=\frac{M_c}{n}=\frac{2^{21}-2}{31}=2\frac{2^{20}-1}{2^{5}-1}=\sum_{i=0}^{3} 2^{5i+1}= 2^1+2^6+2^{11}+2^{16}=6750.
\end{align}

\end{example}

The next example would be for $n=127$, which we also simulated in \figref{fig:simober}. Unfortunately, the next Mersenne
prime is only at $8191$, which might be not anymore considered as a short-packet length.  However, the authors want to
emphasize that other CPC constructions exist which  do not require Mersenne prime lengths $n$ or even binary alphabets.
In \cite{L-NR14} the authors showed construction of CPCs with alphabet size $q$, given as a power of a prime, and block
length $n=q^m-1$ for any positive integer $m$. Hence, for $q=2$ and a binary cyclic code $(n=2^m-1,k,\dmin)$ which
satisfy \cite[Thm.1]{L-NR14} will result in $(2^k-1)/n$ CPC codewords. An example is given for $n=15=2^4-1$ and
$k=8,\dmin=4$ where $n$ is not a Mersenne prime number and the maximal bound \eqref{eq:optimalCC} of $(2^8-1)/15=17$ is
achieved, allowing to encode $B=4$ bits and correct $2$ bit errors.  which is very close to the BCH code $(15,7)$ with
$2$bit error correction.

Let us mention the shortest non-trivial CPC, which can be addressed without cyclic code construction, but also with no
error correction capabilities.

\begin{example}
For $n=3$ we get only have $2^3=8$ binary words of length $3$ which have $4$ different cyclic permutable codewords
%
%\begin{align}
%  \vc_1=\begin{pmatrix} 1\\ 1\\ 1\end{pmatrix}, 
%  \vc_2=\begin{pmatrix} 1\\ 1\\ 0\end{pmatrix},
%  \vc_3=\begin{pmatrix} 1\\ 0\\ 0\end{pmatrix},
%  \vc_4=\begin{pmatrix} 0\\ 0\\ 0\end{pmatrix}
$\vc_1=\eins, \vc_2=( 1, 1, 0),\ 
  \vc_3=(1,0,0),\ 
  \vc_4=\zero$
%\end{align}
%
allowing to encode $B=2$bits of information with no error correction.
However, this code allows an TO estimation, as well as a CIR estimation.  
For a super short control signal, this might be therefore interesting. 
By omitting $\vc_1$ and $\vc_4$, i.e., by dropping one bit of information, we can even estimate with the DiZeT oversampling decoder
any possible CFO.  
\end{example}

%% file: frequencyoffsetDiZeT2.pdf_tex
%% Creator: Inkscape inkscape 0.92.1, www.inkscape.org
%% PDF/EPS/PS + LaTeX output extension by Johan Engelen, 2010
%% Accompanies image file 'frequencyoffsetDiZeT2.pdf' (pdf, eps, ps)
%%
%% To include the image in your LaTeX document, write
%%   \input{<filename>.pdf_tex}
%%  instead of
%%   \includegraphics{<filename>.pdf}
%% To scale the image, write
%%   \def\svgwidth{<desired width>}
%%   \input{<filename>.pdf_tex}
%%  instead of
%%   \includegraphics[width=<desired width>]{<filename>.pdf}
%%
%% Images with a different path to the parent latex file can
%% be accessed with the `import' package (which may need to be
%% installed) using
%%   \usepackage{import}
%% in the preamble, and then including the image with
%%   \import{<path to file>}{<filename>.pdf_tex}
%% Alternatively, one can specify
%%   \graphicspath{{<path to file>/}}
%% 
%% For more information, please see info/svg-inkscape on CTAN:
%%   http://tug.ctan.org/tex-archive/info/svg-inkscape
%%
\begingroup%
  \makeatletter%
  \providecommand\color[2][]{%
    \errmessage{(Inkscape) Color is used for the text in Inkscape, but the package 'color.sty' is not loaded}%
    \renewcommand\color[2][]{}%
  }%
  \providecommand\transparent[1]{%
    \errmessage{(Inkscape) Transparency is used (non-zero) for the text in Inkscape, but the package 'transparent.sty' is not loaded}%
    \renewcommand\transparent[1]{}%
  }%
  \providecommand\rotatebox[2]{#2}%
  \ifx\svgwidth\undefined%
    \setlength{\unitlength}{557.11291785bp}%
    \ifx\svgscale\undefined%
      \relax%
    \else%
      \setlength{\unitlength}{\unitlength * \real{\svgscale}}%
    \fi%
  \else%
    \setlength{\unitlength}{\svgwidth}%
  \fi%
  \global\let\svgwidth\undefined%
  \global\let\svgscale\undefined%
  \makeatother%
  \begin{picture}(1,0.80113081)%
    \put(0,0){\includegraphics[width=\unitlength,page=1]{frequencyoffsetDiZeT2.pdf}}%
    \put(0.75463356,0.32937643){\color[rgb]{0,0,0}\makebox(0,0)[lb]{\smash{$R$}}}%
    \put(0.64276471,0.69767502){\color[rgb]{0,0,0}\makebox(0,0)[lb]{\smash{$\alp_2$}}}%
    \put(0.26796442,0.72449228){\color[rgb]{0,0,0}\makebox(0,0)[lb]{\smash{3}}}%
    \put(0.03825287,0.39897568){\color[rgb]{0,0,0}\makebox(0,0)[lb]{\smash{4}}}%
    \put(0.65515476,0.04493417){\color[rgb]{0,0,0}\makebox(0,0)[lb]{\smash{$K=6$}}}%
    \put(0,0){\includegraphics[width=\unitlength,page=2]{frequencyoffsetDiZeT2.pdf}}%
    \put(0.19965304,0.04182965){\color[rgb]{0,0,0}\makebox(0,0)[lb]{\smash{5}}}%
    \put(0,0){\includegraphics[width=\unitlength,page=3]{frequencyoffsetDiZeT2.pdf}}%
    \put(0.54885041,0.74452954){\color[rgb]{0,0,0}\makebox(0,0)[lb]{\smash{$\talp_2$}}}%
    \put(0,0){\includegraphics[width=\unitlength,page=4]{frequencyoffsetDiZeT2.pdf}}%
    \put(0.85644461,0.46051787){\color[rgb]{0,0,0}\makebox(0,0)[lb]{\smash{$2\phi_Q$}}}%
    \put(0,0){\includegraphics[width=\unitlength,page=5]{frequencyoffsetDiZeT2.pdf}}%
    \put(0.53870755,0.37556403){\color[rgb]{0,0,0}\makebox(0,0)[lb]{\smash{$\phi$}}}%
    \put(0.70108585,0.38095173){\color[rgb]{0,0,0}\makebox(0,0)[lb]{\smash{$\phi_Q$}}}%
    \put(0,0){\includegraphics[width=\unitlength,page=6]{frequencyoffsetDiZeT2.pdf}}%
    \put(0.59438631,0.45994141){\color[rgb]{0,0,0}\makebox(0,0)[lb]{\smash{$\tht_6$}}}%
    \put(0,0){\includegraphics[width=\unitlength,page=7]{frequencyoffsetDiZeT2.pdf}}%
    \put(0.81615075,0.33410141){\color[rgb]{0,0,0}\makebox(0,0)[lb]{\smash{$\alp_1$}}}%
    \put(0,0){\includegraphics[width=\unitlength,page=8]{frequencyoffsetDiZeT2.pdf}}%
    \put(0.71248322,0.65823108){\color[rgb]{0,0,0}\makebox(0,0)[lb]{\smash{$Q\phi_Q$}}}%
    \put(0.67432951,0.32822107){\color[rgb]{0,0,0}\makebox(0,0)[lb]{\smash{$1$}}}%
    \put(0.38085368,0.6101178){\color[rgb]{0,0,0}\makebox(0,0)[lb]{\smash{$\im$}}}%
  \end{picture}%
\endgroup%

%% file: simulations.tex
\section{Simulations}

\begin{figure}[t]
\centering \def\svgwidth{0.6\textwidth} \footnotesize{ 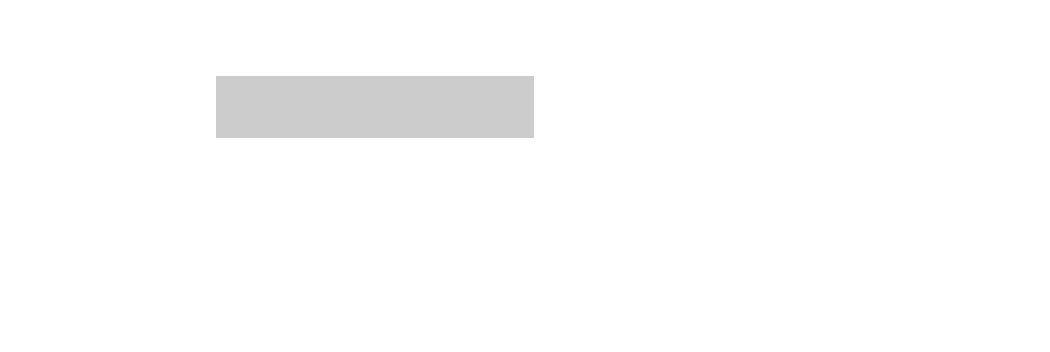 }
\caption{Comparison to two successive OFDM-DPSK blocks, one OFDM-IM, a pilot and QPSK, and a BMOCZ block in time-domain.
The solid bars denote the used subcarriers/zeros.}
  \label{fig:ofdmdpsk}
\end{figure}

%We simulated for data block lengths $K=31=2^5-1$ the BER performance over $\Ebno$ for BMOCZ at different code rates
%(for error correction), see \figref{fig:ber31}. 
We determined by Monte-Carlo simulations with MatLab 2017a the bit-error-rate (BER) over averaged $\Ebno$ and $\rSNR$
under various channel settings and block lengths $K$.  The transmit and receive time will be in all schemes $N=K+L$ over
which the CIR $\vh$ of length $L$ is assumed to be static, see \figref{fig:ofdmdpsk}.  Therefore, the energy per bit is
$E_b=B/N$ with $B=\Norm{\vm}_1$ which is the inverse of the spectral efficiency $\rho=N/B$.  In each simulation run, the
CIR coefficients are redrawn according to the channel statistic given by the decay exponent $p\leq 1$.
The CFO $\phi$ is drawn uniformly from $[0,2\pi)$ and
  the timing-offset $\tau_0$ uniformly form $[0,N]$.  Furthermore, the additive distortion is also drawn from an i.i.d.
  Gaussian distribution $\CN(0,N_0)$, which we will scale to obtain various received SNR and $\Ebno$ values,
  \figref{fig:afo_bersnr2}. 
%
%  After simulating all SNR points, a new channel, carrier frequency offset, and noise vector will be drawn. Hence, each
%  run of an SNR point will see a new realization of the channel, CFO, T0, and the noise. 

Due to the embedded error correcting and a complete failure if a wrong CPC codeword is detected, the BER and BLER
(block-error-rate) are
almost identical over received SNR for BMOCZ-ACPC, see \figref{fig:afo_blersnr2}.

Of course, it would be also possible to detect a wrong decoding and request a retransmit.
However, this is in contrast to our system aspects since we want to address sporadic communication of single packets.

\subsection{Multiple Receive Antennas}

For a receiver with $M$ antennas, we may exploit receive antenna diversity, since each antenna will receive the transmit
signal over an independent CIR realization (best case). Due to the short wavelength in the mmWave band large antenna
arrays with $\lam/2$ spacing can be easily installed on small devices.   We assume
in all simulations:
\begin{itemize}
  \item The $B$ information bits $\vm\in\Ftwo^B$ are drawn uniformly.
  \item All signals arrive with the same timing-offset $\tau_0$ at the $M$ receive antennas (dense
    antenna array, fixed relative antenna positions (no movements)).
  \item The clock-rate for all $M$ antennas is identical, hence all  received signals have same CFO.
  \item The maximal CIR length $L$, sparsity level $S$, and PDP $p$ are the same for all antennas.
  \item Each received signal experience an independent noise and CIR realization, with 
    sparsity pattern $\vs_m\in\{0,1\}^L$ for $|\supp{\vs_m}|=S$ and  $\vh_m\in\C^L$ where $h_{m,l}=\CN(0,s_{m,l}p^l)$. 
\end{itemize}
The DiZeT decoder \eqref{eq:dizetdecoder} for BMOCZ without TO and CFO, can be implemented straight forward to a
\emph{single-input-multiple-output} SIMO antenna system: 
\begin{align}
  \hm_{\nx} = \begin{cases} 1&, \sum_{m=1}^M|\vy_m\vtD_{\uvR} \Fmatrixa|^2_{Q(k-1)}<
    R^{2K-2}\sum_{m=1}^M|\vy_m\vtD_{\uvR^{-1}} \Fmatrixa|^2_{Q(k-1)}\\
    0 &, \text{ else}
  \end{cases}.
\end{align}
%for $k=1,\dots,K$.

\subsection{CFO Estimation for Multiple Receive Antennas}

As in \secref{sec:fractionalCFO} we can estimate the
fractional CFO by \eqref{eq:fractionalCFOestimater} for $M$ received signals $\vy_m$
\begin{align}
  \hat{q}= \arg\min_{q\in[Q]} \sum_{k=1}^K \min\left\{\sum_{m=1}^M| \vty_m \vtD_R \Fmatrix_{QK}|_{Qk\oplus_{\tN} q},
  \sum_{m=1}^M |\cc{\vty^-}_m \vtD_R \Fmatrix_{QK}|_{Qk\oplus_{\tN} q}\right\}\label{eq:fractionalCFOestimaterSIMO}.
\end{align}

\if0
In \figref{fig:varchannel_berebno} we simulated the BER performance over \Ebno\ for various channel scenarios. It can be
seen that for a very sparse and fast decaying power delay profile, the performance loss will be $1-4$dB, especially if the
channel length becomes in the order of the signal length.  
\fi

\begin{figure}[t]
\centering 
  \begin{subfigure}[b]{0.49\textwidth}
    \hspace{-2ex}
    \includegraphics[width=1.05\textwidth]{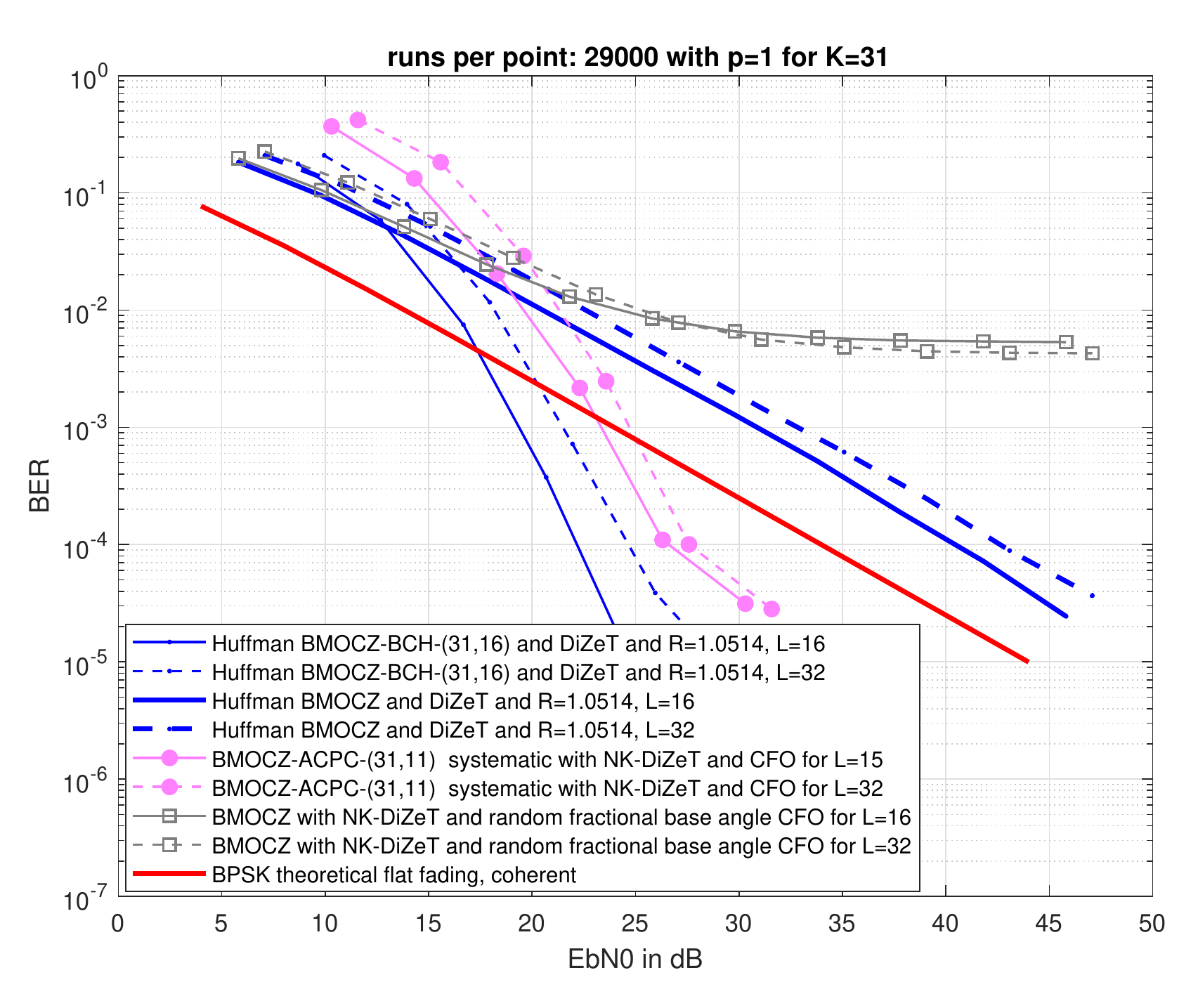}
    \vspace{-1cm}
   \caption{}
  \label{fig:afo_bersnr2}
  \end{subfigure}
  \begin{subfigure}[b]{0.49\textwidth}
  \hspace{1ex}
  %  \centering 
    \includegraphics[width=1.03\textwidth]{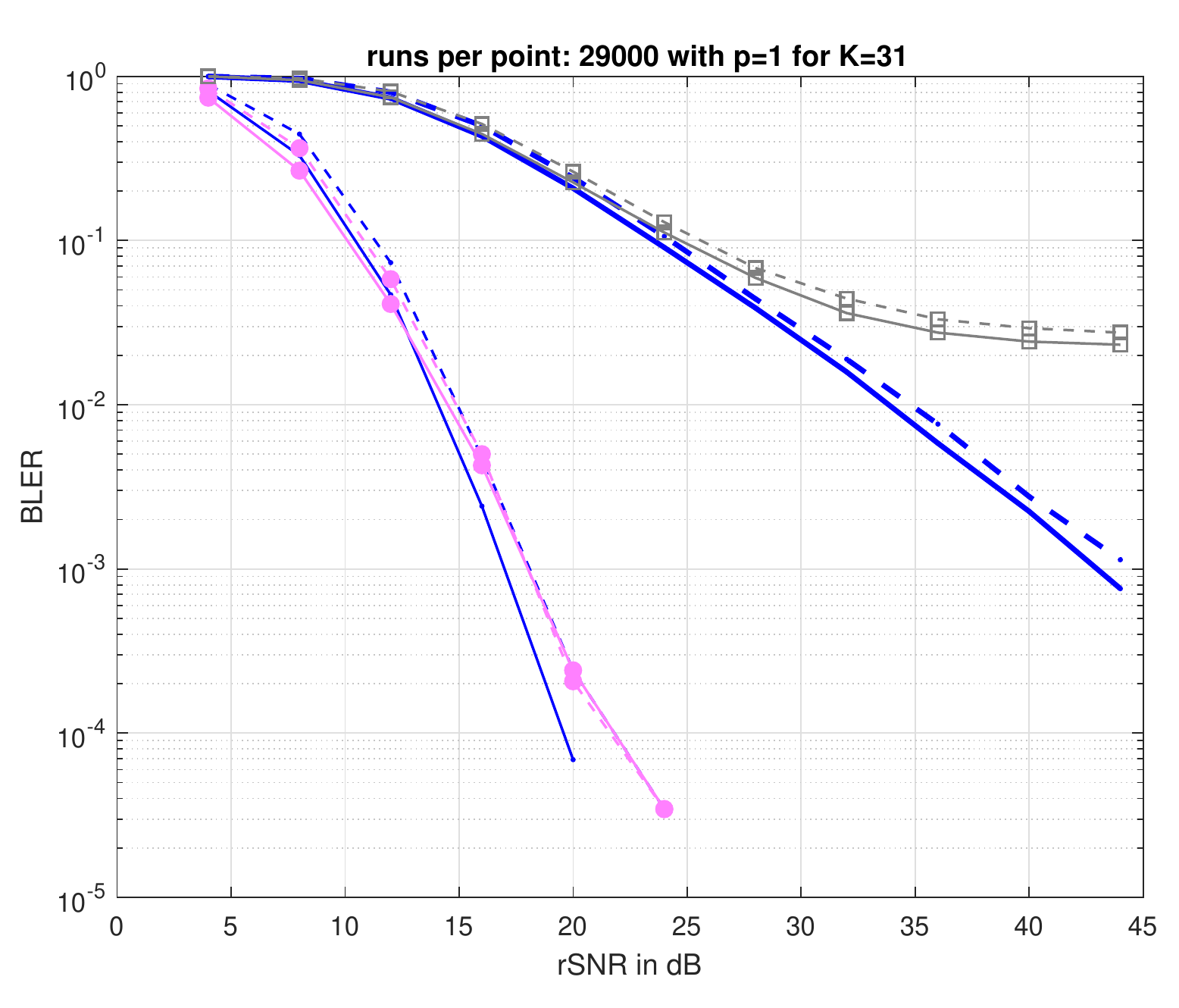}
    \vspace{-1cm}
  \caption{}
  \label{fig:afo_blersnr2}
  \end{subfigure}
  \vspace{-2ex}
  \caption{Simulations for arbitrary CFO with ACPC-(31,11) for $B=11$bits. The $E_b/N_0$ averaged over the fading
  channel is defined here as $1/BN_0$ for $B$ information bits. }
  \label{fig:ber31}
\end{figure}

The effect of different coding rates on BER and \emph{block error rate} (BLER) for the BMOCZ-ACPC-$(K,B)$ scheme with
$K=31$ transmitted zeros and channel lengths $L=16,32$ with flat power profile $p=1$ is shown in
\figref{fig:ber31}. If the coding rate is decreased to $B/K=6/31\simeq1/5$, allowing up to $5$ bit error corrections,
we achieve a BLER of $10^{-1}$ at almost $6$dB received SNR, which is $6$dB better as for the coding rate $16/31\simeq
1/2$. Hence, if power is an issue, the coding rate can be decreased accordingly to approach a low SNR regime, at the
cost of data rate. 

\begin{figure}[t]
%  \centering 
  \vspace{-3ex}
  \includegraphics[width=\textwidth]{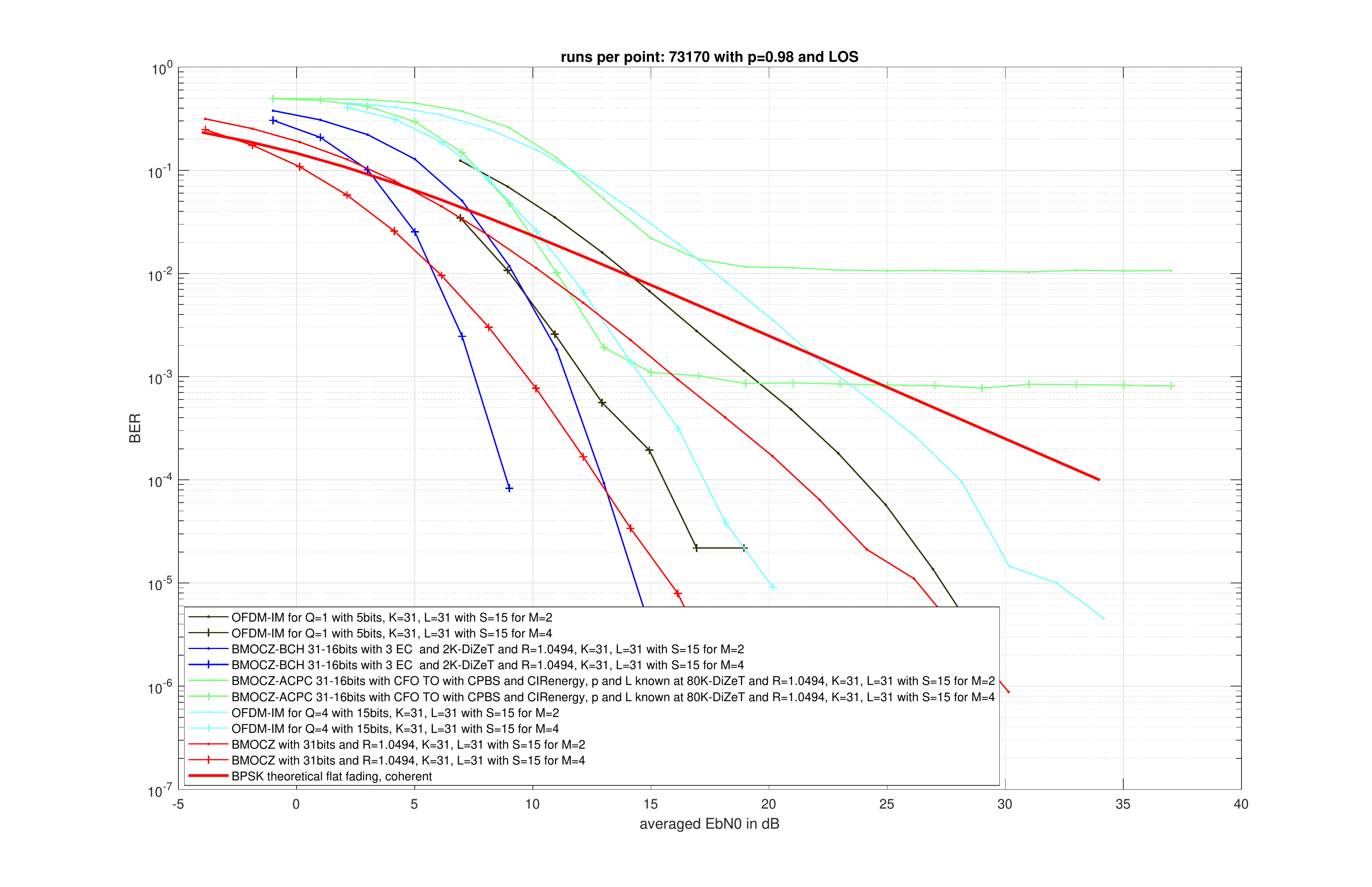}
  \vspace{-1.5cm}
\caption{BER over \Ebno\ with $M=2$ and $4$ antennas for BMOCZ-ACPC-$(31,16)$ distorted by TO and CFO (green curves). 
OFDM-IM is without CFO and TO.}
\label{fig:OFDMIM_ACPC_TO_CFO}
\end{figure}

\subsection{Timing and Effective Channel Length Estimation}\label{sec:TOundCutsim}

We chose a random integer timing offset in $\tau_0\in\{0,1,2\dots,N-1\}$ in \figref{fig:OFDMIM_ACPC_TO_CFO} and with an
exponentially power delay profile exponent $\pdp=0.98$ and dominant LOS path.  In the CPBS \algref{alg:cpbs} and CIRenergy
\algref{alg:CIRenergy} we used the sum of all received antenna samples and the sum of the noise powers 
\begin{align}
  |r_n|^2=\sum_{m=1}^M |r_{m,n}|^2 \quad \text{ and }\quad  \sigma^2=MN_0.
\end{align}
Indeed, if the CIR length is larger with
exponential decay, a wrong TO estimation yields to less performance degradation as for shorter lengths, since the last
channel tap will be much smaller in average power. 
We simulated by choosing randomly for each simulation, consisting of $D$  different noise powers $N_0$,  
Each binary plain message $\vm$ will result in a codeword $\vc\in\Ftwo^K$ which corresponds to a normalized BMOCZ symbol
$\vx\in\C^{K+1}$. We will normalize the CIR at the transmitter to $\vth=\vh/\Expect{\Norm{\vh}^2}$,
where the average energy  of the CIR is given by \eqref{eq:averageCIRpower} as the expected power delay profile  in
$\vs$
\begin{align}
  E_{S,h}= \Expect{\Norm{\vh}^2}= \sum_{l=0}^{L-1} s_l p^l.
\end{align}
%
\if0
\begin{figure}[hb]
\begin{center}
\begin{subfigure}[b]{0.49\textwidth}
  \includegraphics[width=1.2\textwidth]{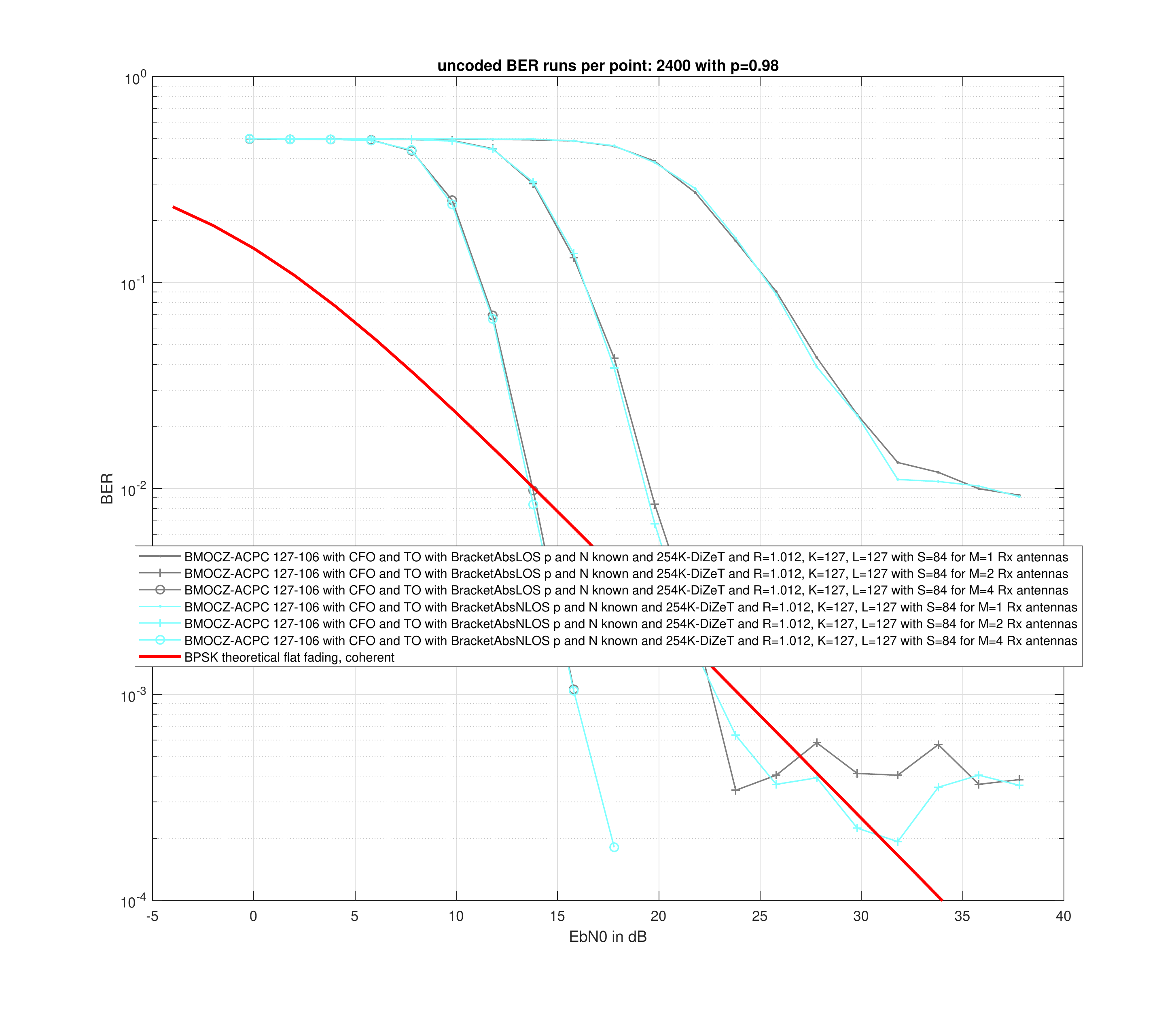}
  \caption{} \label{fig:BERoverEbno_S83_LOS}
\end{subfigure}
\begin{subfigure}[b]{0.49\textwidth}
  \includegraphics[width=\textwidth]{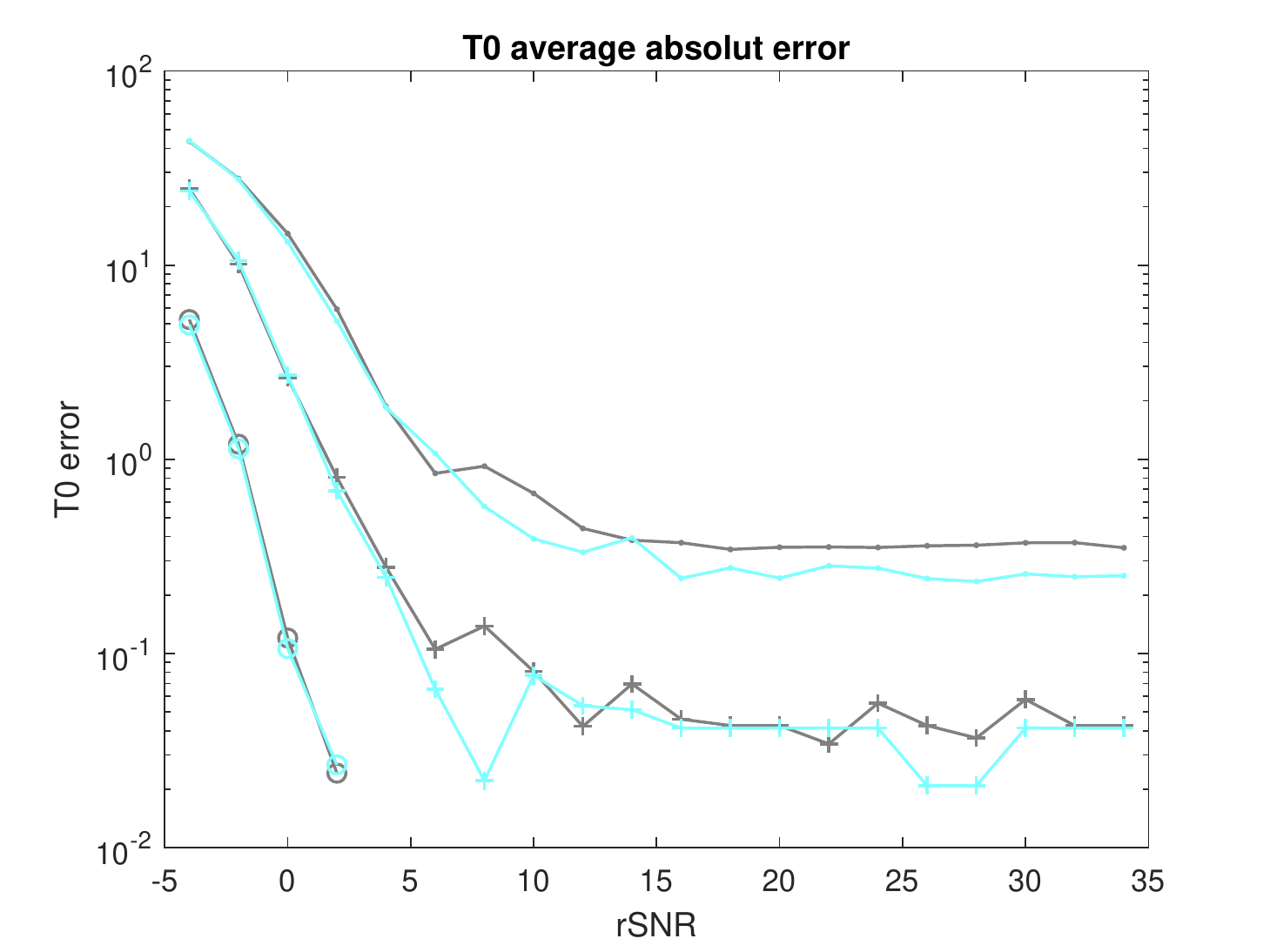}
  \caption{} \label{fig:T0error_S83_LOS}
\end{subfigure}
\end{center}
\vspace{-3ex} \caption{BER over $E_b/N_0$ for $p=0.98, L=127$ and $S=83$ with LOS for $M=1,2$ and $4$ receive antennas. 
  Right figure shows timing-offset error for $2400$ runs for each SNR point.}
\label{fig:ebnoLOSlong} 
\end{figure}
\fi
This ensures that for each selected sparsity pattern of the CIR, we will have normalized CIR energy if selecting random
channel taps by the law of large numbers. By averaging over the sparsity pattern, this would result in a large deviation of
the CIR energy and would require many more simulations, therefore we calculated the average power with knowledge of the
sparsity patterns, i.e., by knowing the support realization.
%\paragraph{LOS path simulations}
%
%In many cases, we will have a LOS path, which will be the path with shortest delay and dominant in power, i.e. $s_0=1$.  
%We therefore set  $|\tilh_0|^2=1/2$  and $\tilh_l=h_l/\sqrt{E_{S,h}-1}$.
%However, this will result in most cases with a very strong LOS path, since half the energy is spread over $S$ taps,
%although exponential decaying. We therefore simulated with $\tilh_0$ as the tap with strongest power $|h_s|^2$ in $S$ and
%scale it by $p^{-s}$, to reflect the power delay profile. Note, this will be in average give $|h_0|^2=1$ and therefore
%result in a normalized CIR if we set $\tilh_l=h_l/\sqrt{E_{S,h}}$ for $l>0$ as before. 

\begin{figure}[H]
  \vspace{-3ex}
  \begin{subfigure}[b]{0.47\textwidth}
  \hspace{-1ex}
  \includegraphics[width=1.25\textwidth]{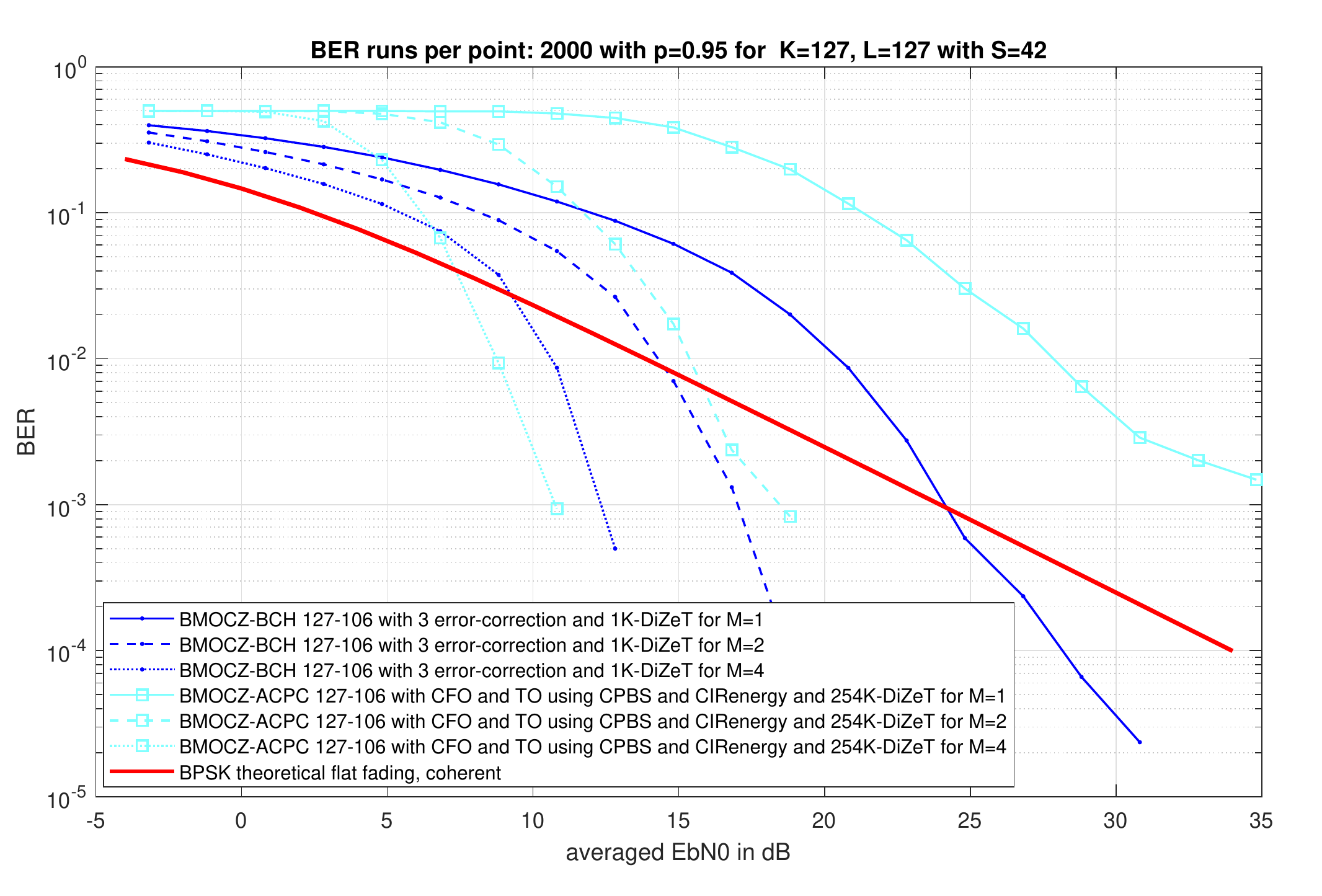} 
  \end{subfigure}
  \hspace{7ex}
  \begin{subfigure}[b]{0.47\textwidth}
  \includegraphics[width=0.98\textwidth]{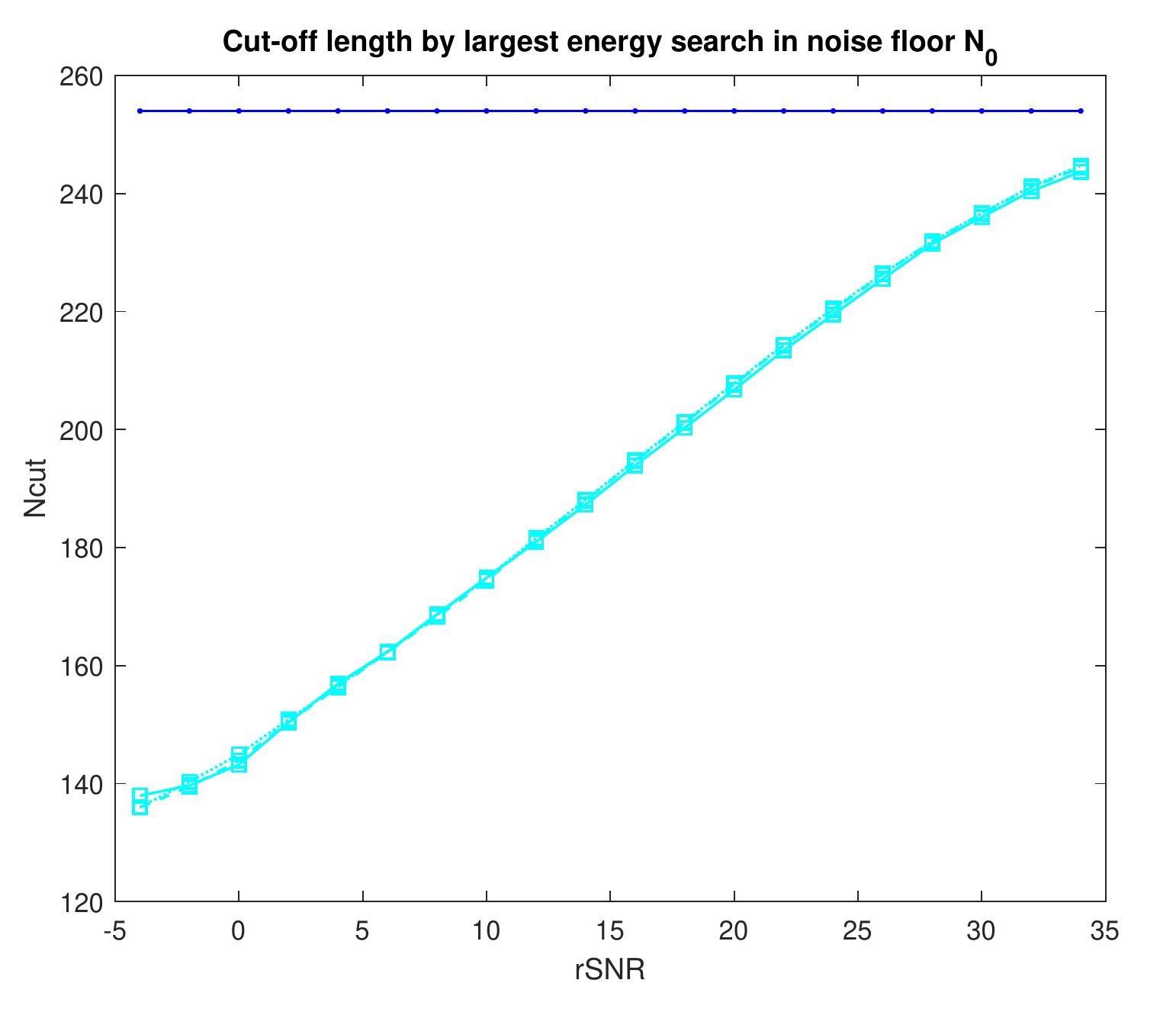}
  \end{subfigure}
  \vspace{-5ex}
  \caption{SIMO for $M=1,2,4$ receive antennas over $2$k runs per point at $p=0.95$ and $S=42$. BCH with no CFO and TO and
    ACPC with CFO and TO using \algref{alg:cpbs}+\ref{alg:CIRenergy}.}\label{fig:simober}
  \vspace{-2ex}
\end{figure}

\begin{figure}[H]
\vspace{-4ex}
\begin{subfigure}[b]{0.47\textwidth}
  \hspace{-2ex}\includegraphics[width=1.15\textwidth]{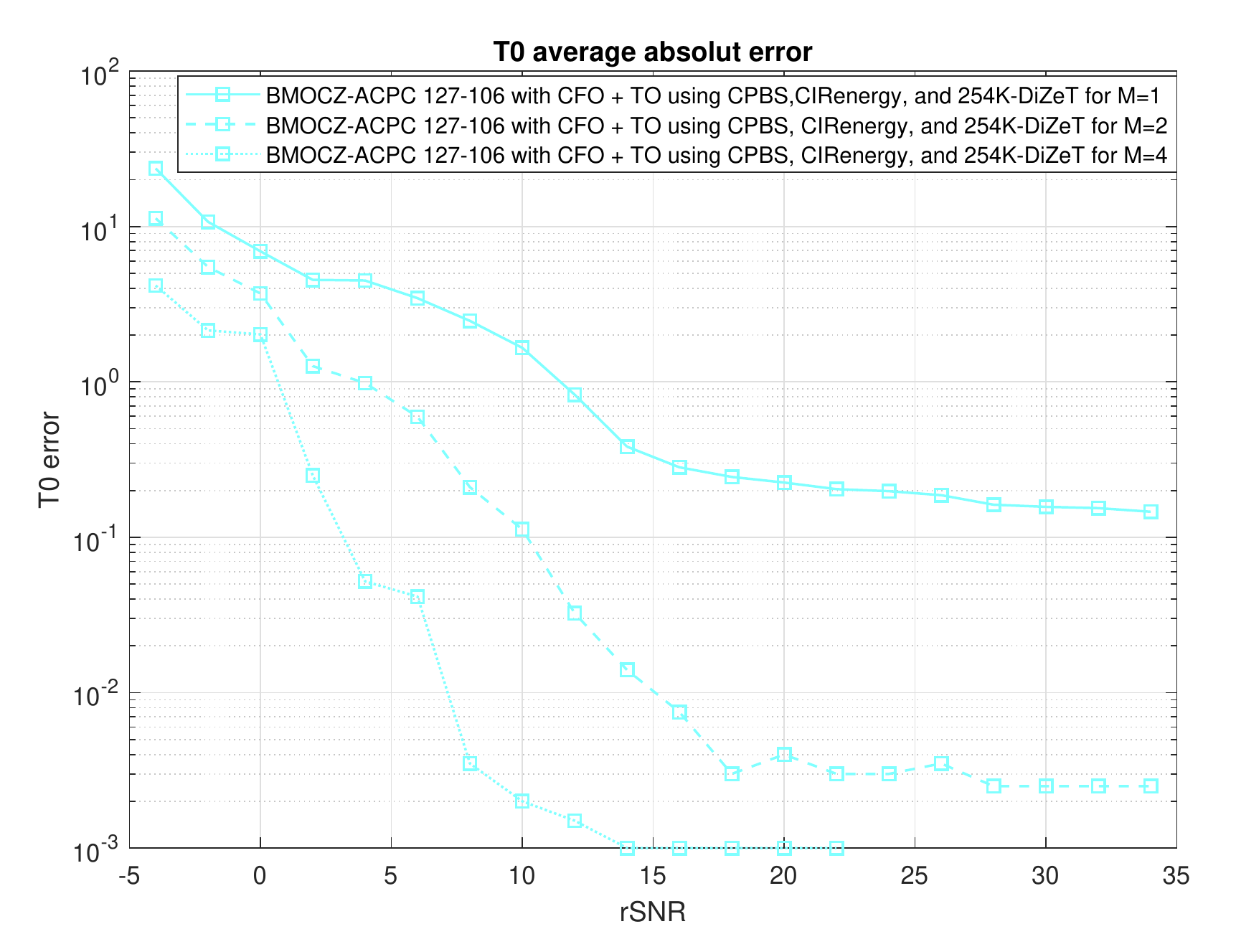} 
\end{subfigure}
\vspace{-3ex}
\hspace{2ex}
\begin{subfigure}[b]{0.47\textwidth}
\includegraphics[width=1.12\textwidth]{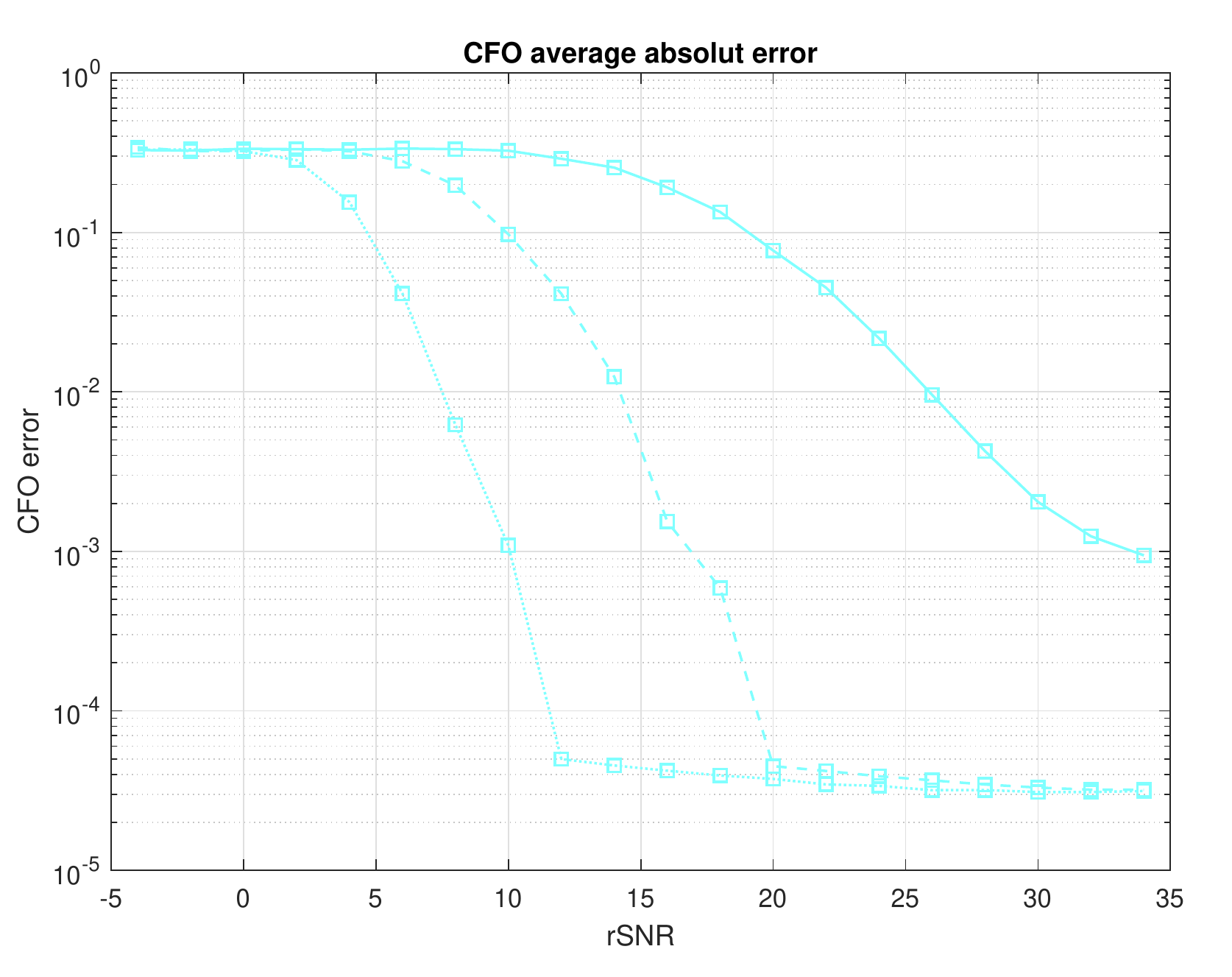}
\end{subfigure}
\caption{TO and CFO  estimation errors for $M=1,2$ and $M=4$ receive antennas.}
\label{fig:T0Lcut}
\end{figure}

\subsection{Comparison to Noncoherent Schemes}

\begin{figure}[t]
\begin{subfigure}[b]{\textwidth}
  \centering 
  \hspace{-1ex}\includegraphics[width=1.05\textwidth]{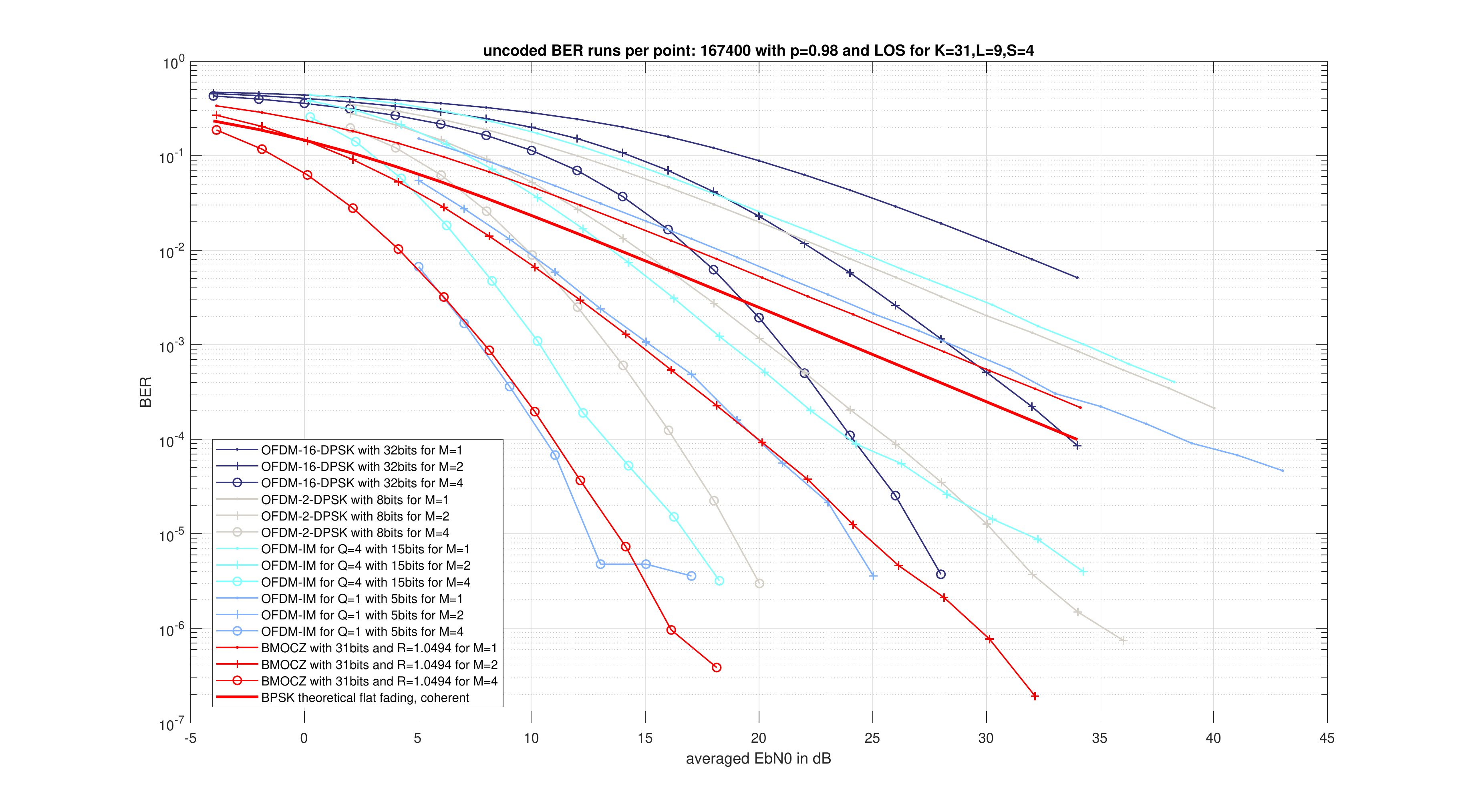}
  \vspace{-7ex}
  \caption{ }
  \label{fig:BMOCZ_IM_DPSK_L9}
\end{subfigure}
\begin{subfigure}[b]{\textwidth}
  \centering
  \includegraphics[width=0.9\textwidth]{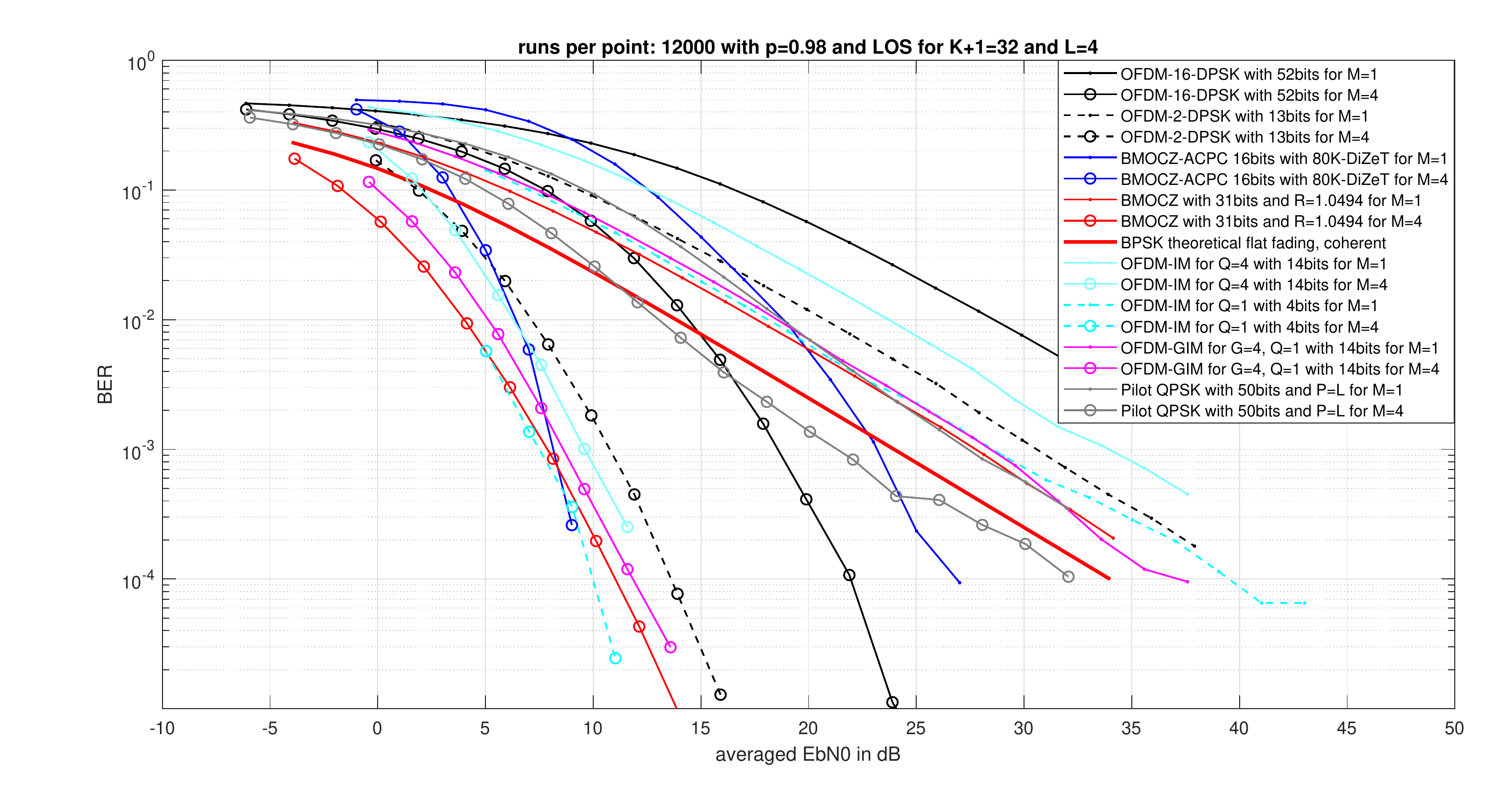}
  \vspace{-3.0ex}
  \caption{}
  \label{fig:BMOCZ_IM_DPSK_L4}
  \vspace{-1.5ex}
\end{subfigure}
  \caption{BMOCZ comparison to OFDM-IM, -GIM, -DPSK with $2$ OFDM symbols, and Pilot-QPSK for $1$ and $4$ antennas at
    maximal CIR length $L=9$ in \subfigref{fig:BMOCZ_IM_DPSK_L9} and $L=4$ in \subfigref{fig:BMOCZ_IM_DPSK_L4}.}
\end{figure}

First, we will compare without TO and CFO BMOCZ to OFDM and pilot based schemes, as in \cite{WJH18b}, in
\figref{fig:quadrigaCIR} with a single antenna where the CIR is generated from the simulation framework Quadriga
\cite{JRBT14} and finally in \figref{fig:BMOCZ_IM_DPSK_L9} with multiple-receive antennas. 

\paragraph{Pilot and SC-FDE with QPSK} Here we use a pilot impulse $\sqrt{\frac{E}{2}}\vdel_0$ of length $P=L$ to
determine the CIR and $K+1-L$ symbols to transmit data via QPSK in a single-carrier (SC) modulation by a
 \emph{frequency-domain-equalization} (FDE). Applying FDE, we can then decode the QPSK or
QAM modulated OFDM subcarriers $K+1-L$, see \cite{WJH18a}. The energy $E$ is split evenly between the pilots and data
symbols, which results in better BER performance for high SNR, see \figref{fig:BMOCZ_IM_DPSK_L4} (we not use SNR
knowledge at the transmitter) 

\paragraph{OFDM-Index-Modulation (IM)} We also compare to: OFDM-IM with $Q=1$ and $Q=4$ active subcarriers out of
$\NIM=\Nalp+1$ and to \emph{OFDM-Group-Index-Modulation} (GIM) with $Q=1$ active subcarriers in each group of size
$G=4$. To obtain an OFDM symbol we need to add a cyclic-prefix, which requires $\NIM\geq L$.  For OFDM systems, the CFO
will result in a circular shift of the $\NIM$ subcarriers and hence create the same confusion as for BMOCZ. The only
difference is, that OFDM operates only on the unit circle, whereas BMOCZ operates on two circles inside and outside the
unit circle.  Note, in OFDM-IM a cyclic permutable code is not applicable, since the information for example with $Q=1$
is a cyclic shift, which is exactly what the CFO introduces. For more active subcarriers $Q>1$ and grouping the carriers
in groups, ICI free IM schemes can be deployed, see for example \cite{WCY17}. A group size of $G=4$ seems to perform the
best for OFDM-IM. However, this will require $L\ll K$ which is not our proposed regime for MOCZ.

\paragraph{OFDM-Differential-Phase-Shift-keying (DPSK)}
We will use two successive OFDM blocks to encode differentially the bits via $Q$-PSK over $\Nxdpsk$ subcarriers. To ensure
the same transmit and receive lengths as for BMOCZ, we will split the BMOCZ symbol length $\Nx$ in two OFDM symbols with
cyclic prefix $\vxcpone$ and $\vxcptwo$ of equal length\footnote{Note, that $K$ has to be odd to divide by $2$.} 
%
%\begin{align}
%  \Ndpsk=\frac{\Nx}{2}\quad,\quad \Nxdpsk=\Ndpsk-(\Nh-1).\\
$\Ndpsk=N/2$,
%\end{align}
%
where we have to chose $N=K+L$ to be even. Furthermore, to include a CP of length $L-1$ in each OFDM block, we need
$\Ndpsk=(K+L)/2\geq 2L-1$ resulting in the requirement $K\geq 3L-2$.  If $L$ is even and $K=nL$ for some $2<n\in\N$ we
get $\Nxdpsk=\Ndpsk-L+1=(n-1)L/2+1$ subcarriers in each OFDM block. The shortest transmission time  is then given for
even $L$ with $n=3$ by $N=4L$, resulting in $\Nxdpsk=L+1$ subcarriers. Modulating them with $Q$-PSK allows to transmit
$(L+1)\log Q$ bits differentially. To match the spectral-efficiency of BMOCZ as best as possible, we select $Q=8$ to
encode $3$bits per subcarrier and hence $B=(L+1)3=(K/3+1)3=K+3$ message bits, which is $3$bits more than BMOCZ.
%$L-1 = (K+1)/4$ we get $\Nxdpsk=\Ndpsk-\Nh+1=(\Nx)/4$, which requires $Q=16$ to transmit $\Nx$ bits via $16$-PSK over
%$(\Nx)/4$ independent subcarriers.  This is $3$bit more than BMOCZ and allows a maximal channel length of $L=\Nxdpsk+1$
%to add the cyclic prefix. Since $K$ has to be odd, it is no possible to match perfectly the spectral efficiency to
%BMOCZ.
%
The encoding of the DPSK is done relative to the first OFDM block $i=1$, which will transmit PSK constellation points as
 with phase zero $s_k^{(1)}=1$ respectively data phases $s_k^{(2)}=e^{\im 2 \pi q_k/Q}$ with
$q_k=bi2de(m_{(k-1)\log(Q)+1},\dots,m_{k\log (Q)})$ for $k=1,\dots,\Nxdpsk$.  Hence, in time domain, we obtain 
\begin{align}
  \vx=(\vxcpone, \vxcptwo),\quad
  \vx_{\cp}^{(i)}=(\cp^{(i)}, \vx^{(i)}),\quad
  \cp^{(i)}=(x_{\Ndpsk-L+1}^{(i)},\dots,x_{\Ndpsk-1}^{(i)}),\quad  \vx^{(i)}= \Fmatrixa
  \vs^{(i)}
\end{align}
for $i=1,2$.
Here $\vx$ will be also normalized. 
After removing the CP at the receiver the received data symbols in frequency domain via the $m$th antenna for the $k$th
carrier is given by
\begin{align}
  R_{m,k}^{(i)}=H_{m,k} s^{(i)}_k + W_{m,k}^{(i)}
\end{align}
where we consider $\{W_{m,k}^{(i)}\}$ as independent circularly symmetric Gaussian random variables and $H_{m,k}\in\C$
the channel coefficient of the $k$th subcarrier. Hence, each subcarrier can be seen as a Rayleigh flat fading channel
and we will use the decision variable  for a hard-decoding of $M$ antennas
%
%\begin{align}
%  \hat{s}_k = \min_{s\in[Q]} \sum_{m=1}^M 
%              \left|\arg\left( \frac{R^{(2)}_{m,k}}{R^{(1)}_{m,k}}\right)- \frac{2\pi s}{Q}\right|^2,
%\end{align}
%
\begin{align}
  \hat{q}_k = \argmin_{q\in[Q]} \left|\frac{1}{M}\sum_{m=1}^M 
  R^{(2)}_{m,k}\cc{R^{(1)}_{m,k}}- e^{\im \frac{2\pi q}{Q}}\right|^2
  = \text{int}\left(\frac{Q}{2\pi} \angle\left(\frac{1}{M}\sum_{m=1}^M R^{(2)}_{m,k}\cc{R^{(1)}_{m,k}}\right)
\!\!  \mod Q\right)
\end{align}
see for example \cite{BAE-HH17} (multiple users) or \cite[Sec.8.1]{Jaf05} (single antenna). Here $\text{int}(\cdot)$ rounds to
the nearest integer. We ignore here a possible weighting by knowledge of SNR.

\subsection{Simulations with Quadriga Channel Simulator}  

\begin{figure}[t]
  \centering 
\begin{subfigure}[b]{\textwidth}
\vspace{-2ex}
\centering
\includegraphics[width=\textwidth]{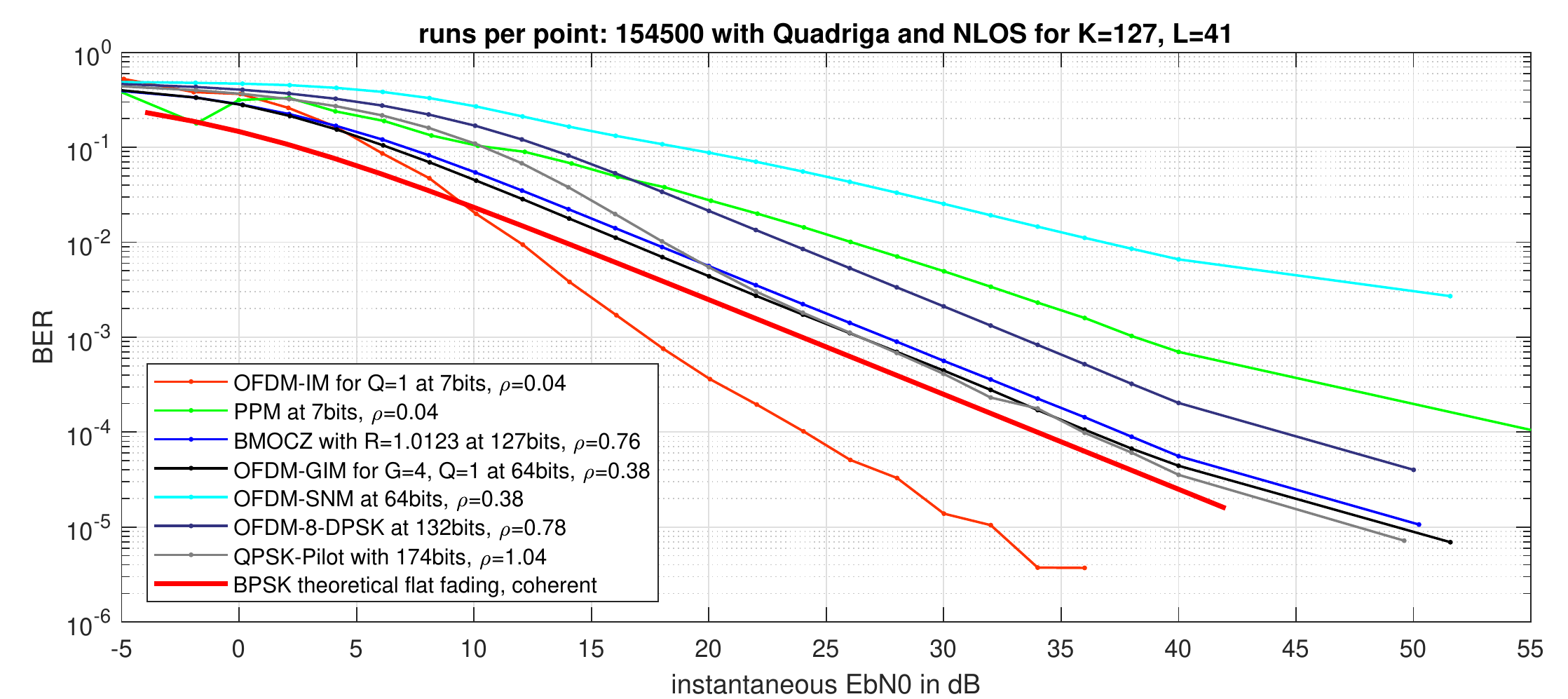}
    \vspace{-5ex}
  \caption{}\label{fig:quadrigaBER}
\end{subfigure}
\vspace{-3ex}
\hspace{2ex}
\begin{subfigure}[b]{\textwidth}
  \centering
  \includegraphics[width=0.8\textwidth]{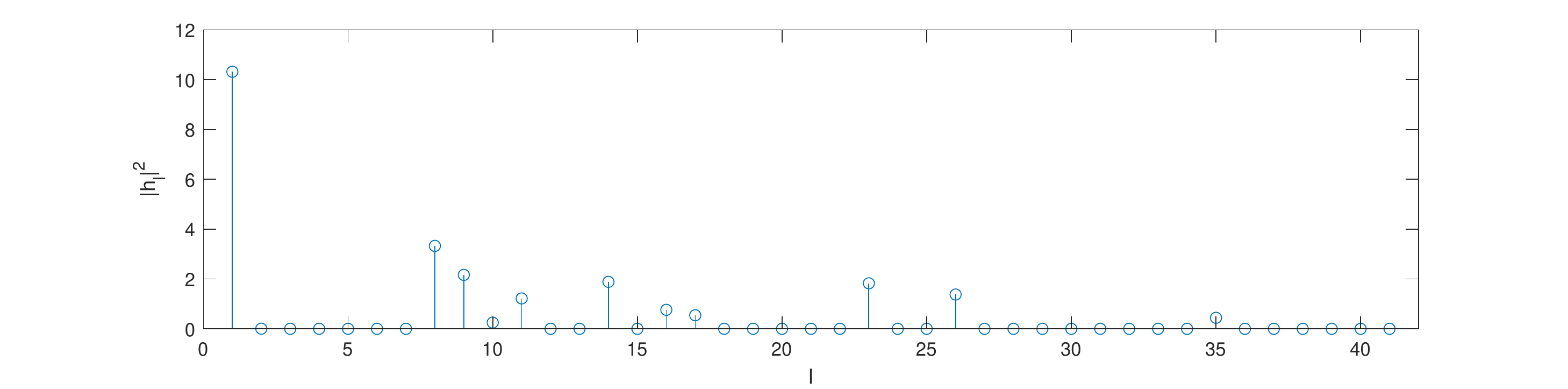}
  \vspace{-1ex}
  \caption{}
  \label{fig:quadrigaCIR}
\end{subfigure}
\caption{\subfigref{fig:quadrigaBER} BER of blind schemes over instantaneous SNR corrected by spectral efficiency 
and \subfigref{fig:quadrigaCIR} random channel realizations with Quadriga at maximal delay spread $L=41$.} \vspace{-2ex}
\end{figure}
We used the version 2.0 of the Quadriga channel simulator\footnote{Can be obtained from
  \url{http://quadriga-channel-model.de/}, see \cite{JRBT14}}
 \cite{JRBT14}, to generate random CIRs for the Berlin outdoor
 scenario (``BERLIN\_UMa\_NLOS''),
 with NLOS at a carrier frequency $f_c= 4$Ghz and bandwith $W=150$Mhz, see \figref{fig:quadrigaBER}.
 Transmitter and receiver are stationary using omnidirectional
 antennas ($\lambda/2$). The transmitter might be a base station mounted at $10$m
 altitude and the receiver might be a ground user with ground distance
 $20$m. The LOS distance is then $\approx 22$m.
%The $M=4$ receive antennas are $\lam/2$ separated on an array. The support of the
%CIR are the same, but not the realizations of the coefficients $h_l$. See \figref{fig:quadrigaCIR} for the PDP of a CIR
%realization.

%% file: OFDMDPSK_BMOCZ2.pdf_tex
%% Creator: Inkscape inkscape 0.92.1, www.inkscape.org
%% PDF/EPS/PS + LaTeX output extension by Johan Engelen, 2010
%% Accompanies image file 'OFDMDPSK_BMOCZ2.pdf' (pdf, eps, ps)
%%
%% To include the image in your LaTeX document, write
%%   \input{<filename>.pdf_tex}
%%  instead of
%%   \includegraphics{<filename>.pdf}
%% To scale the image, write
%%   \def\svgwidth{<desired width>}
%%   \input{<filename>.pdf_tex}
%%  instead of
%%   \includegraphics[width=<desired width>]{<filename>.pdf}
%%
%% Images with a different path to the parent latex file can
%% be accessed with the `import' package (which may need to be
%% installed) using
%%   \usepackage{import}
%% in the preamble, and then including the image with
%%   \import{<path to file>}{<filename>.pdf_tex}
%% Alternatively, one can specify
%%   \graphicspath{{<path to file>/}}
%% 
%% For more information, please see info/svg-inkscape on CTAN:
%%   http://tug.ctan.org/tex-archive/info/svg-inkscape
%%
\begingroup%
  \makeatletter%
  \providecommand\color[2][]{%
    \errmessage{(Inkscape) Color is used for the text in Inkscape, but the package 'color.sty' is not loaded}%
    \renewcommand\color[2][]{}%
  }%
  \providecommand\transparent[1]{%
    \errmessage{(Inkscape) Transparency is used (non-zero) for the text in Inkscape, but the package 'transparent.sty' is not loaded}%
    \renewcommand\transparent[1]{}%
  }%
  \providecommand\rotatebox[2]{#2}%
  \ifx\svgwidth\undefined%
    \setlength{\unitlength}{302.37763657bp}%
    \ifx\svgscale\undefined%
      \relax%
    \else%
      \setlength{\unitlength}{\unitlength * \real{\svgscale}}%
    \fi%
  \else%
    \setlength{\unitlength}{\svgwidth}%
  \fi%
  \global\let\svgwidth\undefined%
  \global\let\svgscale\undefined%
  \makeatother%
  \begin{picture}(1,0.33447055)%
    \put(0,0){\includegraphics[width=\unitlength,page=1]{OFDMDPSK_BMOCZ2.pdf}}%
    \put(0.21645064,0.22630495){\color[rgb]{0,0,0}\makebox(0,0)[lb]{\smash{$\pilot$}}}%
    \put(0,0){\includegraphics[width=\unitlength,page=2]{OFDMDPSK_BMOCZ2.pdf}}%
    \put(0.21855178,0.08027174){\color[rgb]{0,0,0}\makebox(0,0)[lb]{\smash{$\cpone$}}}%
    \put(0,0){\includegraphics[width=\unitlength,page=3]{OFDMDPSK_BMOCZ2.pdf}}%
    \put(0.52097357,0.08055722){\color[rgb]{0,0,0}\makebox(0,0)[lb]{\smash{$\cptwo$}}}%
    \put(0,0){\includegraphics[width=\unitlength,page=4]{OFDMDPSK_BMOCZ2.pdf}}%
    \put(0.19660348,0.00365724){\color[rgb]{0,0,0}\makebox(0,0)[lb]{\smash{$\tau_0$}}}%
    \put(0.80814073,0.00365724){\color[rgb]{0,0,0}\makebox(0,0)[lb]{\smash{$\tau_0\!+\!N\!-\!1$}}}%
    \put(0.23073223,0.02093479){\color[rgb]{0,0,0}\makebox(0,0)[lb]{\smash{$\!L\!-\!1$}}}%
    \put(0.37417957,0.02043514){\color[rgb]{0,0,0}\makebox(0,0)[lb]{\smash{$\Nxdpsk$}}}%
    \put(0,0){\includegraphics[width=\unitlength,page=5]{OFDMDPSK_BMOCZ2.pdf}}%
    \put(0.22062219,-0.99451599){\color[rgb]{0,0,0}\makebox(0,0)[lt]{\begin{minipage}{3.10842805\unitlength}\raggedright \end{minipage}}}%
    \put(-0.00143879,0.08195369){\color[rgb]{0,0,0}\makebox(0,0)[lb]{\smash{OFDM-DPSK}}}%
    \put(0,0){\includegraphics[width=\unitlength,page=6]{OFDMDPSK_BMOCZ2.pdf}}%
    \put(0.21748517,0.15279661){\color[rgb]{0,0,0}\makebox(0,0)[lb]{\smash{$\cp$}}}%
    \put(0,0){\includegraphics[width=\unitlength,page=7]{OFDMDPSK_BMOCZ2.pdf}}%
    \put(-0.00143879,0.15364036){\color[rgb]{0,0,0}\makebox(0,0)[lb]{\smash{OFDM-IM}}}%
    \put(0,0){\includegraphics[width=\unitlength,page=8]{OFDMDPSK_BMOCZ2.pdf}}%
    \put(-0.00217999,0.29701369){\color[rgb]{0,0,0}\makebox(0,0)[lb]{\smash{BMOCZ}}}%
    \put(0.96001902,0.04396013){\color[rgb]{0,0,0}\makebox(0,0)[lb]{\smash{$t$}}}%
    \put(0,0){\includegraphics[width=\unitlength,page=9]{OFDMDPSK_BMOCZ2.pdf}}%
    \put(0.73351789,0.29940212){\color[rgb]{0,0,0}\makebox(0,0)[lb]{\smash{$\guard$}}}%
    \put(0,0){\includegraphics[width=\unitlength,page=10]{OFDMDPSK_BMOCZ2.pdf}}%
    \put(-0.00127893,0.2259737){\color[rgb]{0,0,0}\makebox(0,0)[lb]{\smash{Pilot-QPSK}}}%
    \put(0.73354163,0.22864775){\color[rgb]{0,0,0}\makebox(0,0)[lb]{\smash{$\guard$}}}%
    \put(0,0){\includegraphics[width=\unitlength,page=11]{OFDMDPSK_BMOCZ2.pdf}}%
    \put(0.73980347,0.02093478){\color[rgb]{0,0,0}\makebox(0,0)[lb]{\smash{$L\!-\!1$}}}%
    \put(0,0){\includegraphics[width=\unitlength,page=12]{OFDMDPSK_BMOCZ2.pdf}}%
    \put(0.52918788,0.02093478){\color[rgb]{0,0,0}\makebox(0,0)[lb]{\smash{$L\!-\!1$}}}%
  \end{picture}%
\endgroup%